\documentclass[lettersize,journal]{IEEEtran}
\usepackage[noadjust]{cite}
\usepackage{srcltx}
\usepackage{graphicx}
\usepackage{epstopdf}
\usepackage{psfrag}
\usepackage{epsfig}
\usepackage[tbtags,sumlimits,nointlimits,reqno]{amsmath}
\usepackage{amssymb}
\usepackage{xcolor}
\usepackage{float}
\usepackage{bm}
\usepackage{subfigure}
\usepackage[most]{tcolorbox}
\usepackage{soul}
\usepackage{footnote}
\usepackage{amsmath}
\usepackage{color}

\usepackage[framemethod=TikZ]{mdframed}
\usepackage{lipsum}
\mdfdefinestyle{MyFrame}{%
	linecolor=blue,
	outerlinewidth=2pt,
	roundcorner=20pt,
	innertopmargin=\baselineskip,
	innerbottommargin=\baselineskip,
	innerrightmargin=20pt,
	innerleftmargin=20pt,
	backgroundcolor=gray!50!white}

\begin{document}
	\title{Designing Space-Time Metamaterials: \\The Central Role of Dispersion Engineering}
	\author{Sajjad Taravati,~\IEEEmembership{Senior Member,~IEEE}
		\thanks{S. Taravati is with the Faculty of Engineering and Physical Sciences, University of Southampton, Southampton SO17 1BJ, UK (e-mail: s.taravati@soton.ac.uk).}%
		\thanks{}}
	
	\markboth{}%
	{* \MakeLowercase{\textit{et al.}}: Bare Demo of IEEEtran.cls for IEEE Journals}

	\maketitle
	
	\begin{abstract}
Space-time metamaterials are redefining wave engineering by enabling fully dynamic four-dimensional control of electromagnetic fields, allowing simultaneous manipulation of frequency, amplitude, momentum, and propagation direction. This unified functionality moves well beyond reciprocity-breaking mechanisms, marking a fundamental transition from static media to polychromatic, energy-efficient wave processors. This article establishes dispersion engineering as the core design paradigm for these dynamic systems. We show that the dispersion relation, linking frequency and wavenumber, serves as a master blueprint governing exotic wave phenomena such as nonreciprocity, beam splitting, asymmetric frequency conversion, amplification, spatial decomposition, and momentum bandgaps. By analyzing analytical dispersion surfaces and isofrequency contours in subluminal, luminal, and superluminal modulation regimes, we reveal how tailored spatiotemporal modulation orchestrates controlled energy flow among harmonic modes. We further demonstrate how this framework directly informs practical device operation, highlighting advanced implementations including angular–frequency beam multiplexing in superconducting Josephson junction arrays. Combining insights from wave theory, numerical modeling, and experimental realization, this work provides a comprehensive roadmap for leveraging dispersion engineering to design next-generation metamaterials for wireless communication, quantum technologies, and integrated photonics.
	\end{abstract}
	
	
	\IEEEpeerreviewmaketitle
	
	\section{Introduction}\label{sec:introduction}
\IEEEPARstart{D}ispersion diagrams serve as the fundamental cartography for wave propagation in engineered media~\cite{Taravati_Kishk_MicMag_2019,Oliner_PIEEE_1963,Cassedy_PIEEE_1967,Chu1972,oudich2023tailoring}. By charting the relationship between frequency and wavevector, these diagrams unveil the core physics governing phase and group velocities, bandgaps, and the very modes a structure can support~\cite{chu1972wave,Taravati_PRAp_2018}. In static photonic crystals and metamaterials, dispersion analysis is a powerful diagnostic tool~\cite{caloz2011metamaterial,pu2019methodologies}. However, with the advent of space-time metamaterials, media whose properties are modulated periodically in both spatial and temporal domains—dispersion engineering transitions from a passive diagnostic to an active design principle. The ability to dynamically sculpt the dispersion landscape in four dimensions (frequency and three spatial momenta) unlocks unprecedented control over wave-matter interaction~\cite{taravati20234d}, enabling phenomena that defy conventional physical constraints, such as magnet-free nonreciprocity, parametric amplification, and nonreciprocal quantum transitions.

The field of linear space-time modulation has already demonstrated remarkable potential for revolutionizing electromagnetic systems~\cite{Taravati_Kishk_MicMag_2019,darvish2024modern,valizadeh2024analytical,taravati2024spatiotemporal,taravati2025finite,taravati2025light,das2025electrical}. By modulating material parameters like permittivity or conductivity in a traveling-wave fashion \cite{taravati20234d,amra2024linear,Taravati_NC_2021,tiukuvaara2021floquet}, researchers have realized nonreciprocal components \cite{Taravati_Kishk_TAP_2019,taravati2020full,wu2024analysis,taravati2024nonlinear,nagulu2024synthetic,boshgazi2025magnet}, frequency converters \cite{Tien_JAP_1958,Taravati_PRB_Mixer_2018,taravati2021pure,moreno2024space,taravati2024efficient,valizadeh2025integrated,taravati2025_entangle}, parametric amplification~\cite{Cullen_NAT_1958,tien1958parametric,Taravati_Kishk_PRB_2018,li2019nonreciprocal,pendry2020new,sumetsky2025transformation}, and advanced beam-steering, nonreciprocal and target recognition metasurfaces \cite{taravati2020full,Taravati_ACSP_2022,wang2023pseudorandom,wang2023pseudorandom,sisler2024electrically,phi2024controlled}, multifunctional antennas~\cite{Taravati_AMA_PRApp_2020,zang2024spatiotemporally,amini2025flexible,yang2025harmonic}, and quantum technologies~\cite{taravati2025space}.

This article is dedicated to establishing dispersion engineering as the central methodology for designing space-time metamaterials. We demonstrate that a deep understanding of dispersion and isofrequency contours, across subluminal, luminal, and superluminal modulation regimes, provides a universal framework for predicting and controlling wave phenomena. The article systematically explores how tailored spatiotemporal modulation orchestrates the flow of energy between harmonic modes, from foundational linear systems to the latest nonlinear and quantum applications. By synthesizing insights from analytical theory~\cite{cassedy1963dispersion,tamir1964wave,Fan_NPH_2009,Taravati_PRB_2017,Taravati_Kishk_TAP_2019,taravati_PRApp_2019,Taravati_AMA_PRApp_2020}, full-wave simulation~\cite{taravati2025finite}, and experimental realization~\cite{taravati2016mixer,Taravati_PRB_2017,Taravati_PRB_Mixer_2018,taravati2017self,taravati2020full}, this work provides a comprehensive roadmap for harnessing dispersion engineering to create next-generation devices for wireless communications, integrated photonics, and quantum information science.

This article is structured to guide the reader from fundamental principles to advanced applications. Section II delves into the theoretical implications, beginning with a general framework for linear periodic space-time metamaterials and their dispersion characteristics, followed by a detailed analysis of the ubiquitous sinusoidal modulation case. Section III presents a series of illustrative examples that showcase dispersion engineering in action. This includes achieving nonreciprocity in the sonic regime~\cite{Peng1969,Taravati_PRB_2017}, creating exotic phenomena in equilibrated media, and designing devices based on asymmetric photonic bandgaps, leaky-wave transitions, and phase-engineered time loops. This section also covers the realization of key components like isolators, nonreciprocal phase shifters, beam-steering metasurfaces, and circulators. The discussion then extends to more complex systems, including aperiodic metamaterials for pure frequency conversion and, ultimately, nonlinear systems. Section IV is dedicated to the frontier of nonlinear space-time metamaterials, focusing on their application in angular-frequency beam multiplexing and groundbreaking quantum effects such as nonreciprocal photon absorption and quantum state transfer. Finally, Section V concludes the paper by summarizing the findings and outlining future perspectives in this rapidly evolving field.

\section{Theoretical Implications}	
\subsection{General Periodic Linear Space-Time Metamaterials}

\begin{figure}
	\begin{center}
		\includegraphics[width=1\columnwidth]{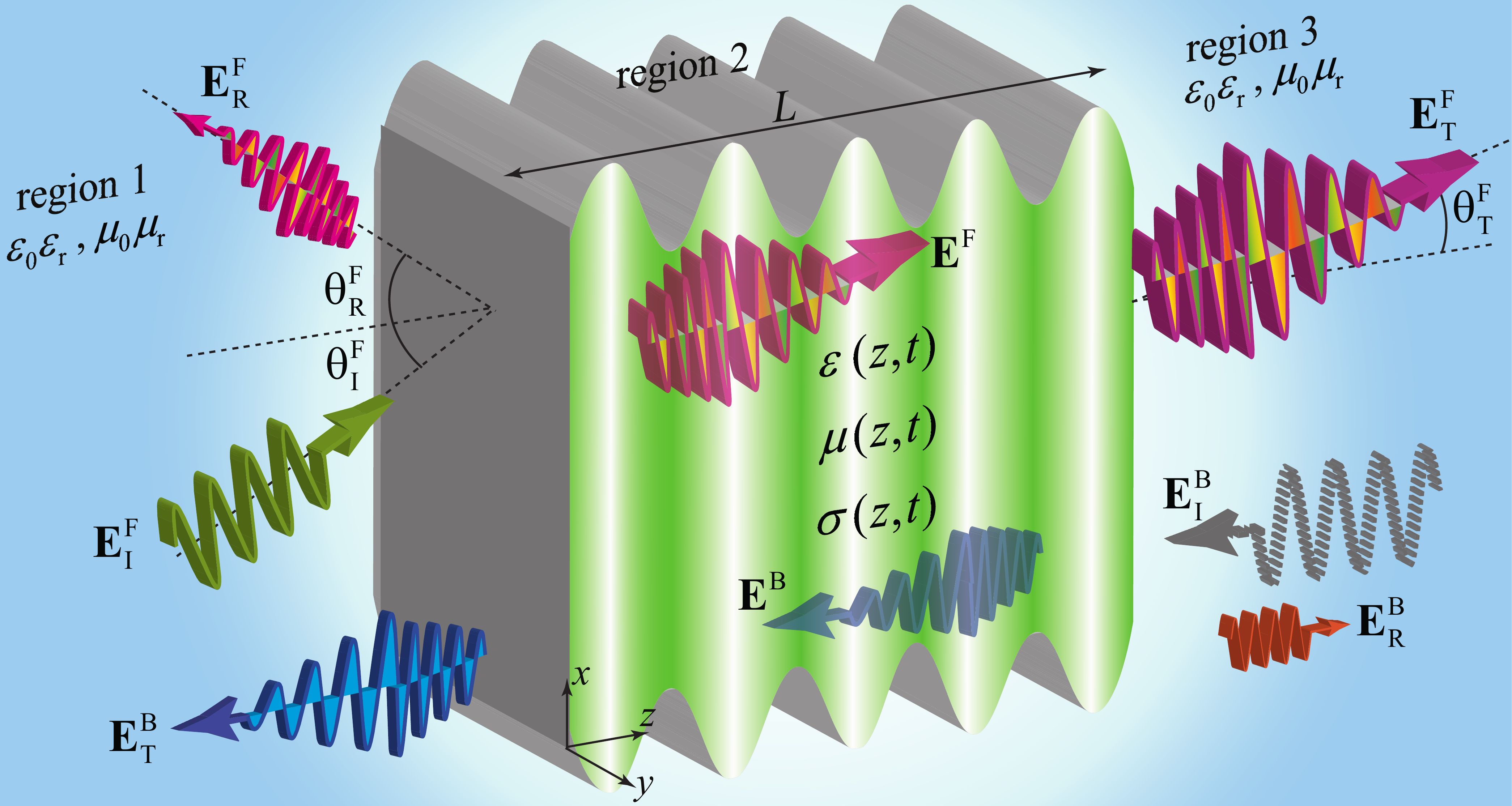}
		\caption{Conceptual schematic of electromagnetic waves transformation by a space-time metamaterial, illustrating key effects such as frequency conversion, nonreciprocal transmission, power amplification, and spatial wavefront decomposition.}
		\label{Fig:slab_sch}
	\end{center}
\end{figure}
Figure~\ref{Fig:slab_sch} illustrates a generic configuration of electromagnetic wave transmission and reflection through a periodic space–time–modulated metamaterial slab of thickness $L$, characterized by spatially and temporally varying electric permittivity $\epsilon(z,t)$, magnetic permeability $\mu(z,t)$, and electric conductivity $\sigma(z,t)$. The slab is bounded by two semi-infinite, homogeneous, and unmodulated media. Here, the problem with the incident wave propagating towards the $+z$-direction, shown in the top of Fig.~\ref{Fig:slab_sch}, is called the forward problem ``F'', whereas the problem with the incident wave propagating towards the $-z$-direction, shown in the bottom of Fig.~\ref{Fig:slab_sch}, is called the backward problem ``B''. The space-time modulation is characterized by the temporal frequency $\omega_\text{m}$ and spatial frequency $\beta_\text{m}$. The modulation phase velocity, i.e., $v_\text{m}=\omega_\text{m}/\beta_\text{m}$, may be smaller or greater than the phase velocity of the background medium, $v_\text{b}=c/\sqrt{\epsilon_\text{r}\mu_\text{r}}$, with $c=1/\sqrt{\mu_0 \epsilon_\text{0}}$ being the speed of light in vacuum. The ratio between the modulation and background phase velocities, i.e. $\gamma=v_\text{m}/v_\text{b}$, represents the space-time velocity ratio~\cite{Taravati_PRB_2017}. As a consequence, the pump wave spatial frequency reads
\begin{equation}
	\beta_\text{m}=\frac{\omega_\text{m} \sqrt{\epsilon_\text{r}\mu_\text{r}}}{c \gamma}.
	\label{eq:beta_m}
\end{equation}

The constitutive parameters $\epsilon(z,t)$, $\mu(z,t)$, and $\sigma(z,t)$ may be expanded in a space--time Fourier series as
\begin{subequations}\label{eqa:constitutive_Fourier}
	\begin{align}
		\epsilon(z,t) &= \sum_{m=-M}^{M} \sum_{n=-N}^{N} \epsilon_{m,n}\,
		e^{\,j(m\beta_\text{m}z - n\omega_\text{m} t)}, \\
		\mu(z,t) &= \sum_{m=-M}^{M} \sum_{n=-N}^{N} \mu_{m,n}\,
		e^{\,j(m\beta_\text{m}z - n\omega_\text{m} t)}, \\
		\sigma(z,t) &= \sum_{m=-M}^{M} \sum_{n=-N}^{N} \sigma_{m,n}\,
		e^{\,j(m\beta_\text{m}z - n\omega_\text{m} t)},
	\end{align}
\end{subequations}
where $\beta_\text{m}=2\pi/\Lambda_\text{m}$ and $\omega_\text{m}=2\pi/T_\text{m}$ denote the spatial and temporal modulation frequencies, respectively. The coefficients
$\epsilon_{m,n}$, $\mu_{m,n}$, and $\sigma_{m,n}$ are given by the Fourier integrals
\begin{subequations}\label{eqa:Fourier_coeffs_def}
	\begin{align}
		\epsilon_{m,n} &= \frac{1}{\Lambda_\text{m} T_\text{m}}
		\int_{0}^{T_\text{m}}\!\int_{0}^{\Lambda_\text{m}} \epsilon(z,t)\,
		e^{-j(m\beta_\text{m}z - n\omega_\text{m} t)}\,dz\,dt, \\
		\mu_{m,n} &= \frac{1}{\Lambda_\text{m} T_\text{m}}
		\int_{0}^{T_\text{m}}\!\int_{0}^{\Lambda_\text{m}} \mu(z,t)\,
		e^{-j(m\beta_\text{m}z - n\omega_\text{m} t)}\,dz\,dt, \\
		\sigma_{m,n} &= \frac{1}{\Lambda_\text{m} T_\text{m}}
		\int_{0}^{T_\text{m}}\!\int_{0}^{\Lambda_\text{m}} \sigma(z,t)\,
		e^{-j(m\beta_\text{m}z - n\omega_\text{m} t)}\,dz\,dt.
	\end{align}
\end{subequations}

For purely temporal modulation (no spatial dependence), these reduce to the one-dimensional expansions
$\epsilon(t) = \sum_{n}\epsilon_{n} e^{-j n\Omega t}$,
and likewise for $\mu(t)$ and $\sigma(t)$. We consider general oblique incidence of a $y$-polarized electric filed under angle of incidence of $\theta_{\text{i}}$ to the slab, as

\begin{equation}\label{eqa:Ei}
	\mathbf{E}_\text{I} (x,z,t)= \mathbf{\hat{y}} E_0 e^{i\left( k_x x +\beta_{z,0} z- \omega _0 t  \right) },
\end{equation}
where $E_0$ is the amplitude of the incident wave, and $\omega_0$ and $k_\text{I}=\sqrt{k_x^2 +\beta_{z,0}^2}$ are respectively temporal spatial frequencies of the incident wave. 

Given the temporal periodicity of the slab, the electric field inside the slab may be represented based on temporal Bloch-Floquet decomposition as
\begin{subequations}
	\begin{equation}\label{eqa:Em}
		\begin{split}
			\mathbf{E} (x,z,t)&=\mathbf{\hat{y}} \sum_{n=-\infty}^{\infty} \textbf{E}_{n}(x,z) e^{-i \omega_{n} t},
		\end{split}
	\end{equation}
	where $\omega_{n}=\omega_{0}+n\Omega$, and 
	\begin{equation}\label{eqa:Fourier-E}
		\textbf{E}_n(x,z)= \sum_{m=-M}^{M} \textbf{E}_{mn} \exp[i(k_x x+m\varphi z/L)],
	\end{equation}
\end{subequations}
where $\textbf{E}_{mn}$ is the unknown electric field coefficient matrix, $\varphi$ is the unknown spatial frequency between $0$ and $2\pi$, to be known given the spatial variation of the slab, $k_x=k_0 \sin(\theta_{\text{i}})=\omega_0 \sqrt{\epsilon_\text{av} \mu_\text{av}}\sin(\theta_{\text{i}})/c$, with $\epsilon_\text{av}$ and $\mu_\text{av}$ being the average permittivity and permeability of the slab. The magnetic field inside the slab may be represented as
\begin{subequations}
		\begin{equation}\label{eqa:Hm}
			\begin{split}
				\mathbf{H} (x,&z,t)=\dfrac{1}{\eta} \mathbf{\hat{k}} \times\mathbf{E} (x,z,t)\\
				&=\dfrac{1}{\eta} \left( \mathbf{\hat{x}} k_{x} +\mathbf{\hat{z}} \beta_{z,n}  \right) \times\mathbf{E} (x,z,t)\\
				&=\dfrac{1}{k_n} \sum_{n=-\infty}^{\infty} \left( -\mathbf{\hat{x}} \beta_{z,n} +\mathbf{\hat{z}} k_{x} \right) \textbf{H}_{n}(x,z) e^{-i \omega_{n} t}.
			\end{split}
		\end{equation}
	where the spatially variant magnetic field coefficient $\textbf{H}_{n}(x,z)$ reads
	\begin{equation}\label{eqa:Fourier-H}
		\textbf{H}_n(x,z)= \sum_{m=-M}^{M} \textbf{{H}}_{mn} \exp[i(k_x x+m\varphi z/L)].
	\end{equation}

The wavenumber $k_n$ in~\eqref{eqa:Hm} is represented by
\begin{equation}\label{eqa:k_n}
	k_n=\sqrt{k_{x}^2+\beta_{z,n}^2},
\end{equation}
\end{subequations}
and $\beta_{z,n}=\beta_{z,0}+n \beta_\text{m}$.
The unknown field coefficient matrices, $\textbf{E}_{mn}$, $\textbf{H}_{mn}$, and $\beta_{z,n}$ are to be found through satisfying Maxwell equations, so that
\begin{subequations}
	\begin{equation}\label{eqa:Max1}
		\nabla\times\textbf{E} (x,z,t)=-\dfrac{\partial [\mu(z,t) \textbf{H} (x,z,t)]}{\partial t},
	\end{equation}
	\begin{equation}\label{eqa:Max2}
			\nabla\times\textbf{H} (x,z,t)
		=\dfrac{\partial [\epsilon (z,t)\textbf{E}(x,z,t)]}{\partial t}
		+ \sigma(z,t)\textbf{E}(x,z,t),
	\end{equation}
\end{subequations}
 yielding
\begin{subequations}\label{eqa:coupled_equations}
	\begin{equation}\label{eqa:matrcoup1}
		\textbf{E}(x,z)=\textbf{Z}(z) \textbf{H}(x,z),
	\end{equation}
and
	\begin{equation}\label{eqa:matrcoup2}
		\textbf{H}(x,z) =\textbf{Y}(z) \textbf{E}(x,z),
	\end{equation}
where the Fourier coefficients of the constitutive coupling read
	\begin{equation}\label{eqa:Y_coeff_compact}
		\mathbf{Y}_m^{(n)} \;=\; \dfrac{\omega_n}{k_n}\!\left(\epsilon_m+\dfrac{\sigma_m}{j\omega_n}\right),
	\end{equation}
	\begin{equation}\label{eqa:Z_coeff_compact}
		\mathbf{Z}_m^{(n)} \;=\; \dfrac{\omega_n}{k_n}\,\mu_m,
	\end{equation}
\end{subequations}
with $\epsilon_m,\mu_m,\sigma_m$ being the spatial Fourier coefficients of $\epsilon(z,t),\mu(z,t),\sigma(z,t)$ for the considered temporal harmonic and where the superscript \((n)\) on $\mathbf{Y}_m^{(n)}$ emphasizes the harmonic dependence due to the $1/\omega_n$ factor.  

Equations~\eqref{eqa:matrcoup1} and~\eqref{eqa:matrcoup2} form the coupled matrix equation of the general STM medium in Fig. 1. To solve this coupled matrix equation, we express the aperiodic/periodic spatially-variant $Z(z)$ and $Y(z)$ matrices based on the series expansion as 
\begin{subequations}\label{eqa:Fourier-ZY}
	\begin{equation}\label{eqa:Fourier-Z}
		\textbf{Z}(z)=\sum_{m=-M}^{M} \textbf{Z}_m \exp[(im\varphi z)/L],
	\end{equation}
	\begin{equation}\label{eqa:Fourier-Y}
		\textbf{Y}(z)=\sum_{m=-M}^{M} \textbf{Y}_m \exp[(im\varphi z)/L],
	\end{equation}
\end{subequations}
where $\textbf{Z}_m$ and $\textbf{Y}_m$ are unknown coefficients to be determined given the spatial variation of the slab. Inserting the series expansions of the electric and magnetic fields, in~\eqref{eqa:Fourier-E} and ~\eqref{eqa:Fourier-H}, and those of the permittivity and permeability in~\eqref{eqa:Fourier-ZY}, into the coupled matrix equation in~\eqref{eqa:matrcoup1} and~\eqref{eqa:matrcoup2}, yields
\begin{subequations}\label{eqa:ST_inv_eq}
	\begin{equation}\label{eqa:Fourier-Em+1}
		\textbf{E}_{m} = \bm{L}_m \sum_{p=-M}^{M} \textbf{Z}_{m-p} \textbf{H}_p,
	\end{equation}
	\begin{equation}\label{eqa:Fourier-Hm+1}
		\textbf{H}_{m} = \bm{L}_m \sum_{p=-M}^{M} \textbf{Y}_{m-p} \textbf{E}_p 
	\end{equation}
	where
	\begin{equation}\label{eqa:Fourier-lambda}
		\bm{L}_m = 
		\text{diag} \left\lbrace  \left[ \left(\kappa_{-N}+im\varphi/L \right) ^{-1}
		\cdots  \left(\kappa_{N}+im\varphi/L \right) ^{-1} \right] \right\rbrace,
	\end{equation}		
\end{subequations}
for $m=-M,\cdots,0,\cdots,M$. 

The coupled ST-invariant matrix equation in~\eqref{eqa:ST_inv_eq} may be expressed as
\begin{subequations}
	\begin{equation}\label{eqa:cc}
		\overrightarrow{\textbf{E}}_m =\overrightarrow{\bm{L}}_m  \overrightarrow{\textbf{Z}} \overrightarrow{\textbf{H}}_m,
	\end{equation}
	\begin{equation}\label{eqa:cc2}
		\overrightarrow{\textbf{H}}_m =\overrightarrow{\bm{L}}_m  \overrightarrow{\textbf{Y}} \overrightarrow{\textbf{E}}_m.
	\end{equation}

Inserting Eq.~\eqref{eqa:cc2} into~\eqref{eqa:cc}, yields
\begin{equation}\label{eqa:cc6}
	\left[ \textbf{A} \right]  \overrightarrow{\textbf{E}_m}=0,
\end{equation}
where
\begin{equation}\label{eqa:A}
\textbf{K} =\left[ \overrightarrow{\bm{L}}_m  \overrightarrow{\textbf{Z}} \overrightarrow{\bm{L}}_m  \overrightarrow{\textbf{Y}}
- \textbf{I} \right], 
\end{equation}
where $\textbf{I}$ is the identity matrix. Equation~\eqref{eqa:cc6} represents the wave equation inside the general slab, providing the electric field coefficients through satisfying initial condition at $z=0$, i.e.,
\begin{equation}\label{eqa:}
	\textbf{E}(0)=\sum_{m=-M}^{M} \textbf{E}_m =\textbf{E}_\text{0}.
\end{equation}

After computing the electric field coefficients using~\eqref{eqa:cc6}, those of the magnetic field may be achieved using~\eqref{eqa:cc2}. The Equation~\eqref{eqa:cc6} has nontrivial solution if  
\begin{equation}\label{eqa:disp_rel}
	\text{det} \left[\textbf{K} \right]   =0. 
\end{equation}
\end{subequations}

Equation~\eqref{eqa:disp_rel} represents the dispersion relation of the unbounded general medium.

\subsection{Sinusoidal Linear Space-Time Metamaterials}	
We next consider a sinusoidal forward space-time permittivity as a particular case of the general periodic permittivity in~\eqref{eqa:constitutive_Fourier}, namely~\cite{Oliner_1959,Oliner_PIEEE_1963,Cassedy_PIEEE_1965,Cassedy_PIEEE_1967,Taravati_PRB_2017,Taravati_thesis}
\begin{equation}
	\epsilon(z,t)= \epsilon_\text{r}+\epsilon_\text{m} ~\cos(\beta_\text{m}z-\omega_\text{m}t).
	\label{eqa:sin_perm}
\end{equation}

The sonic interval associated with the sinusoidal permittivity in~\eqref{eqa:sin_perm} is obtained as~\cite{Taravati_PRB_2017}
\begin{equation}\label{eqa:sonic_sin}
	\gamma_\text{s,min}
	=\frac{1}{\sqrt{1+\epsilon_\text{m}/\epsilon_\text{r}}}  \leq \gamma \leq \frac{1}{\sqrt{1-\epsilon_\text{m}/\epsilon_\text{r}}}
	=\gamma_\text{s,max},
\end{equation}
where it is understood that $|\epsilon_\text{m} \cos(\beta_\text{m}z-\omega_\text{m}t)| \leq \epsilon_\text{m}$. For a given set
of modulation parameters ($\epsilon_\text{m}$, $\epsilon_\text{r}$, $\omega_\text{m}$, $\beta_\text{m}$, $\gamma$) and variable
$\omega_0$, the periodic dispersion diagram ($\beta_{0p}$'s) of the system can be constructed using, we find the following analytic form for the dispersion relation of the slab
\begin{subequations}
\begin{equation}
	\begin{split}
		 \frac{1}{ - K_{-1}^\pm + \frac{1}{K_{-2}^\pm+ \frac{1}{ - K_{-3}^\pm +\ldots  } }}+K_{0}^\pm 
		 + \frac{1}{ - K_{1}^\pm + \frac{1}{K_{2}^\pm+ \frac{1}{ - K_{3}^\pm + \ldots} }}=0.
	\end{split}
	\label{eqa:dispers_eq}
\end{equation}
where
\begin{equation}
	K_{n}^\pm=\frac{2 \epsilon_\text{r}}{\epsilon_\text{m}}  \left[1- \left( \frac{ k_x^2+ (\beta_{0}^\pm \pm n \beta_\text{m})^2  }{\left[(\omega_0+n\omega_\text{m})/v_\text{b} \right]^2 }\right)\delta_{nn} \right].
	\label{eqa:K_np}
\end{equation}
\end{subequations}

\begin{figure*}
	\begin{center}
				\subfigure[]{\label{Fig:3D_dispers}
		\includegraphics[width=1.2\columnwidth]{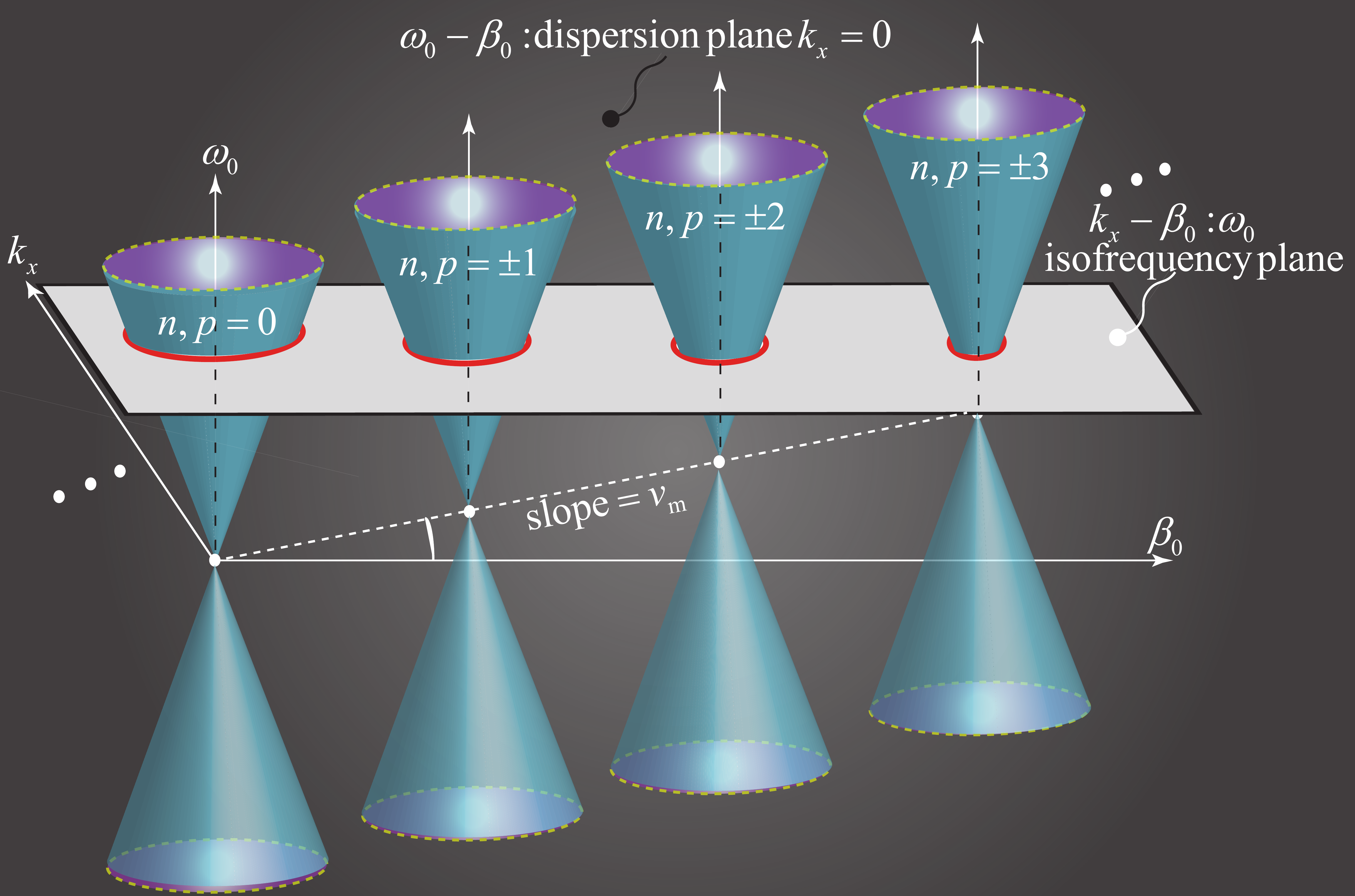}}
						\subfigure[]{\label{Fig:Anal_Disp}
			\includegraphics[width=0.75\columnwidth]{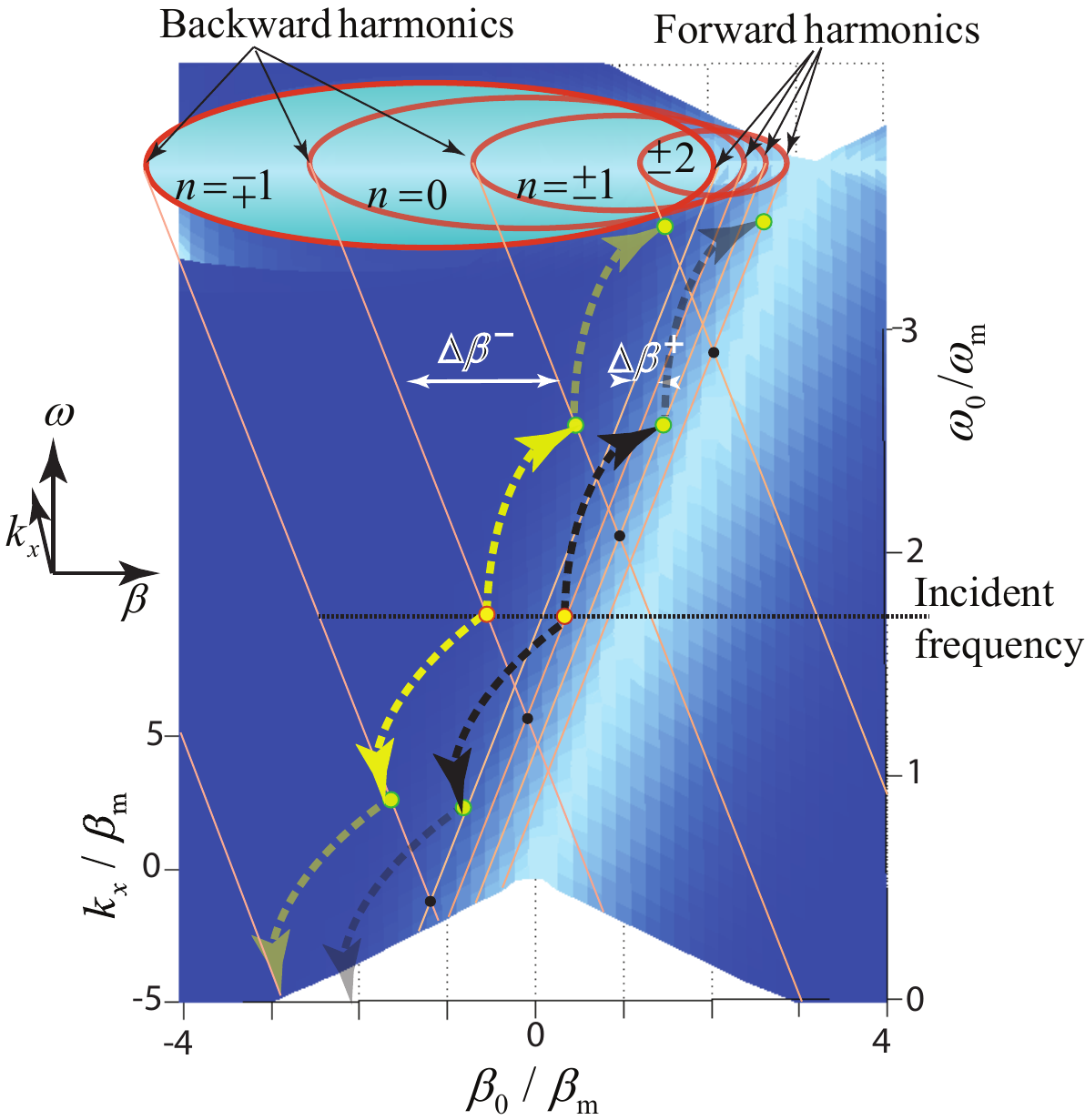}} 	
		\caption{Three-dimensional ($\omega,\beta,k_x$) dispersion diagram of space-time metamaterials. (a)~Visualization of the complete dispersion relationship, showing how conventional 2D diagrams are derived from the 3D dispersion volume of a space-time modulated medium. The band structure is a vertical cut at \(k_x=0\), while the isofrequency contour for excitation \(\omega_0\) is a horizontal cut, together mapping wave propagation in the orthogonal \((\omega, \beta, k_x)\) coordinate system~\cite{Taravati_PRB_2017}. (b)~Analytical dispersion diagram in the subsonic regime (\(\gamma = 0.85\)), showing strong coupling to forward harmonics and weak coupling to backward harmonics~\cite{Taravati_Kishk_TAP_2019}.}
		\label{Fig:3D_dispersion}
	\end{center}
\end{figure*}

In order to gain deeper insight into the wave propagation phenomenology within the space-time modulated slab medium, we next study the dispersion and isofrequency diagrams of the corresponding unbounded medium. In the limiting case of a vanishingly small modulation depth, $\epsilon_\text{m}\rightarrow 0$, the aforementioned equations lead to the closed-form dispersion relation
\begin{equation}
	\left(\frac{k_x}{\beta_\text{m}}\right)^2+\left(\frac{\beta_0^\pm }{\beta_\text{m}} \pm  n \right)^2
	= \gamma^2 \left(\frac{\omega_0}{\omega_\text{m}} +n \right)^2,
	\label{eqa:KKKK2}
\end{equation}
\noindent which represents an infinite periodic set of double cones with apexes at $k_x=0$ and $\beta_0=\pm n \beta_\text{m}$ and slope $v_\text{m}$, as illustrated in Figs.~\ref{Fig:3D_dispers} and~\ref{Fig:Anal_Disp}. A vertical cross section of this 3D diagram at $k_x=0$ produces an infinite periodic set of straight lines in the $\beta_0/\beta_\text{m}-\omega_0/\omega_\text{m}$ plane, and a horizontal cut produces an infinite periodic set of circles centered at $(\beta_0^\pm/\beta_\text{m},k_x/\beta_\text{m})=(\mp n,0)$ with radius $\gamma \left(\omega_0/\omega_\text{m} +n \right)$ in the $\beta_0/\beta_\text{m}-k_x/\beta_\text{m}$ plane.

\begin{figure}
	\begin{center}
	\subfigure[]{\label{Fig:Pa}		
			\includegraphics[width=0.6\columnwidth]{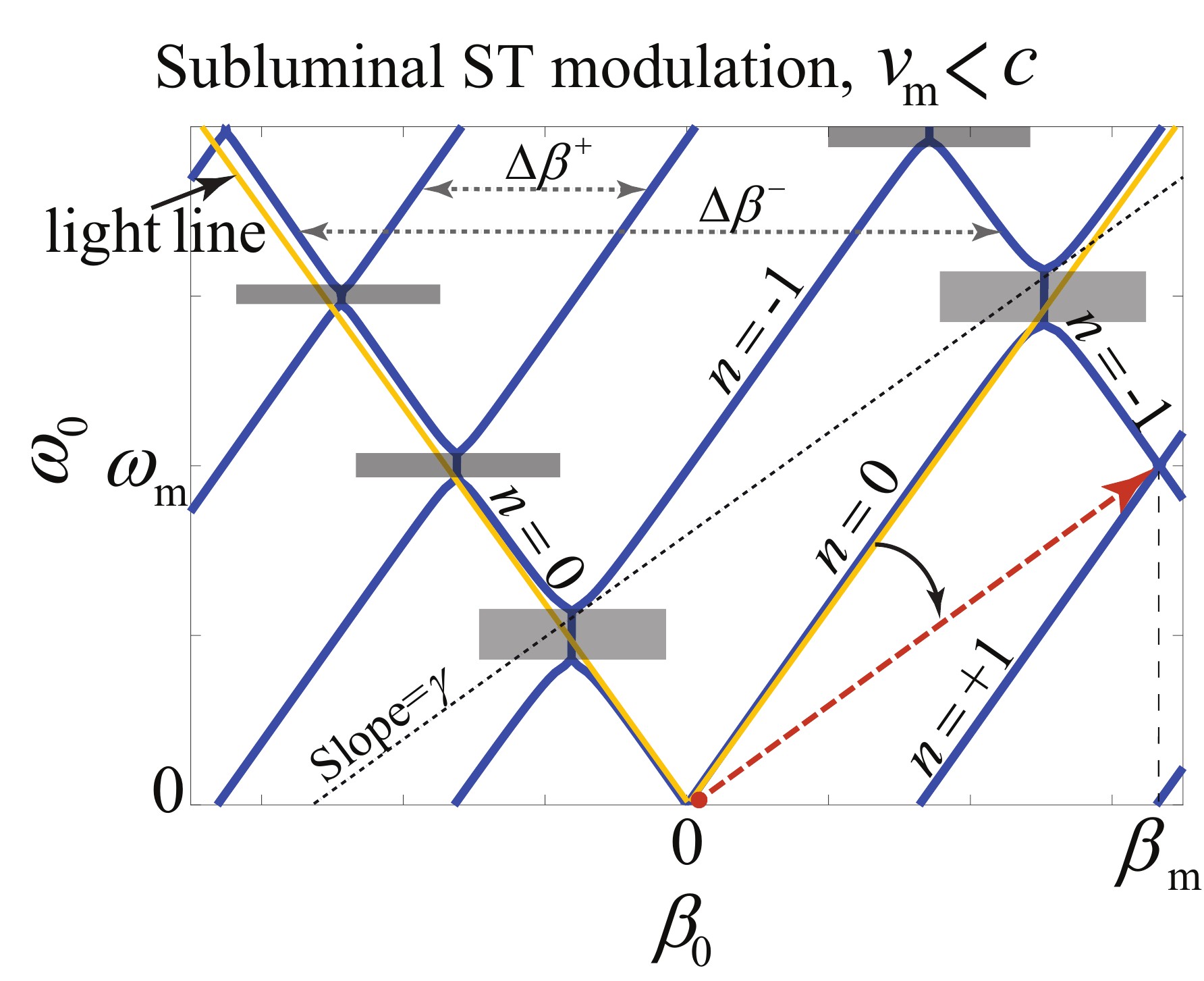}}
	\subfigure[]{\label{Fig:Pb}		
		\includegraphics[width=0.6\columnwidth]{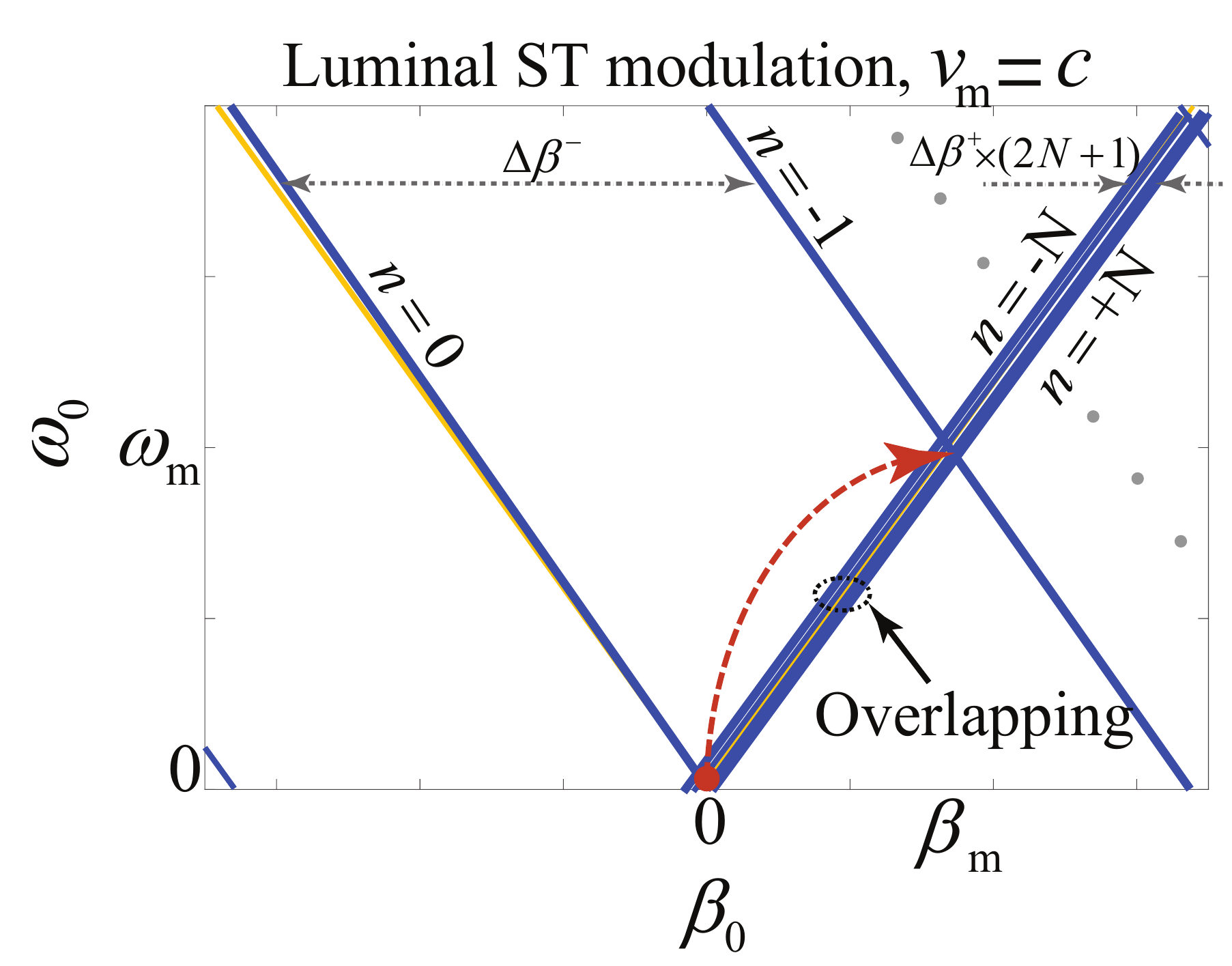}}
	\subfigure[]{\label{Fig:Pc}		
		\includegraphics[width=0.6\columnwidth]{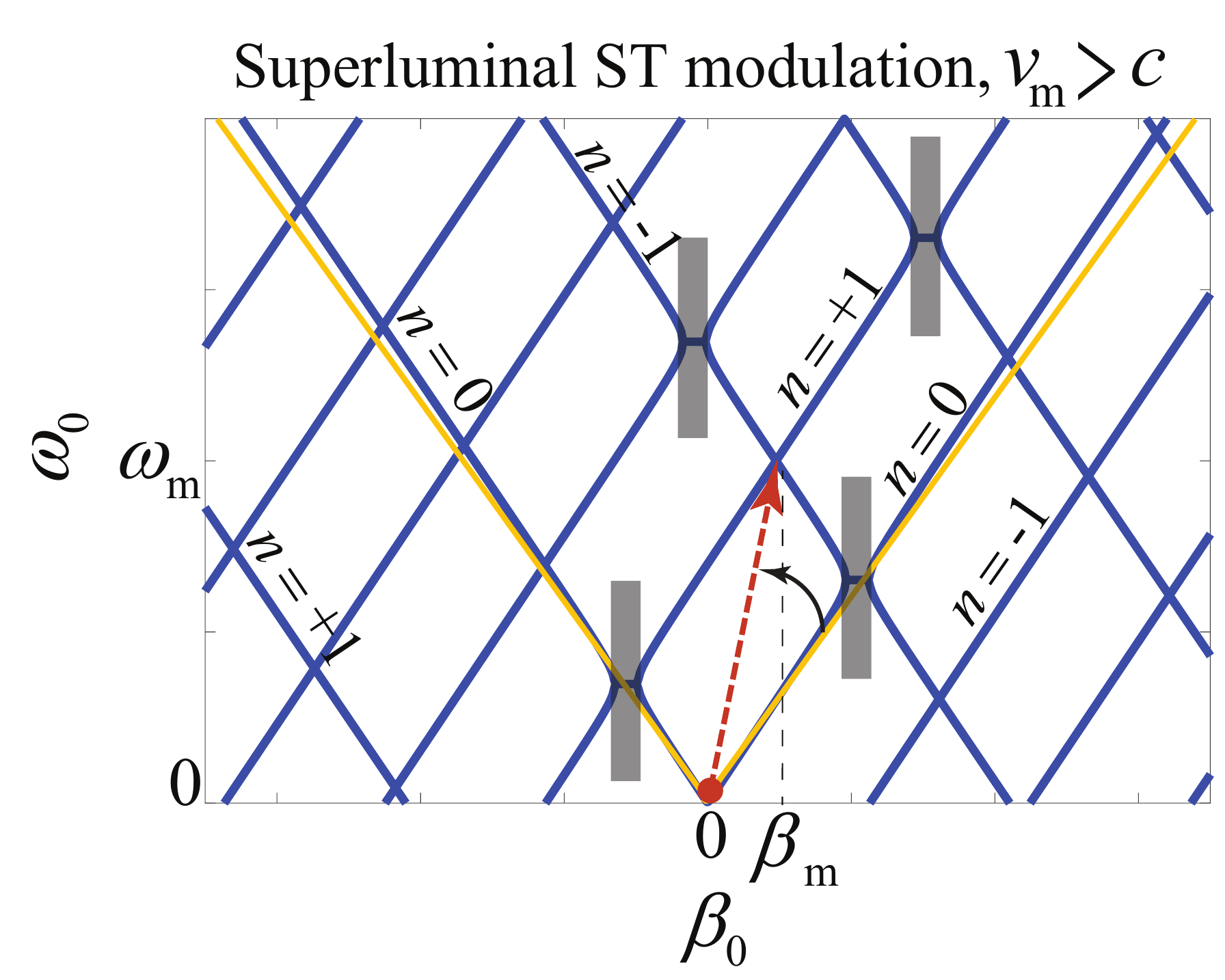}}
		\caption{Analytical ($\omega,\beta$) dispersion diagrams for an unbounded medium with sinusoidal space-time modulation, for (a)~subluminal (\(\gamma = 0.5\)), (b)~luminal (\(\gamma \rightarrow 1\)), and (c)~superluminal (\(\gamma = 3.5\)), for a modulation strength of \(\delta_{\epsilon} = 0.4\) and \(\delta_{\mu} = 0\)~\cite{Taravati_Kishk_MicMag_2019}.} 
		\label{fig:P}
	\end{center}
\end{figure}

Figures~\ref{fig:P}(a) to~\ref{fig:P}(c) present the analytical dispersion diagrams \( (\omega, \beta) \) for an unbounded medium under sinusoidal space-time modulation, contrasting three distinct modulation velocity regimes defined by the parameter \( \gamma = v_m / v_p \), where \( v_m \) is the modulation velocity and \( v_p \) is the phase velocity of the unmodulated medium. In the subluminal modulation (\( \gamma = 0.5 \)) in Fig.~\ref{fig:P}(a), where the modulation travels slower than the wave's natural phase velocity, the dispersion diagram exhibits a series of distinct, well-separated bands. The bandgap openings at the Brillouin zone boundaries are a hallmark of phase-matched interaction between the fundamental wave and its space-time harmonics. This structure is characteristic of a conventional photonic crystal but translated into the space-time domain, leading to phenomena like nonreciprocal bandgaps. In the luminal modulation (\( \gamma \to 1 \)) in Fig.~\ref{fig:P}(b), as the modulation velocity approaches the wave's phase velocity, a dramatic transformation occurs. The dispersion curves coalesce and flatten, indicating a vanishing group velocity \( (v_g = d\omega/d\beta \to 0) \). This 'slow-light' condition is associated with a resonant enhancement of wave-modulation interaction. The diagram suggests the onset of a momentum bandgap, where wave propagation is prohibited for a range of \( \beta \) values, and is critical for achieving strong nonreciprocal effects and parametric amplification. In the superluminal modulation (\( \gamma = 3.5 \)) in Fig.~\ref{fig:P}(c), when the modulation propagates faster than the wave, the dispersion diagram transitions into a complex, wavy structure that is unbounded in \( \beta \). This indicates that for a given frequency \( \omega \), an infinite set of wavenumbers \( \beta \) are supported, a signature of complex Fresnel-Floquet scattering. This regime is essential for realizing unique phenomena such as photon acceleration, frequency comb generation, and exotic nonlocal scattering effects.

\begin{figure}
	\begin{center}
		\subfigure[]{\label{Fig:2D_dispesion_curves_a}
			\includegraphics[width=0.9\columnwidth]{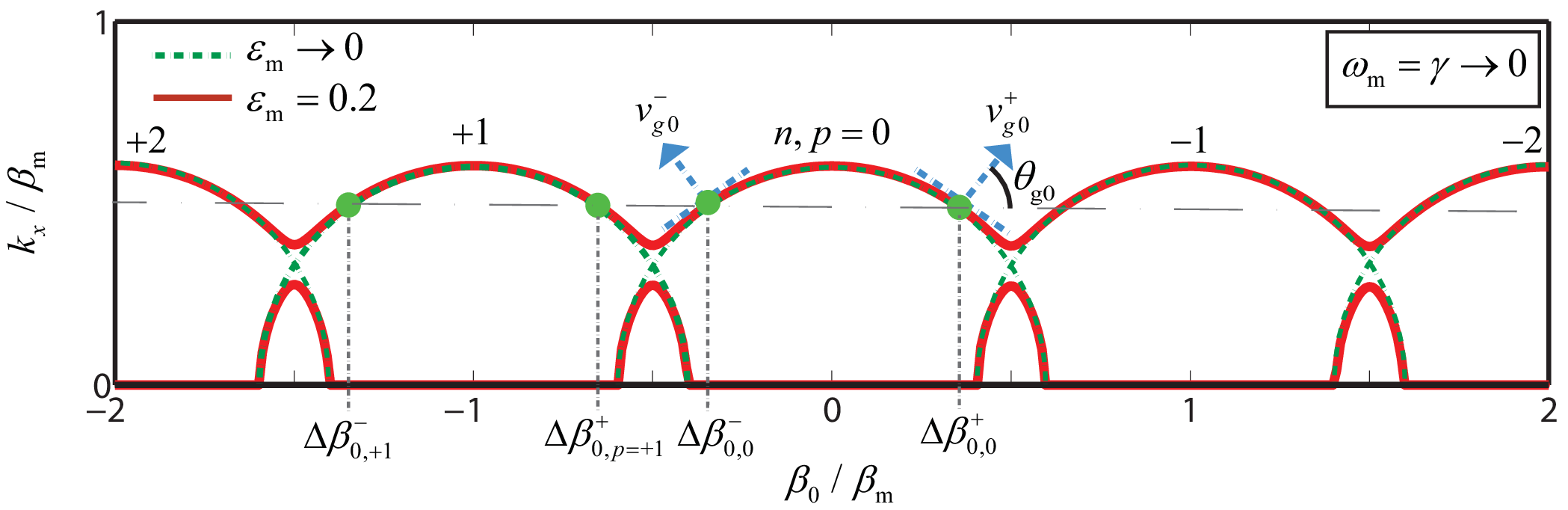}}
		\subfigure[]{\label{Fig:2D_dispesion_curves_b}
			\includegraphics[width=0.9\columnwidth]{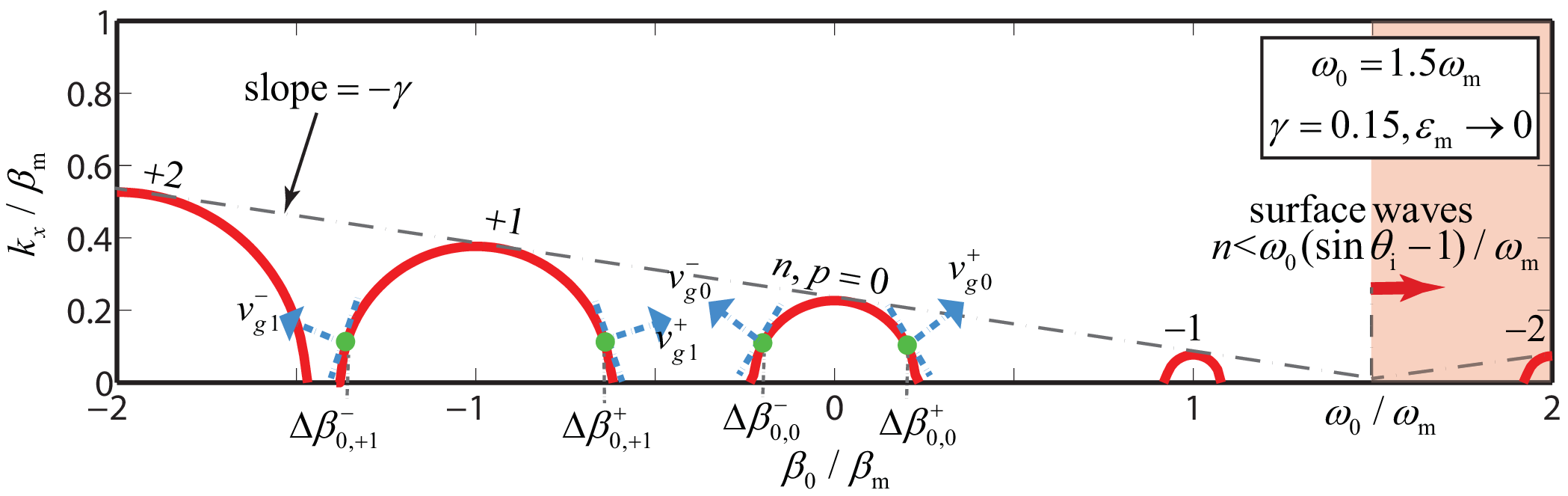}}
		\subfigure[]{\label{Fig:2D_dispesion_curves_c}
			\includegraphics[width=0.9\columnwidth]{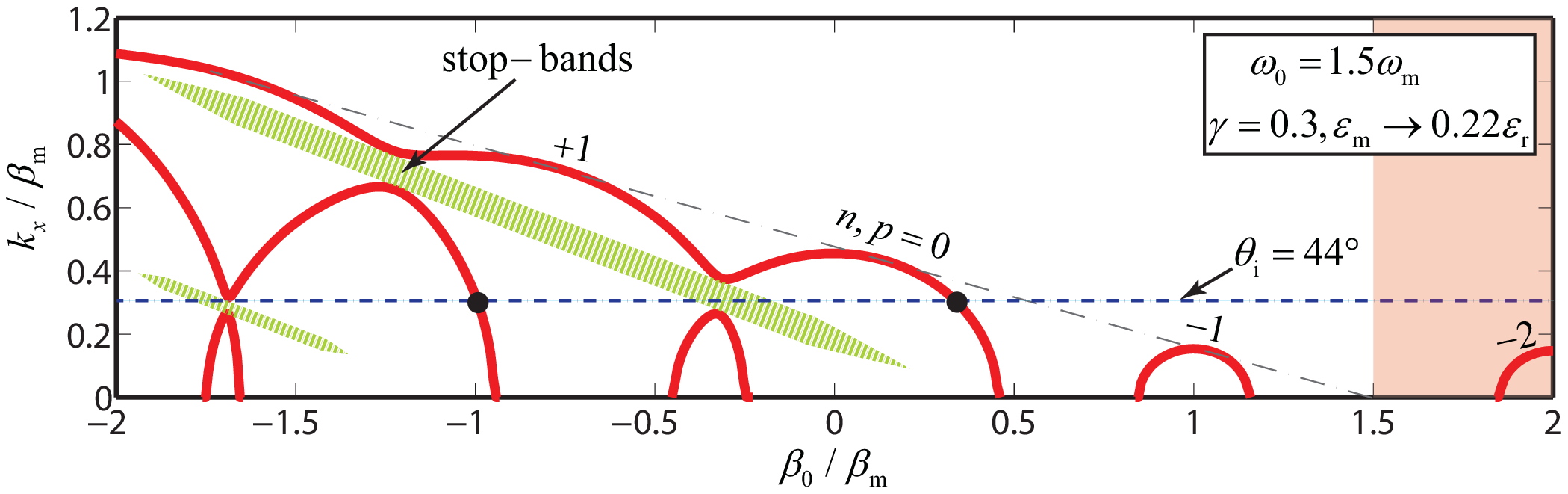}}
		\subfigure[]{\label{Fig:2D_dispesion_curves_d}
			\includegraphics[width=0.9\columnwidth]{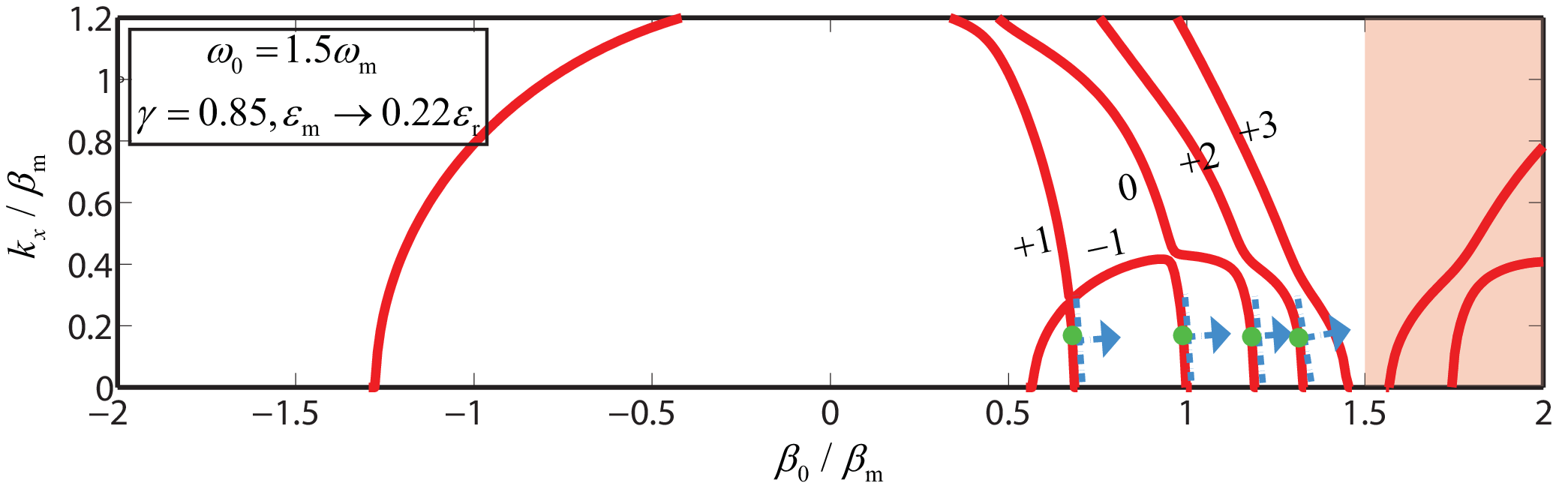}}
		\caption{Isofrequency contours for an unbounded medium with sinusoidal space-time modulation. (a)~Purely spatial modulation (\(\omega_\text{m} = 0\)), corresponding to \(\gamma \rightarrow 0\) with a normalized wave vector of \(0.6\)~\cite{Taravati_PRB_2017}. (b)~Weak space-time modulation (\(\epsilon_\text{m} \rightarrow 0\)) at a low velocity ratio (\(\gamma = 0.15\)) and excitation frequency \(\omega_0 = 1.5\omega_\text{m}\)~\cite{Taravati_PRB_2017}. (c)~Increased modulation strength (\(\epsilon_\text{m} = 0.22\epsilon_\text{r}\)) and velocity ratio (\(\gamma = 0.3\))~\cite{Taravati_PRB_2017}. (d)~Transition to the quasisonic regime (\(\gamma = 0.85\), \(\gamma_\text{s,min}=0.905\)) with the modulation strength held at \(\epsilon_\text{m} = 0.22\epsilon_\text{r}\)~\cite{Taravati_PRB_2017}.}
		\label{Fig:2D_dispesion_curves}
	\end{center}
\end{figure}

The evolution of wave propagation under space-time modulation is further elucidated by the isofrequency contours presented in Figs.~\ref{Fig:2D_dispesion_curves_a} to~\ref{Fig:2D_dispesion_curves_d}, which chart the transition from a conventional static structure to a highly nonreciprocal quasisonic regime. Each panel illustrates the supported wavenumbers \( \beta \) for a fixed excitation frequency \( \omega_0 \), providing a critical map of allowed propagation directions and strengths. Figure~\ref{Fig:2D_dispesion_curves_a} depicts the benchmark case of a purely spatial modulation (\( \omega_\text{m} = 0 \), \( \gamma \to 0 \)). The contour is symmetric and periodic in the wavenumber domain, a direct manifestation of Lorentz reciprocity. This results from the static, grating-like nature of the medium, where forward and backward propagation are fundamentally identical, prohibiting nonreciprocal effects.

Figure~\ref{Fig:2D_dispesion_curves_b} introduces a weak space-time modulation with a low velocity ratio (\( \gamma = 0.15 \)). The symmetry of the contour is broken, signifying the birth of nonreciprocity. The contour distorts, indicating that the phase constants for forward (\( +\beta \)) and backward (\( -\beta \)) propagation are no longer equal. This is the foundational mechanism for devices like isolators and circulators, as it creates a inherent bias in the medium's wave transport properties. Figure~\ref{Fig:2D_dispesion_curves_c} demonstrates the effect of increasing both the modulation strength (\( \epsilon_m \)) and velocity ratio (\( \gamma \)). The nonreciprocal distortion of the isofrequency contour is significantly amplified. The supported wavenumbers for one direction become markedly different from those in the opposite direction, leading to a substantial differential phase shift. This enhanced nonreciprocity is crucial for achieving strong isolation and efficient nonreciprocal phase shifting in practical device implementations. Figure~\ref{Fig:2D_dispesion_curves_d} shows the transition to the quasisonic regime (\( \gamma = 0.85 \)), where the modulation velocity approaches the wave's characteristic speed in the medium. In this regime, the contour becomes highly anisotropic and collapses for a range of wavenumbers, opening a large momentum bandgap. This signifies that waves within a specific momentum range are evanescent and cannot propagate, a powerful tool for controlling and isolating signal paths with exceptional contrast.

\section{Illustrative Dispersion Engineering Examples}
\subsection{Nonreciprocity of Sonic Regime}\label{sec:sonic}

\begin{figure}
	\begin{center}
		\subfigure[]{ \label{Fig:disp_sonic}
			\includegraphics[width=0.8\columnwidth]{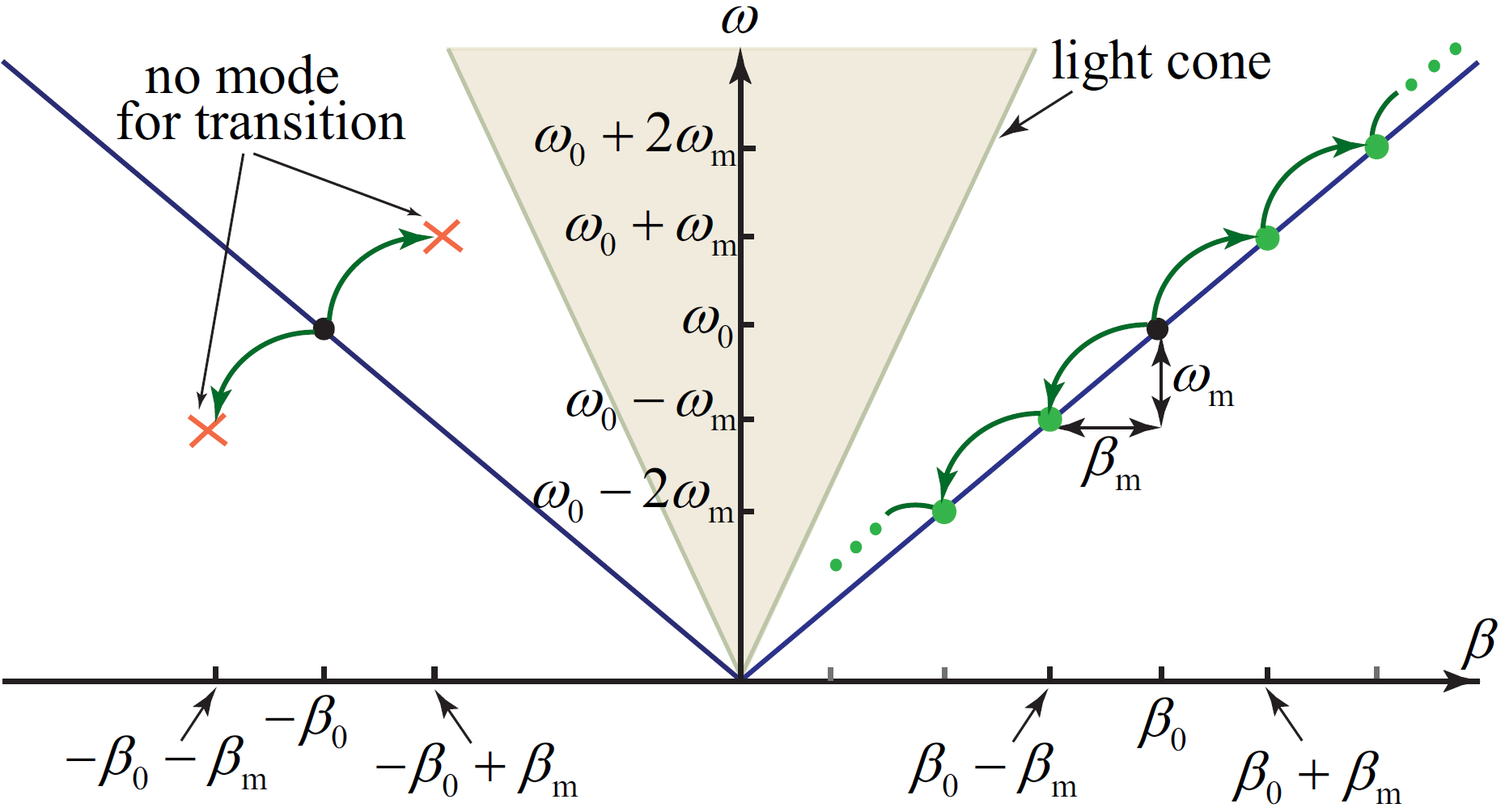} }
				\subfigure[]{ \label{Fig:sonic_isol}
			\includegraphics[width=0.8\columnwidth]{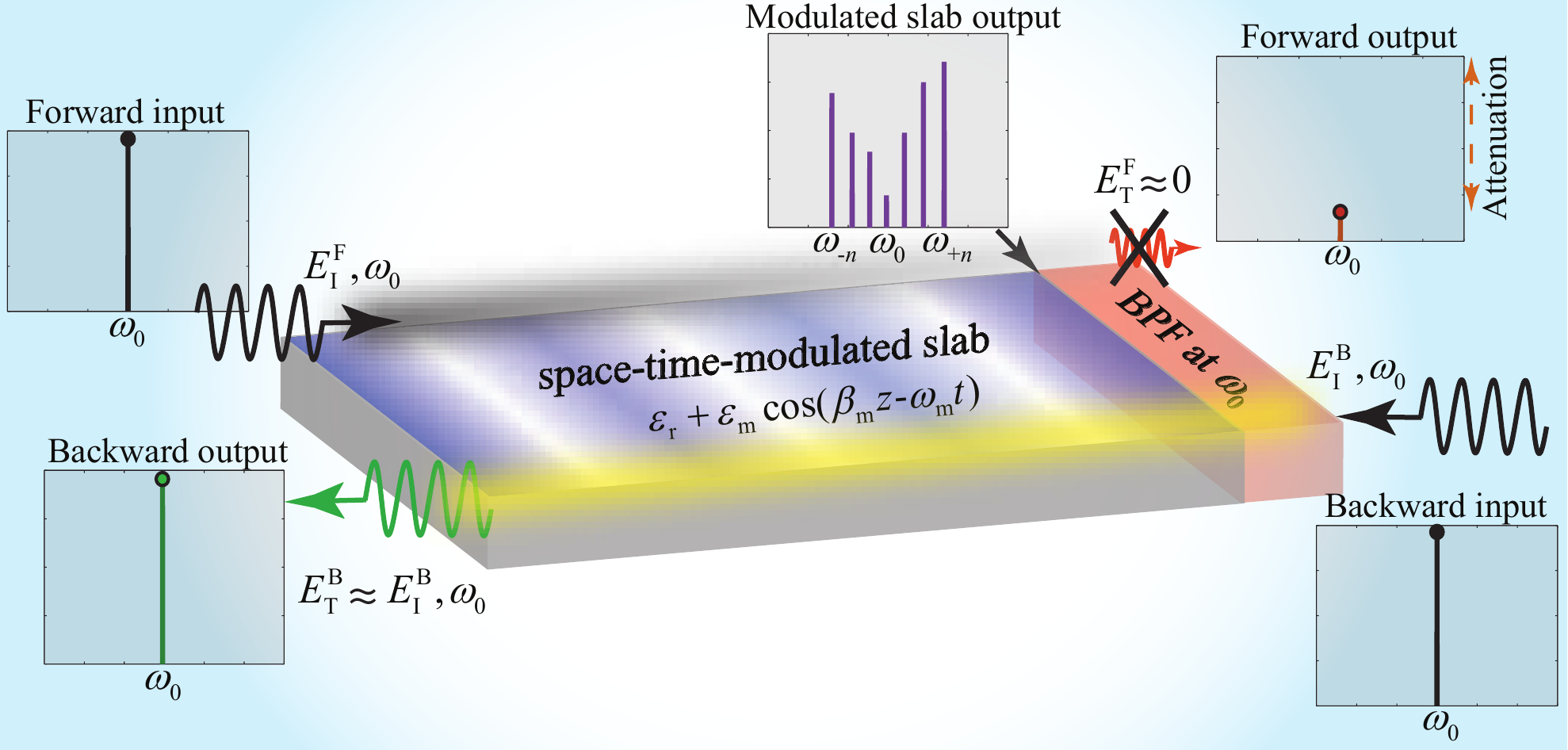} }
		\subfigure[]{ \label{Fig:photo_sonic}
			\includegraphics[width=0.8\columnwidth]{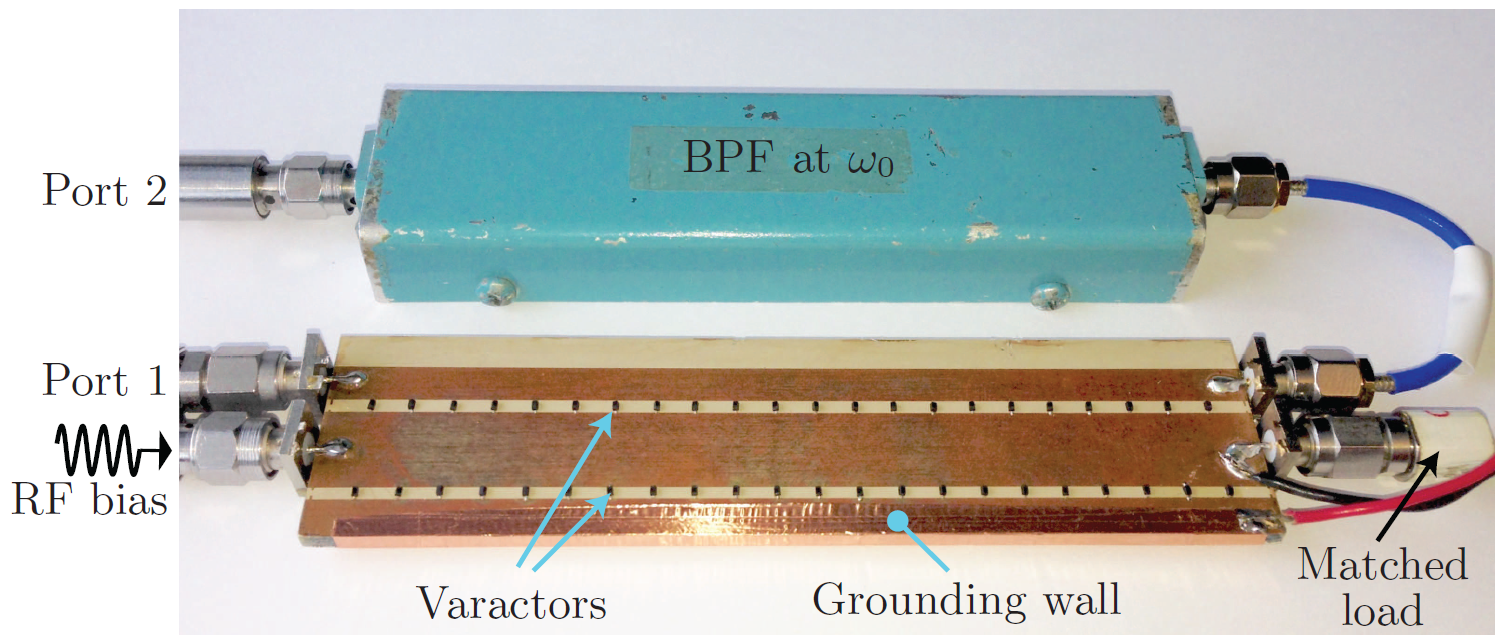} }
		\caption{Leveraging asymmetric dispersion curves at the sonic regime to realization an isolator. (a)~Dispersion diagram highlighting nonreciprocal band transitions~\cite{Taravati_PRB_2017}. (b)~Operational schematic: a forward-traveling wave is converted to higher-order harmonics and suppressed by a band-pass filter, while a backward wave passes without harmonic conversion, realizing isolation~\cite{Taravati_PRB_2017}. (c)~Photograph of the fabricated microwave isolator prototype incorporating an array of subwavelenghtly-spaced space-time-modulated varactors~\cite{Taravati_PRB_2017}.}
	\end{center}
\end{figure}

Figures~\ref{Fig:disp_sonic} to~\ref{Fig:photo_sonic} present the design principle of the sonic isolator is founded on the principle of leveraging nonreciprocal space-time energy transitions to higher-order space-time harmonics, a phenomenon that emerges most strongly when the velocity of the spatiotemporal modulation approaches the velocity of light in the background medium, corresponding to a velocity ratio \(\gamma = v_m/v_b\) approaching unity. In this condition, the dispersion diagram of the unbounded modulated medium reveals a critical asymmetry: the forward propagating waves (those co-directional with the modulation) experience a collapse of their spatial-frequency harmonics, bringing an infinite number of modes into close proximity and strong mutual coupling. Conversely, the backward propagating waves see their harmonics become more widely separated. This fundamental asymmetry in the dispersion landscape means that a wave incident from the forward direction is strongly phase-matched to a dense manifold of higher-order harmonics, facilitating an efficient and distributed transfer of its energy out of the fundamental frequency \(\omega_0\) and into the generated harmonics \(\omega_0 \pm n\omega_\text{m}\). A wave incident from the backward direction, however, encounters a system where such phase-matched coupling is extremely weak, allowing it to pass through the modulated slab largely unaltered at its original frequency.

The operational principle of the isolator is to exploit this stark nonreciprocity in harmonic generation. The design involves a slab whose permittivity is sinusoidally modulated as in Eq.~\eqref{eqa:sin_perm}, operating in the sonic regime. In the forward direction, the incident wave undergoes intense interaction within the slab. Its energy is efficiently and nonreciprocally converted ("pumped" via the external modulation) into a spectrum of higher-order space-time harmonics, leaving the transmitted field at the fundamental frequency \(\omega_0\) significantly attenuated. In the backward direction, the same slab presents a weak perturbation; the incident wave passes through with minimal interaction and thus most of its power remains at \(\omega_0\). To complete the isolator function, this nonreciprocal slab is then cascaded with a bandpass filter tuned to the fundamental frequency \(\omega_0\). In the forward direction, the already-attenuated fundamental signal is blocked by the filter, resulting in high isolation. In the backward direction, the unaltered wave passes through the filter with low insertion loss. Thus, the combination of the sonic space-time modulated slab—which acts as a nonreciprocal harmonic generator—and a subsequent frequency-selective filter creates a path for isolation, allowing transmission in one direction while suppressing it in the opposite. This design was experimentally validated at microwave frequencies, demonstrating over 20 dB of isolation by capitalizing on these engineered nonreciprocal energy transitions in a spatiotemporal medium.

\subsection{Space-time medium at Equilibrium}

\begin{figure}
	\begin{center} 
		\subfigure[]{\label{Fig:Equi_circ_a}
			\includegraphics[width=0.9\columnwidth]{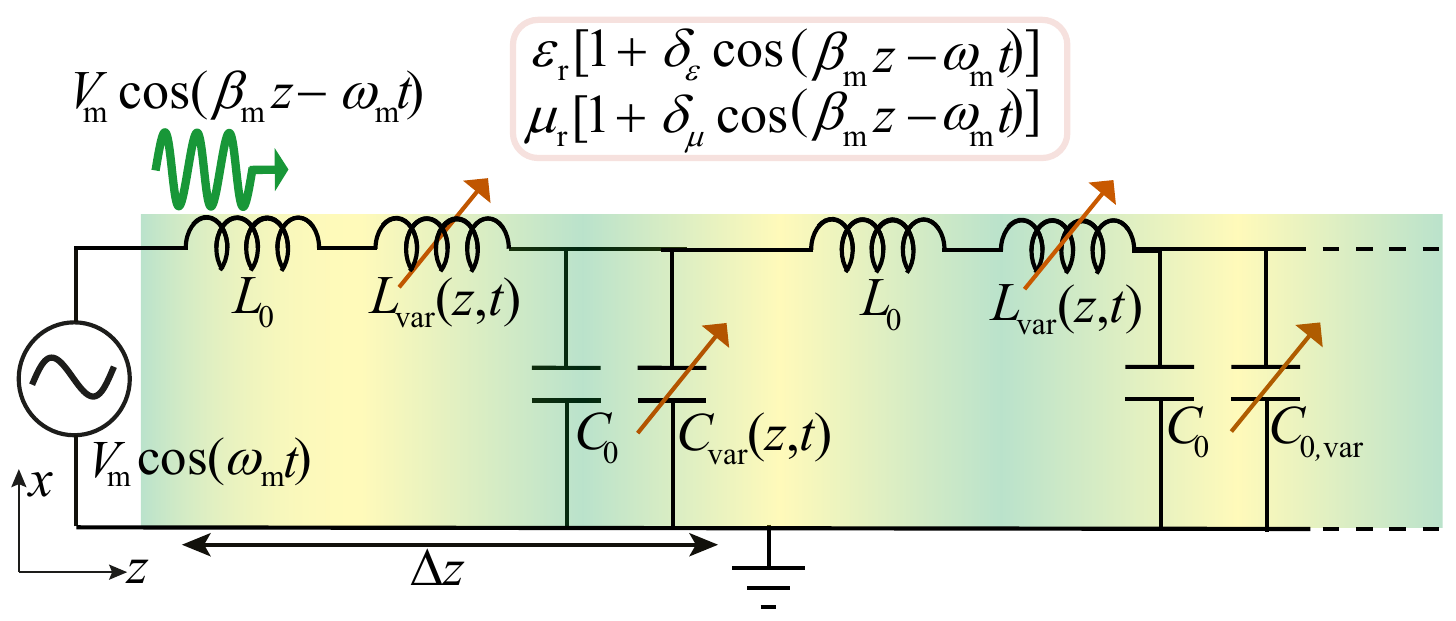} } 
		\subfigure[]{\label{Fig:Equ_disp_b}
			\includegraphics[width=0.48\columnwidth]{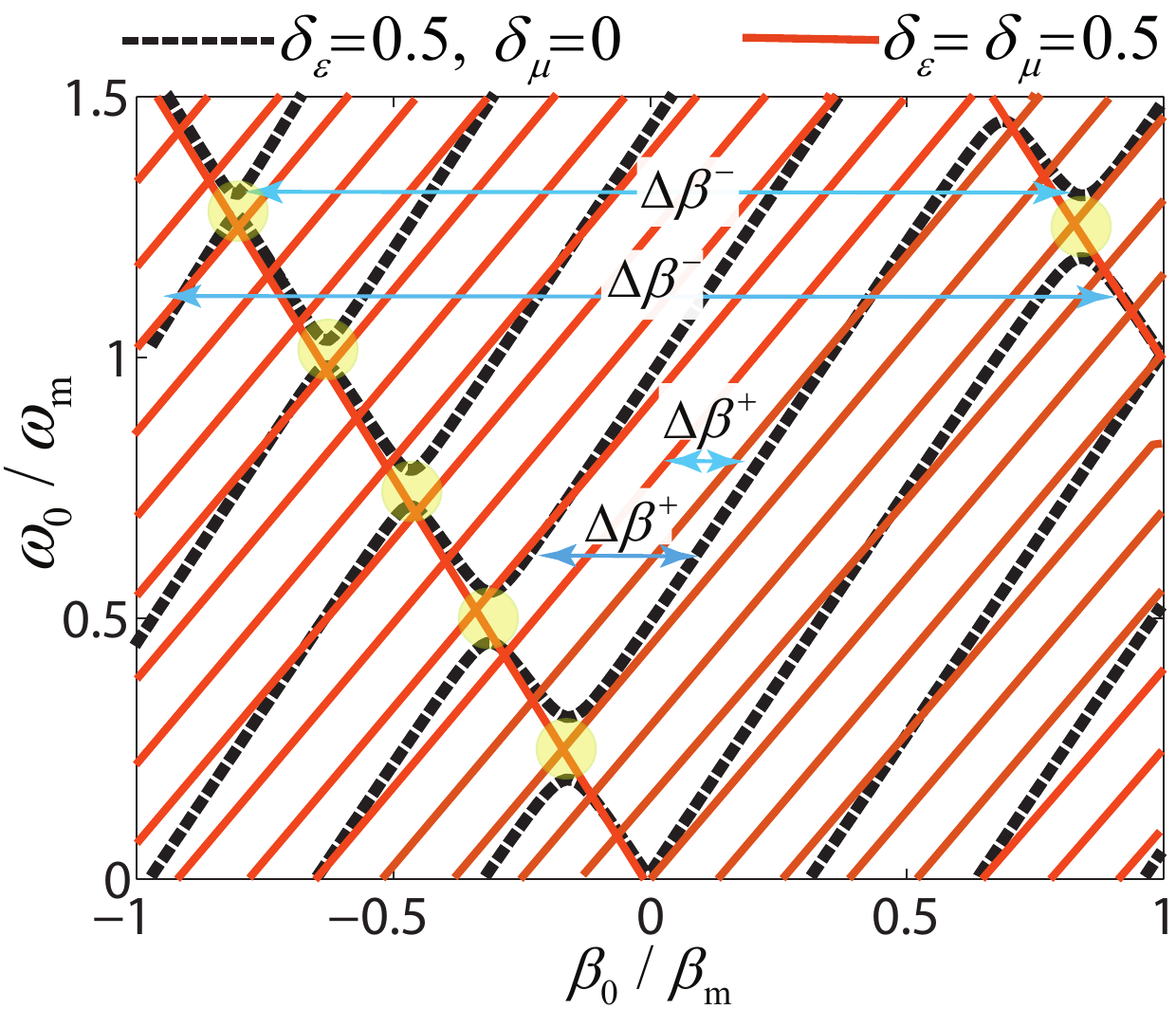}} 
		\subfigure[]{\label{Fig:Equi_disp_c}
			\includegraphics[width=0.48\columnwidth]{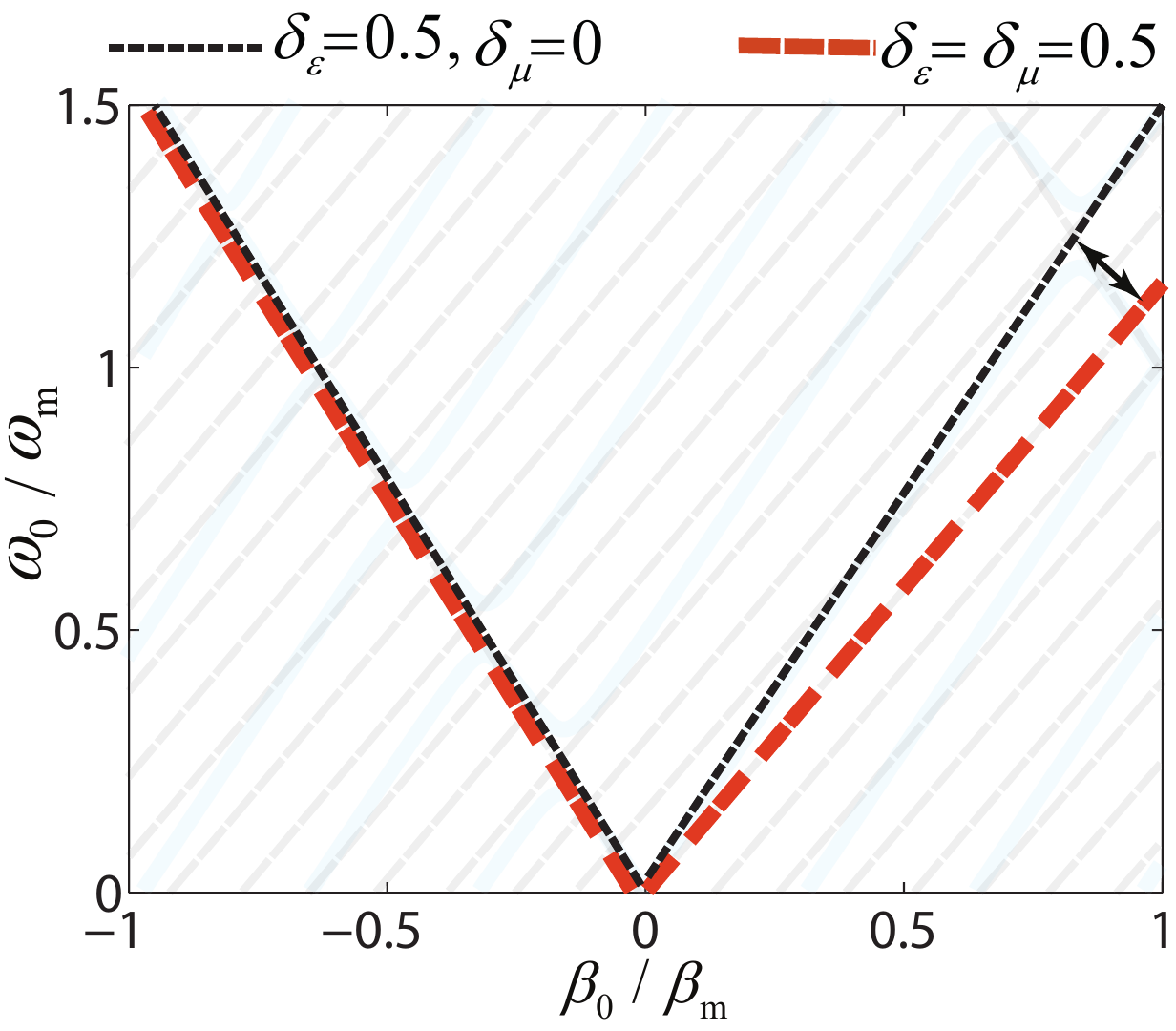}}  
		\caption{Zero dispersion bandgap, enhanced nonreciprocity and zero reflection in an equilibrated space-time metamaterial. (a) Circuit model of a transmission line with simultaneous spatiotemporal modulation of permittivity (via variable capacitors) and permeability (via variable inductors), driven by a pump wave~\cite{Taravati_PRAp_2018}. (b, c) Dispersion diagrams comparing the equilibrated (\(\delta_\mu=\delta_\epsilon=0.5\), \(\gamma_\text{s,l}=0.666\)) and non-equilibrated (\(\delta_\mu=0\), \(\delta_\epsilon=0.5\), \(\gamma_\text{s,l}=0.816\)) cases for a modulation ratio of \(\gamma=0.65\), demonstrating the impact of balanced modulation on the band structure~\cite{Taravati_PRAp_2018}.}
		\label{Fig:wave_scat_b}
	\end{center}
\end{figure}

Figures~\ref{Fig:Equi_circ_a} to~\ref{Fig:Equi_disp_c} demonstrate an equilibrated space-time metamaterials~\cite{Taravati_PhNt_Equi_2018,Taravati_PRAp_2018,huidobro2019fresnel}. The work demonstrates that by imposing identical spatiotemporal modulation on both the permittivity and permeability with $\delta_\mu = \delta_\epsilon$, a remarkable transformation of the photonic band structure occurs. This equilibrium condition yields three groundbreaking phenomena: the complete elimination of photonic band gaps, the achievement of zero spatial and temporal local reflections, and a giant enhancement of linear nonreciprocity, all of which are directly engineered via the medium's dispersion.

A deep insight into this dispersion engineering reveals that the key mechanism is the modification of the coupling strength between space-time Bloch-Floquet harmonics. In a conventional STM medium (only permittivity modulated, $\delta_\mu=0$), the coupling matrix element $A_{n,n+1} = (\delta_\epsilon/2) v_n$ dictates that an excited mode $n$ couples primarily to its two immediate neighbors ($n \pm 1$). In contrast, in an equilibrated STM medium ($\delta_\mu = \delta_\epsilon$), the matrix structure becomes far richer, i.e.,
\begin{equation}
K_{n,n+1} = \frac{\delta_\epsilon}{2} v_n + \frac{\delta_\mu}{2} v_{n+1}, \quad K_{n,n+2} = \frac{\delta_\epsilon \delta_\mu}{4} v_{n+1}.
	\label{eqa:Equi_K}
\end{equation}
which signifies that each mode $n$ now couples strongly to four adjacent modes ($n \pm 1$ and $n \pm 2$), fundamentally altering the energy redistribution pathways within the dispersion diagram. This enhanced multi-harmonic coupling is the origin of the "giant" nonreciprocity, as it facilitates a much more efficient and asymmetric transfer of energy from the fundamental harmonic to higher-order harmonics for forward propagation, while leaving backward propagation largely unaffected.

The most profound consequence of this equilibrium is the complete closure of photonic band gaps. In a nonequilibrated medium, band gaps open at the synchronization points where forward and backward dispersion curves intersect, leading to strong attenuation/reflection. The analysis shows that the band gap width is proportional to $|\delta_\epsilon - \delta_\mu|$. Therefore, by enforcing $\delta_\mu = \delta_\epsilon$, the band gap width collapses to zero. This results in a perfectly transmitting medium for all frequencies in the subsonic regime ($\gamma < 1$), as visually evidenced by the disappearance of the band gaps (magenta circles) in the dispersion diagrams when equilibrium is established.

Furthermore, equilibrium drastically expands the sonic regime interval, defined as the range of velocity ratios $\gamma$ where strong nonreciprocal coupling occurs. For an equilibrated medium, this interval is $\frac{1}{1+\delta_{\epsilon,\mu}} \leq \gamma \leq \frac{1}{1-\delta_{\epsilon,\mu}}$, which is more than twice as wide as that of a nonequilibrated medium. A wider sonic regime provides greater design flexibility, allowing for the achievement of strong nonreciprocity (characterized by $\Delta\beta^+ \to 0$ for forward modes and $\Delta\beta^- \to 2$ for backward modes) with weaker modulation strengths and at a broader range of modulation velocities. This is a crucial practical advantage, as it reduces the required pump energy.

Finally, the equilibrium condition imposes a constant, space-time-independent intrinsic impedance $\eta_{eq} = \eta_0 \eta_r$, which eliminates the infinite series of local reflections that would otherwise occur at the discretized space-time interfaces of a nonequilibrated medium. This "impedance matching" in both space and time, combined with the engineered dispersion that favors unidirectional harmonic generation, enables the realization of a reflectionless, highly nonreciprocal system. The dispersion diagrams visually confirm this by showing a significant "tilt" and reduction in the phase and group velocities specifically for the forward harmonics in the equilibrated case, a direct manifestation of the momentum injection from the modulation being coupled unidirectionally. This sophisticated dispersion engineering, therefore, creates a unique class of media where wave transmission is omnidirectional yet manipulation is fundamentally nonreciprocal.

\begin{figure}
		\subfigure[]{\label{fig:dispiso_3}
		\includegraphics[width=0.49\columnwidth]{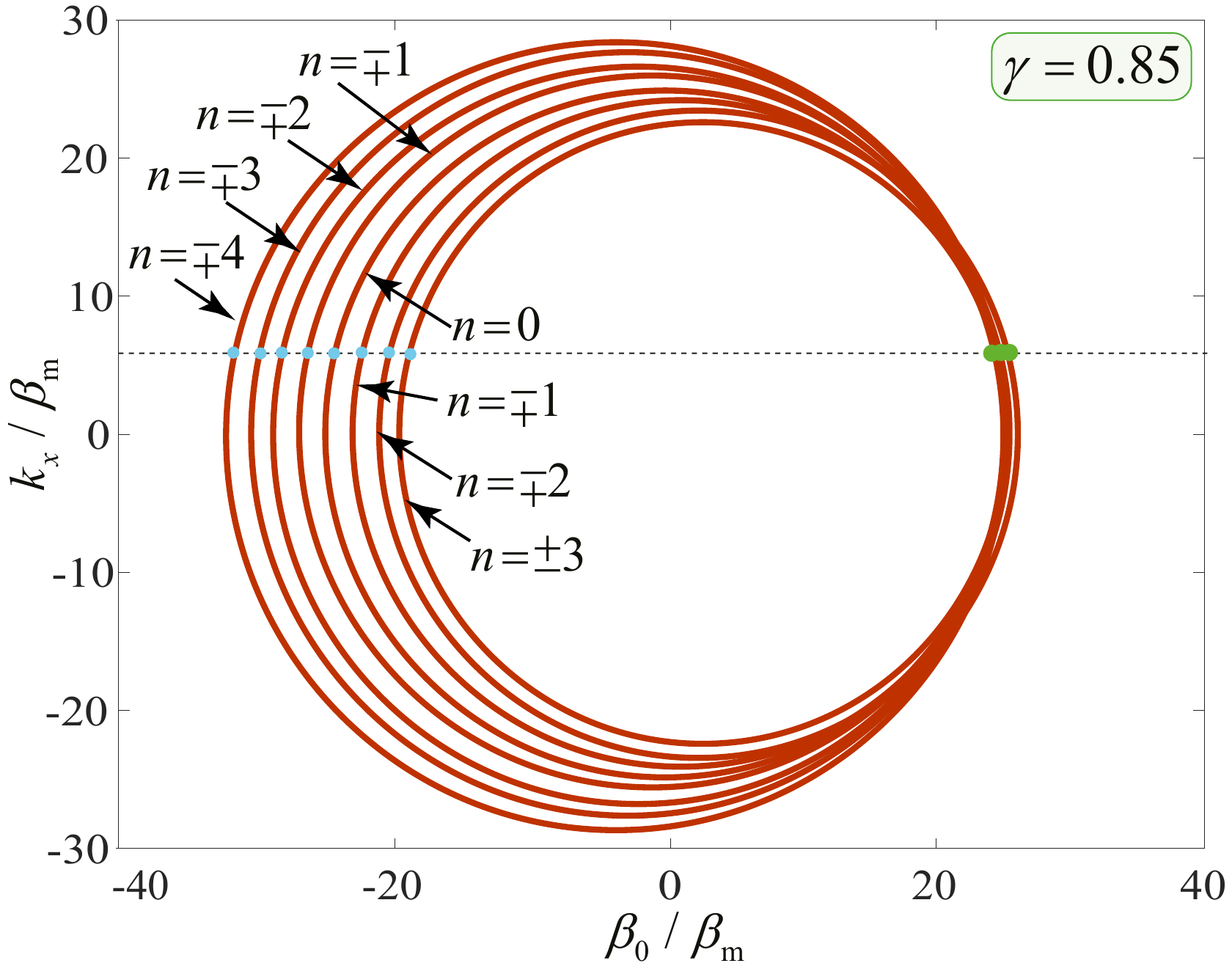}} 
		\subfigure[]{\label{Fig:perm_slab_num_a}
		\includegraphics[width=0.47\columnwidth]{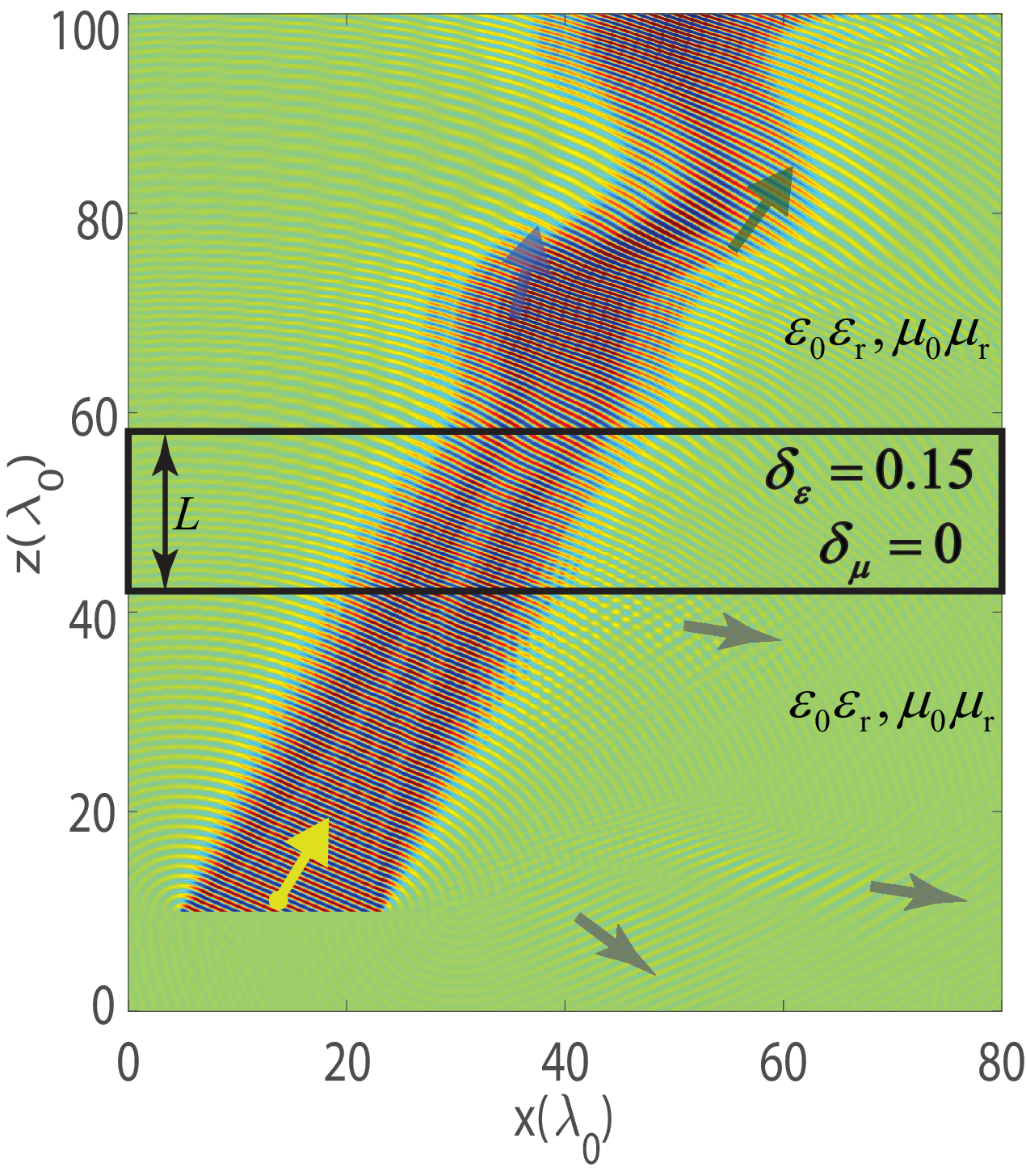}} 
		\subfigure[]{\label{Fig:equi_slab_num_a}
		\includegraphics[width=0.48\columnwidth]{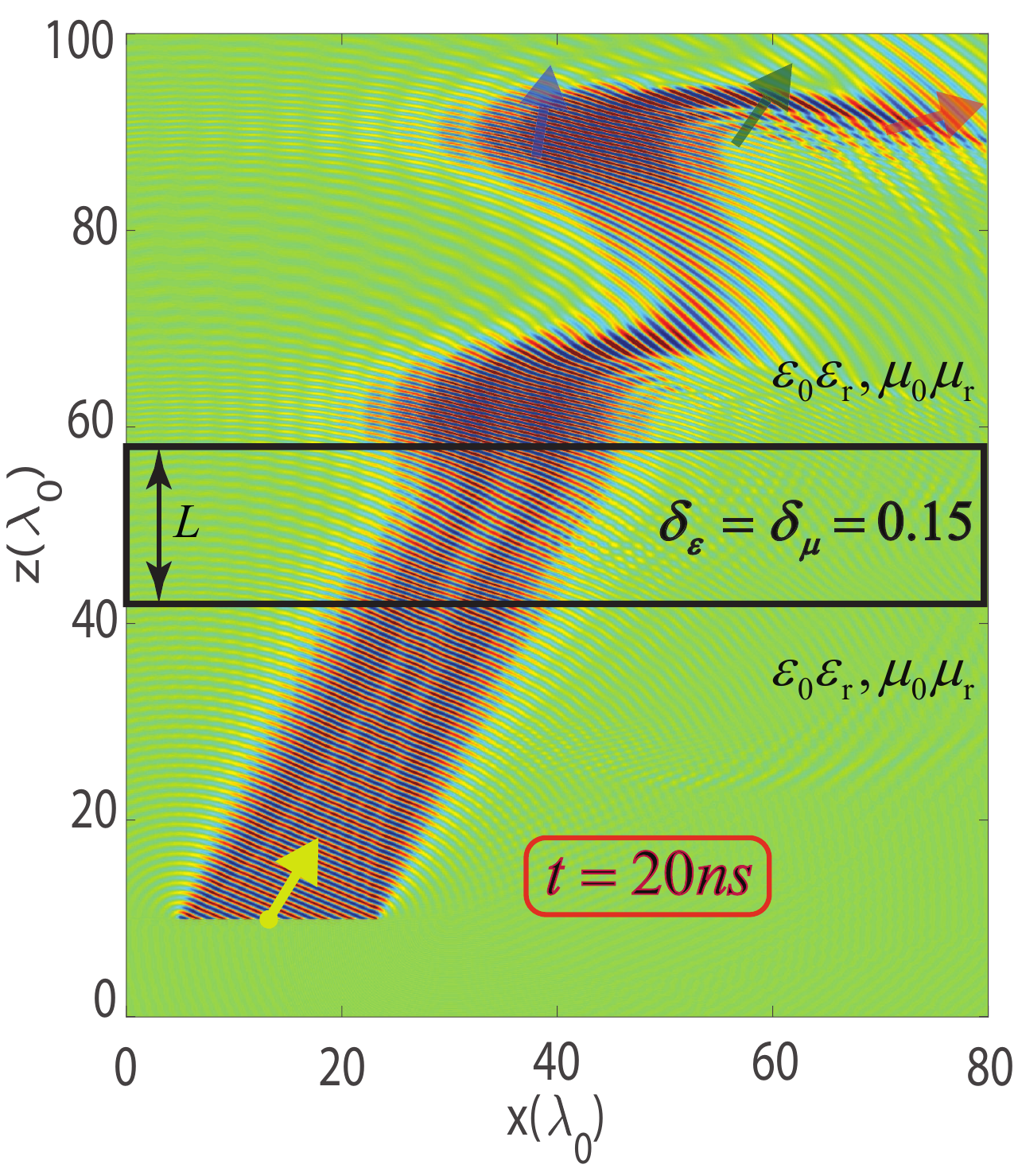}}
	\subfigure[]{\label{Fig:equi_slab_num_bb}
		\includegraphics[width=0.48\columnwidth]{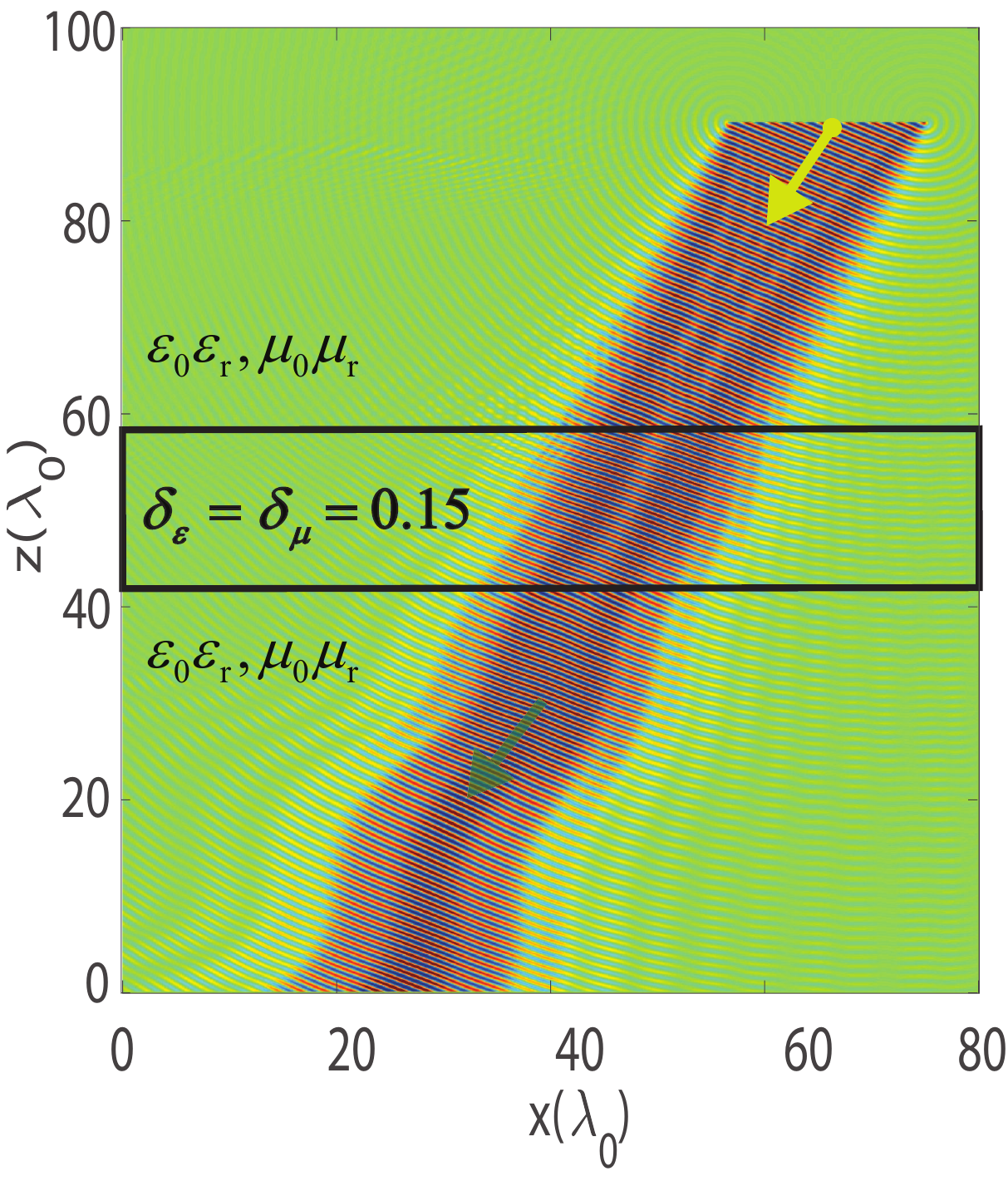}}
	\subfigure[]{\label{Fig:equi_slab_num_c}
		\includegraphics[width=0.48\columnwidth]{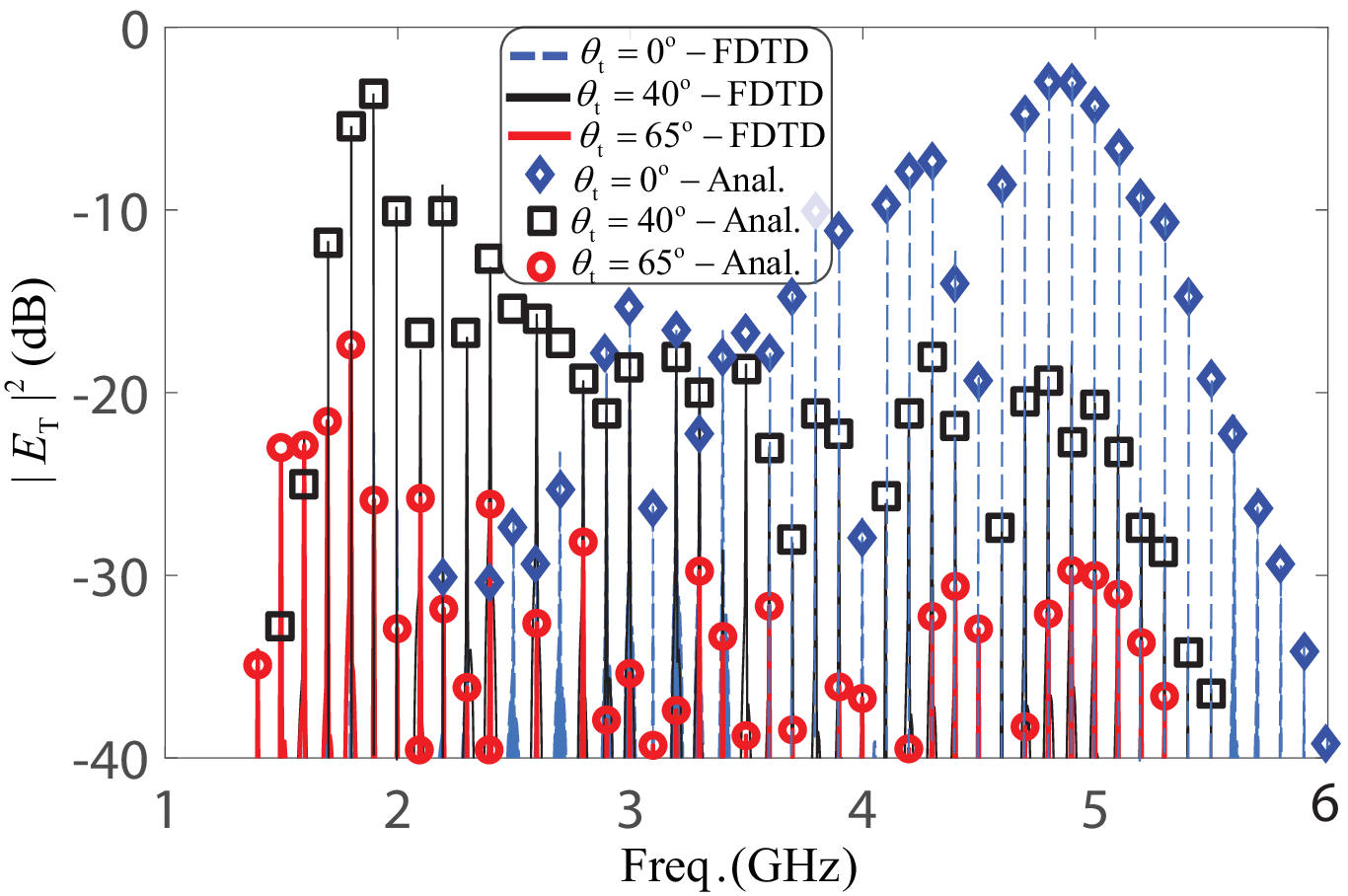}}
		\subfigure[]{\label{Fig:equi_slab_num_bc}
		\includegraphics[width=0.48\columnwidth]{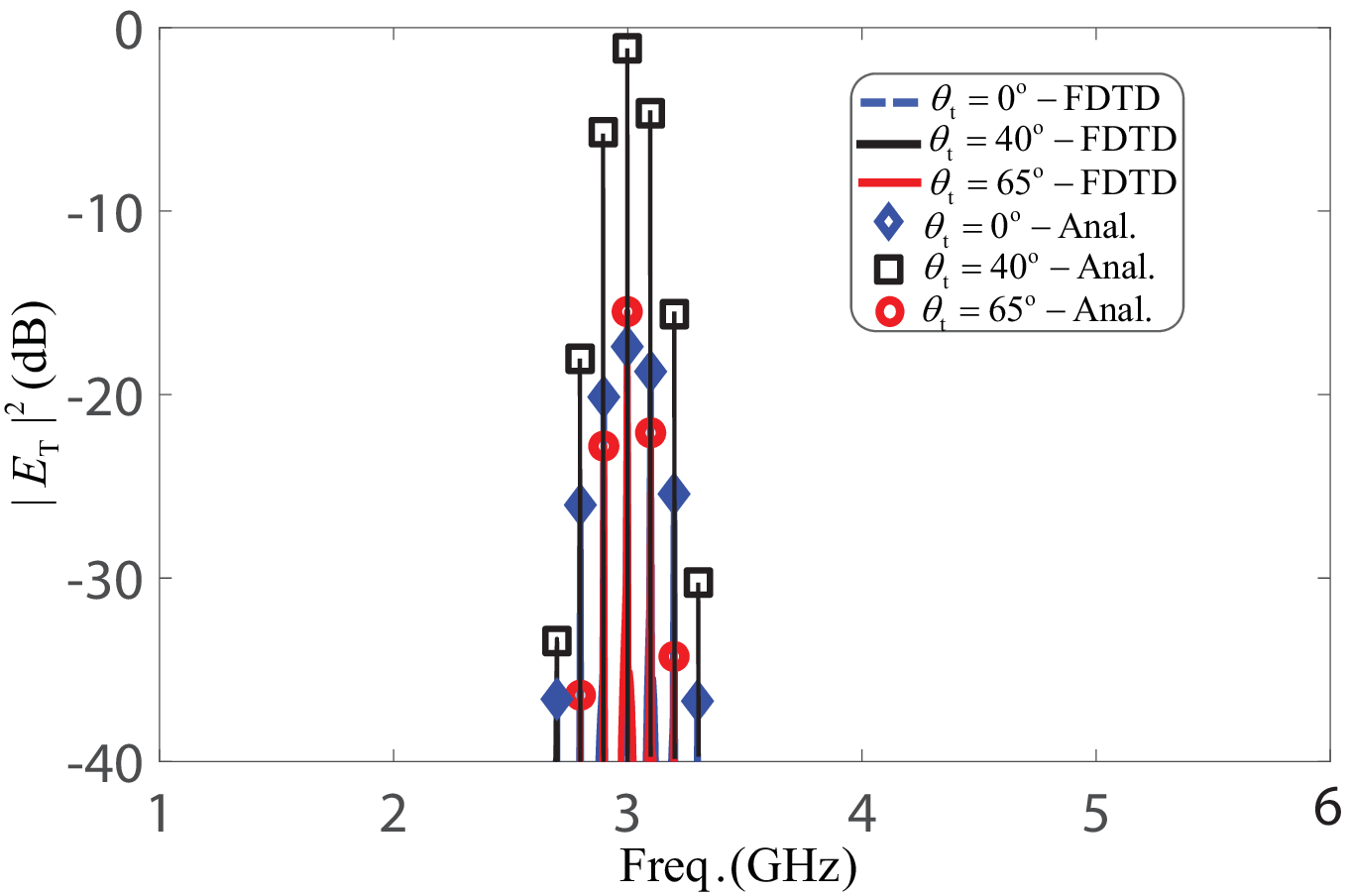}} \caption{Excitation of isofrequency circles at different incidence angles leading to a diverse set of nonreciprocal transmission from a space–time metamaterial slab.
		(a)~Isofrequency diagram at $\omega / \Omega = 3$ for $\delta_{\epsilon} = \delta_{\mu} \rightarrow 0$ in the quasisonic regime ($\gamma = 0.85$).
		(b)~Numerical time-domain simulation of the electric field distribution, $E_y$, for forward and backward oblique excitation of a plane wave with frequency $\omega_0 / \Omega = 30$~GHz, $\gamma = 1$, $L = 16\lambda$, $\theta_\text{i} = 25^\circ$, and modulation strength $\delta_{\epsilon} = 0.15$~\cite{Taravati_Kishk_TAP_2019}.
		(c)~Same as (b) except for equilibrated modulation $\delta_{\mu} = \delta_{\epsilon} = 0.15$, yielding stronger space–time harmonic generation, spatial decomposition, and zero reflection~\cite{Taravati_Kishk_TAP_2019}.
		(d)~Backward incidence and transmission from the equilibrated space–time metamaterial slab in (c), demonstrating nonreciprocal transmission~\cite{Taravati_Kishk_TAP_2019}.
		(e,f)~Frequency spectra corresponding to (c) and (d), respectively, showing nonreciprocal harmonic generation and spatial decomposition~\cite{Taravati_Kishk_TAP_2019}.}
	\label{fig:disper}
\end{figure}

\subsection{Nonreciprocity Based on Asymmetric Photonic Band Gaps}\label{sec:PBI}
\begin{figure}
	\begin{center}
		\subfigure[]{ \label{Fig:disp_asymm}
			\includegraphics[width=0.7\columnwidth]{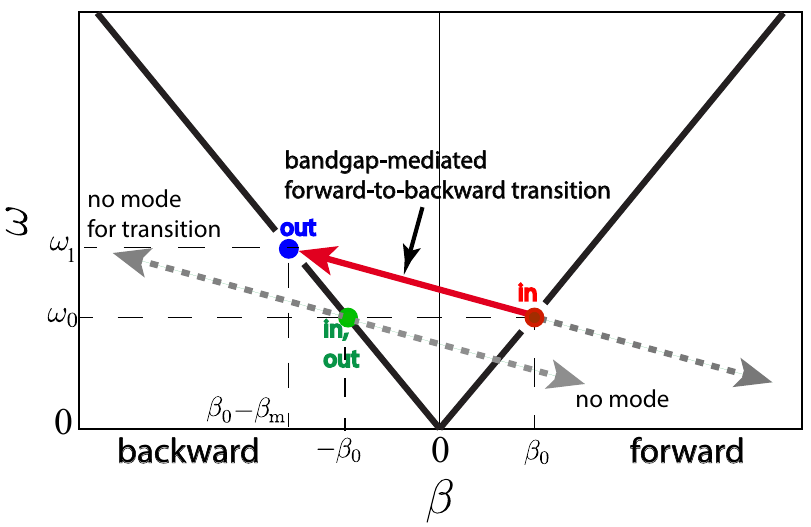} }
		\subfigure[]{ \label{Fig:3D_bragg}
			\includegraphics[width=0.8\columnwidth]{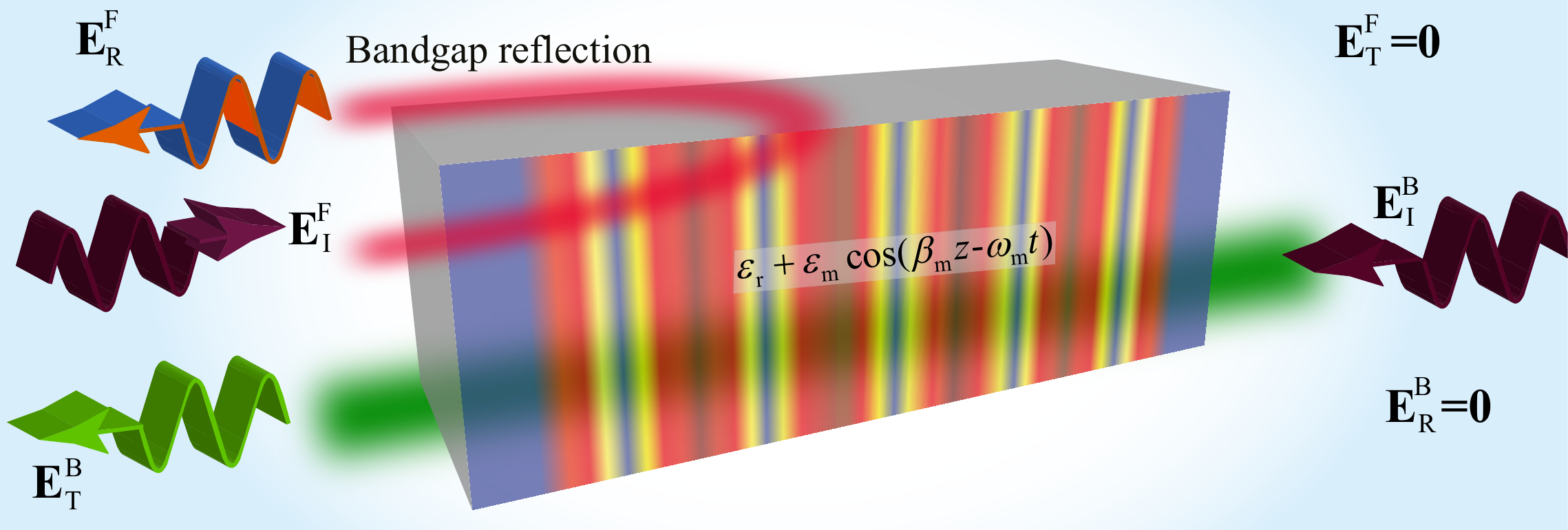} }
					\subfigure[]{ \label{Fig:ass_photo}
				\includegraphics[width=0.8\columnwidth]{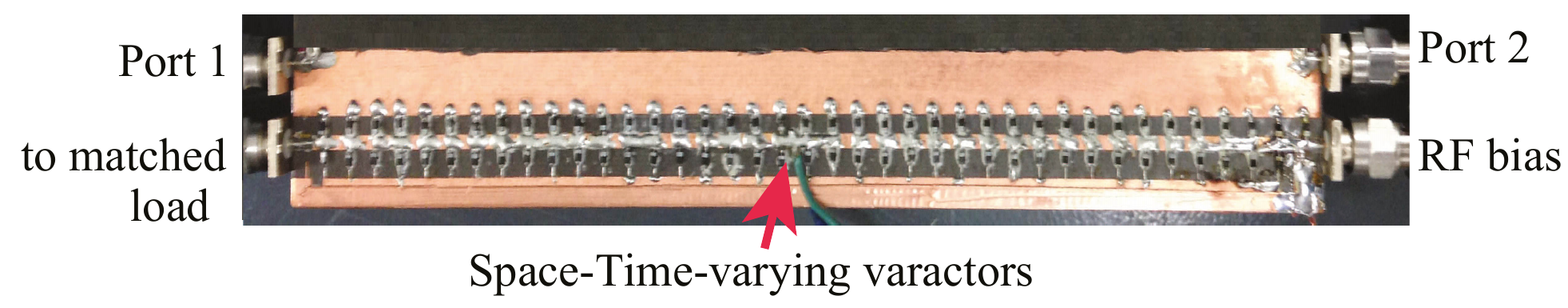} }
		\caption{Leveraging asymmetric electromagnetic band gaps in a space–time–modulated metamaterial to realize an isolator exhibiting nonreciprocal Bragg reflections. (a) Dispersion diagram after~\cite{Chamanara_PRB_2017}. (b) Schematic of the isolator's operating principle, utilizing asymmetric Bragg reflections. (c) Photograph of the fabricated prototype~\cite{Chamanara_PRB_2017}.}
	\end{center}
\end{figure}

Figures~\ref{Fig:disp_asymm} to~\ref{Fig:ass_photo} demonstrate a technique to strategically manipulate the photonic band structure of a space-time-modulated (STM) slab to create asymmetric band gaps, a condition where the dispersion diagram is tilted, leading to different electromagnetic responses for forward and backward propagation directions. This is achieved by introducing a traveling-wave modulation of the permittivity, described by Eq.~\eqref{eqa:sin_perm}, which imparts a momentum $k_{\text{m}}$ and energy $\hbar\omega_{\text{m}}$ to the system. The resulting Brillouin zone is consequently tilted in the ($k$, $\omega$) space, breaking the symmetry between the $+\beta$ and $-\beta$ directions. This tilt is the cornerstone of the dispersion engineering, as it misaligns the band gaps for opposite propagation directions, enabling unidirectional behavior where a wave incident from one side encounters a band gap (and is thus reflected), while a wave from the opposite side lies within a propagation band (and is thus transmitted).

A deep insight into this dispersion engineering reveals that the nonreciprocity stems from the excitation of distinct, linearly independent Bloch-Floquet modes for opposite incidences. In a conventional static periodic structure, the dispersion diagram is symmetric, and modes for opposite directions are simply mirror images, folding onto the same point in the Brillouin zone. In contrast, the space-time modulation lifts this degeneracy. When the slab is excited at a frequency $\omega_0$, it excites an infinite set of space-time harmonics with frequencies $\omega_n = \omega_0 + n\omega_{\text{m}}$. The critical engineering feat is to select modulation parameters such that the dominantly excited mode for forward incidence (e.g., the $p=0$ mode marked in red) is an evanescent gap mode, exponentially decaying within the slab, while the dominantly excited mode for backward incidence is a propagating mode. This modal selectivity is achieved because the phase matching condition, dictated by the alignment of the incident wavevector with the slanted dispersion curves—is different for the two directions due to the broken symmetry. The analysis shows that the impedance of the dominant gap mode harmonics is perfectly matched to the incident medium for both its fundamental ($\omega_0$, forward propagating) and upshifted ($\omega_0+\omega_{\text{m}}$, backward propagating) components, facilitating a complete transfer of the incident energy into an evanescent wave and its subsequent conversion into a reflected wave at a Doppler-shifted frequency.

Furthermore, the dispersion engineering provides control over key device performance metrics. The isolation level can be enhanced by increasing the slab length, allowing the evanescent gap mode to decay sufficiently, while the operational bandwidth is directly proportional to the width of the engineered band gap, which itself scales with the modulation depth $M$. This establishes a fundamental trade-off: weaker modulations allow for higher isolation (due to reduced coupling to spurious modes) and lower required modulation frequencies, but at the cost of a narrower bandwidth and a longer device. This sophisticated manipulation of the photonic band structure via space-time modulation thus enables the creation of magnetless, linear isolators with performance characteristics that are directly tunable through the modulation parameters.

\subsection{Leaky-Wave Space-Time Transition}\label{sec:LW}

By exploiting dispersion engineering in the leaky-wave regime, space-time metamaterials can be designed to function as a unified transceiver, consolidating the three distinct operations of mixing, duplexing, and radiation into a single, multifunctional platform~\cite{Taravati_APS_2015,taravati2016mixer}. The foundation of this integration lies in a multi-layered approach to sculpting the electromagnetic dispersion diagram. It begins with the generation of space-time harmonics, where a periodic modulation scatters an input wave into an infinite set of Floquet-Bloch modes, each at a distinct frequency \(\omega_n = \omega_0 + n\omega_\text{m}\) and wavenumber \(\beta_n = \beta_0 + n\beta_\textbf{m}\). This constitutes the first level of dispersion engineering, as the choice of modulation parameters \((\omega_\text{m}, \beta_\textbf{m})\) allows the designer to precisely place these harmonics, creating new potential channels for signal processing.

\begin{figure}
	\begin{center}
		\subfigure[]{ \label{Fig:disp_LW}
			\includegraphics[width=0.7\columnwidth]{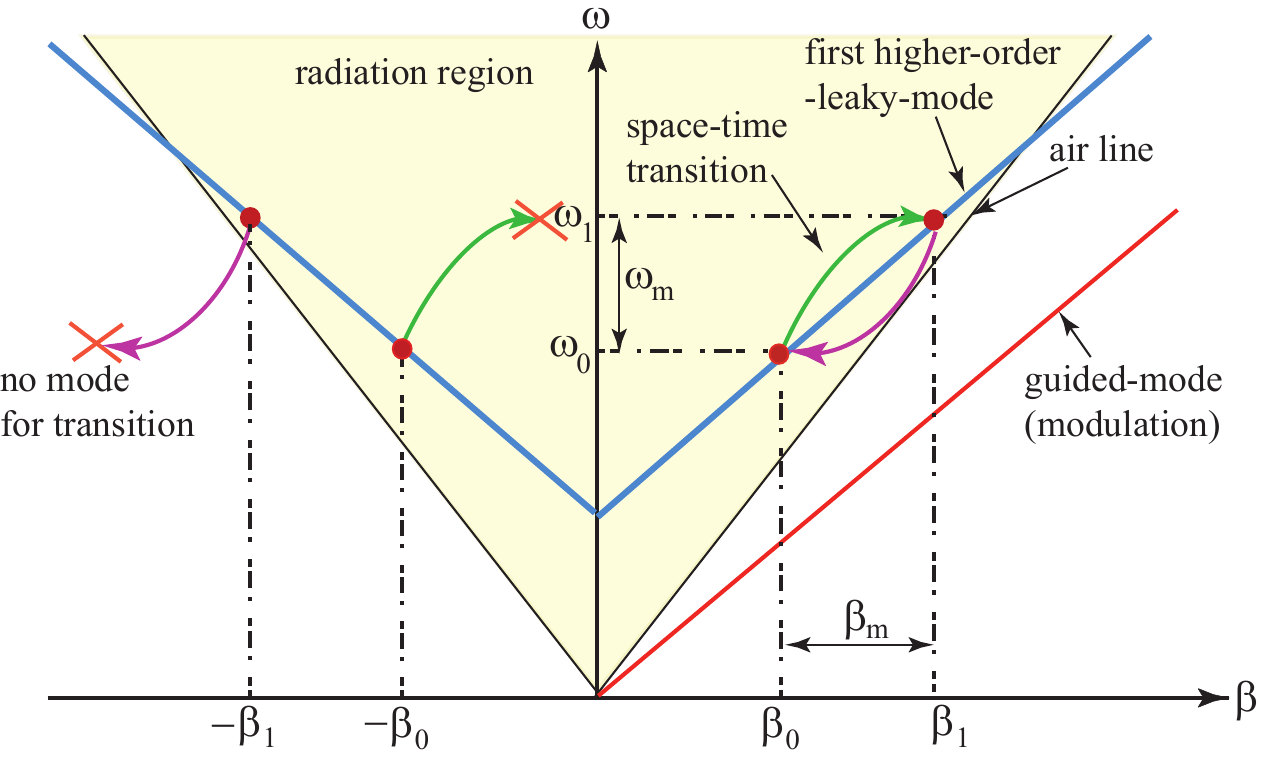} }
		\subfigure[]{ \label{Fig:LW}
			\includegraphics[width=0.7\columnwidth]{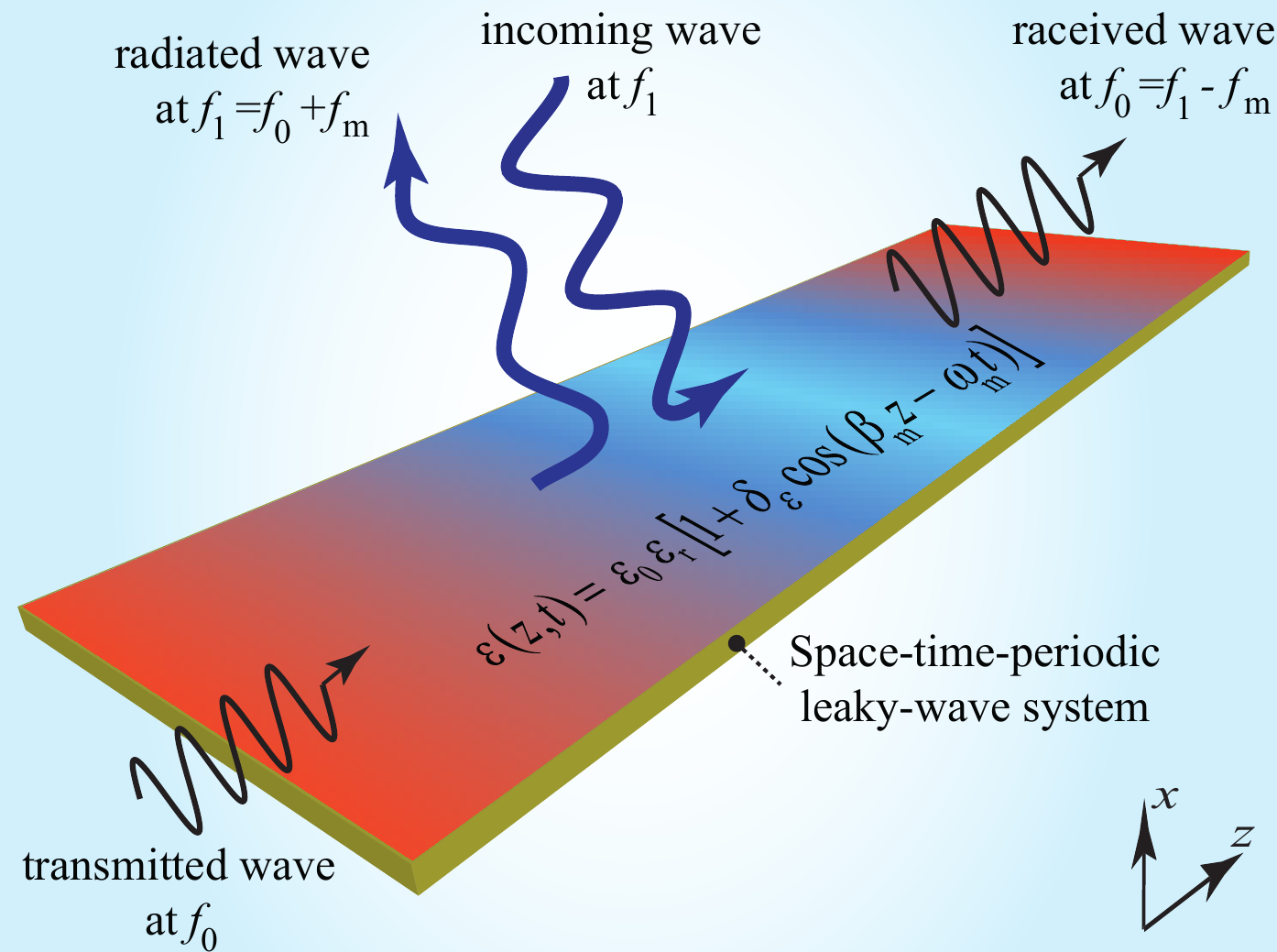} }
		\subfigure[]{ \label{Fig:photo_LW}
			\includegraphics[width=0.8\columnwidth]{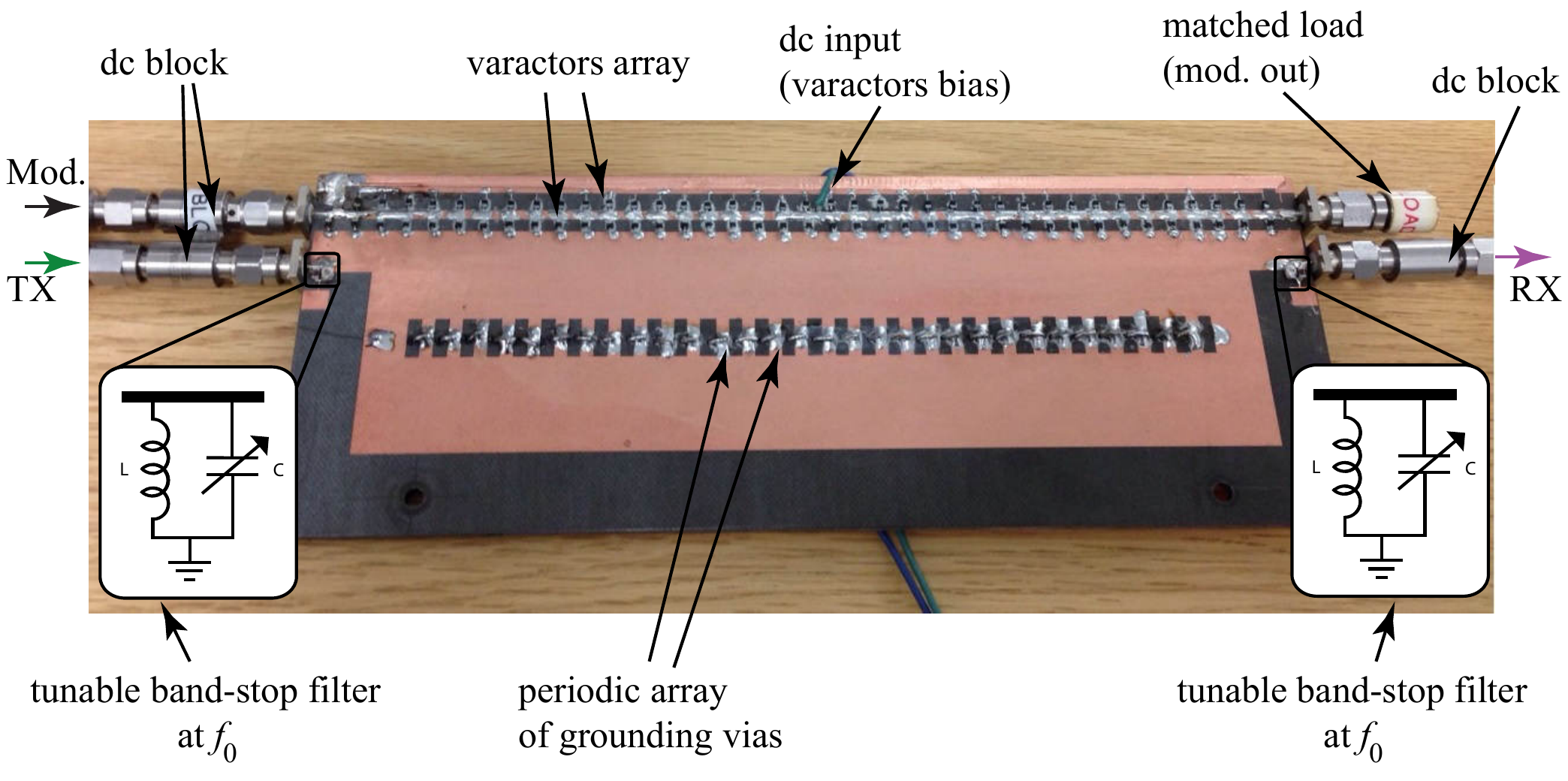} }
		\caption{Space-time dispersion tailoring in the radiation region to realize a leaky-wave transceiver operating as a mixer-duplexer-antenna. (a) Dispersion diagram illustrating nonreciprocal transitions within the radiation region~\cite{taravati2016mixer}. (b) Schematic of the transceiver's operational principle~\cite{taravati2016mixer}. (c) Photograph of the fabricated proof-of-concept prototype~\cite{taravati2016mixer}.}
	\end{center}
\end{figure}

Figures~\ref{Fig:disp_LW} to~\ref{Fig:photo_LW} demonstrate the operation principle and the fabricated space-time leaky-wave transceiver structure. The core functional engineering is achieved through nonreciprocal oblique transitions between these harmonics. Unlike static structures that induce horizontal transitions or purely time-modulated systems that yield vertical ones, the space-time medium creates diagonal pathways that simultaneously shift both frequency and wavenumber. In the uplink, an input signal at \((\omega_0, \beta_0)\) couples to the \(n=+1\) harmonic at \((\omega_1 = \omega_0 + \omega_\text{m}, \beta_1 = \beta_0 + \beta_\text{m})\), intrinsically performing upconversion. Conversely, in the downlink, an incoming wave at \((\omega_1, \beta_1)\) couples back to the fundamental harmonic, performing downconversion. Critically, the unidirectional nature of the modulation ensures these transitions are only permitted in the \(+z\) direction, as a wave propagating backward finds no phase-matched path. This engineered nonreciprocity is the very mechanism that enables the duplexing function, isolating the transmit and receive paths.

For radiation, the system is engineered to ensure the target harmonic operates in the leaky-wave regime. A leaky-wave antenna requires a "fast wave," where the phase constant is less than the free-space wavenumber (\(\beta < k_0\)). The critical design step is to ensure the \(n=+1\) harmonic used for uplink radiation satisfies this condition, \(\beta_1 < k_{01}\). This is achieved by starting with an unmodulated microstrip structure whose intrinsic \(EH_1\) mode is already a fast-wave leaky-mode, characterized by its dispersion \(\beta_{um}(\omega)\). The applied space-time modulation perturbs this fundamental dispersion only slightly under weak modulation, allowing the wavenumber of the radiating harmonic to be predicted as \(\beta_1 \approx \beta_{um}(\omega_0) + \beta_\text{m}\). By carefully selecting the modulation's spatial frequency \(\beta_\text{m}\), the designer can position \(\beta_1\) to satisfy the fast-wave condition at \(\omega_1\), while strategic use of grounding vias suppresses the fundamental mode to ensure efficient coupling into the radiating harmonic.

A powerful consequence of this architecture is dynamic beam steering at a fixed input frequency, a feat enabled by real-time dispersion tuning. In a conventional leaky-wave antenna, beam scanning requires changing the high-frequency input signal \(f_0\). In this space-time system, the beam is scanned by varying the low-frequency modulation signal \(f_\text{m}\). This action directly reconfigured the dispersion diagram: the radiation frequency becomes \(f_1 = f_0 + f_\text{m}\) and the wavenumber becomes \(\beta_1 \approx \beta_{um}(f_0) + \beta_\text{m}(f_\text{m})\). According to the radiation angle equation, \(\theta_1 = \sin^{-1}\left( {c(\beta_0 + \beta_\text{m})}/{(\omega_0 + \omega_\text{m})} \right)\), simultaneously altering the numerator and denominator results in a continuous shift of the beam angle \(\theta_1\). In conclusion, this approach masterfully demonstrates that a space-time modulated metamaterial is not merely a medium with altered properties, but a programmable dispersion engine. By injecting a control wave, one can sculpt the \((\omega, \beta)\) diagram in real-time, creating and destroying pathways for signals to mix, be isolated, and radiate in a highly controlled manner. This represents a paradigm shift from static, fixed-function devices toward dynamic, multi-functional, and software-defined electromagnetic systems.

\subsection{Phase-Engineered Time Transitions}

\begin{figure}
	\begin{center}
			\includegraphics[width=1\columnwidth]{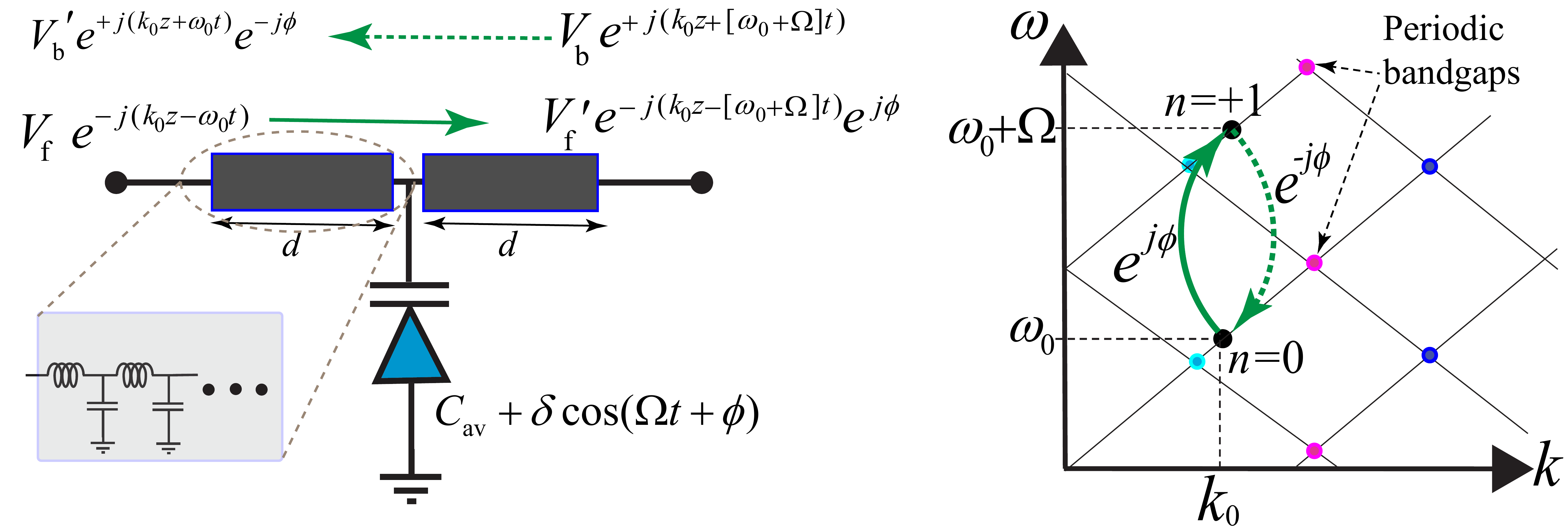}
		\caption{Asymmetric phase shifting in a phase-engineered time-periodic transmission line~\cite{Taravati_AMTech_2021}.}
		\label{Fig:Phase}
	\end{center}
\end{figure}

The principles of nonreciprocal frequency and phase transitions in the phase-engineered temporal transmission line is shown in Fig.~\ref{Fig:Phase}. These principles can be broken down into two key phenomena: the nature of the transitions themselves and the critical role of the modulation phase, which together break Lorentz reciprocity. The core mechanism is a linear, time-varying capacitance
\begin{equation}
 C_{\text{eq}}(t) = C_{\text{av}} + \delta \cos(\Omega t + \phi). 
	\label{eqa:C}
\end{equation}

Unlike a nonlinear element, this component changes its value at every point in space uniformly over time. This temporal periodicity, as dictated by the Floquet theorem, results in a dispersion diagram that is periodic in frequency. This means that a wave with a wavenumber \( k_0 \) at the fundamental frequency \( \omega_0 \) is coupled to space-time harmonics at frequencies \( \omega_0 + n\Omega \), all sharing the same wavenumber \( k_0 \). Graphically, this appears as vertical transitions in the \( \omega-k \) diagram. An input signal does not follow a single dispersion curve; it can make a discrete jump to another curve at the same wavenumber, thus changing its frequency.

The nonreciprocity arises from the fact that the phase of the frequency transition is dictated by the phase of the modulation, \( \phi \), and, crucially, this phase addition depends on the direction of the transition (up- or down-conversion). i)~Forward Path (e.g., Left-to-Right Propagation) corresponds to up-conversion (\( \omega_0 \rightarrow \omega_0 + \Omega \)): When the fundamental harmonic is converted to a higher frequency, the generated harmonic acquires a positive phase shift equal to the modulation phase, \( +\phi \). The analytical solution in the paper (Eq. 6b), \( a_1(z) \propto e^{+i\phi} \), confirms this. Physically, the newly generated harmonic is born with a phase that is synchronized with the modulation pump at that moment. ii)~Backward Path (e.g., Right-to-Left Propagation) corresponds to down-conversion (\( \omega_0 + \Omega \rightarrow \omega_0 \)): When a higher harmonic is converted back down to the fundamental frequency, the process is not the simple inverse of up-conversion. The down-converted fundamental acquires a negative phase shift, \( -\phi \), relative to the pump (Eq. 7b, \( a_0(z) \propto e^{-i\phi} \)).

This asymmetry is the key to nonreciprocity. The phase acquired by a signal depends not just on its frequency, but on its path through the frequency spectrum. A complete round trip through the system (e.g., up-conversion and then down-conversion) results in a net phase difference that is zero \( +\phi + (-\phi) = 0 \). Therefore, by cascading multiple such phase-engineered temporal transmission lines with carefully chosen modulation phases (\( \phi_1, \phi_2 \)) and amplitudes (\( \delta_1, \delta_2 \)), one can create components (as discussed in Sections.~\ref{Sec:Isol},~\ref{Sec:NRPh},~\ref{Sec:NRBM} and~\ref{Sec:Circ}) where the total phase accumulation for a signal traveling from Port 1 to Port 2 leads to constructive interference, while the phase accumulation for a signal traveling from Port 2 to Port 1 leads to destructive interference. 

\subsubsection{Isolator Formed by Two Temporal Loops}~\label{Sec:Isol}

\begin{figure}
	\begin{center}
				\subfigure[]{ \label{Fig:loopIso_disp}
			\includegraphics[width=0.6\columnwidth]{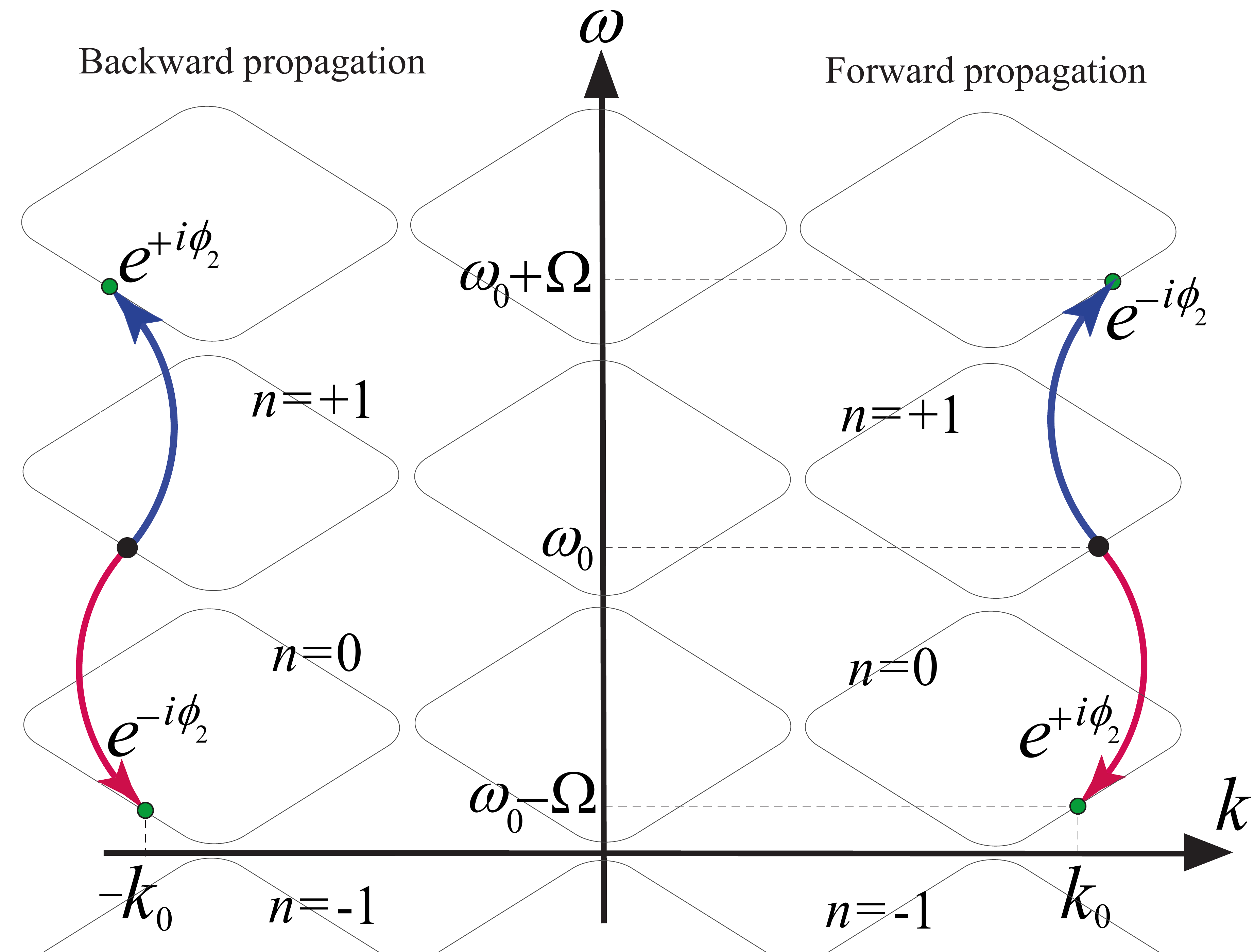}}
		\subfigure[]{ \label{Fig:loopIso_conc}
			\includegraphics[width=0.8\columnwidth]{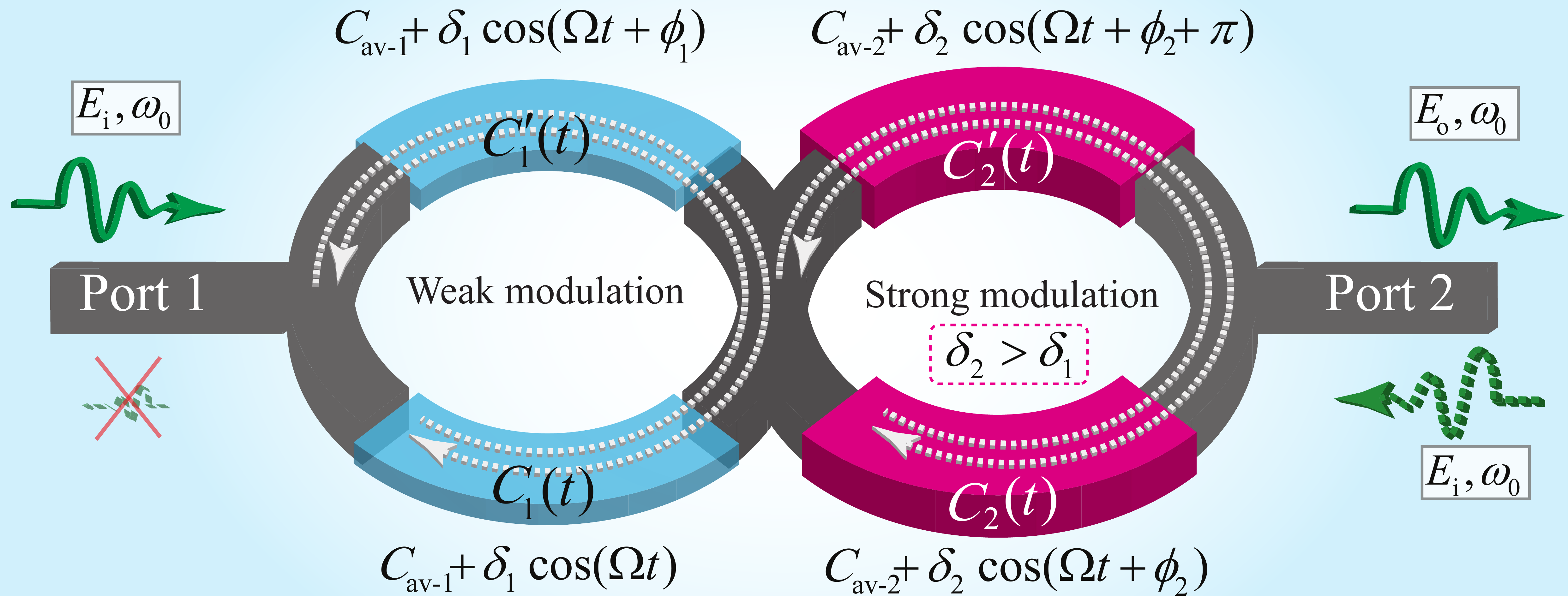}}
		\subfigure[]{ \label{Fig:loopIso_F}
			\includegraphics[width=0.47\columnwidth]{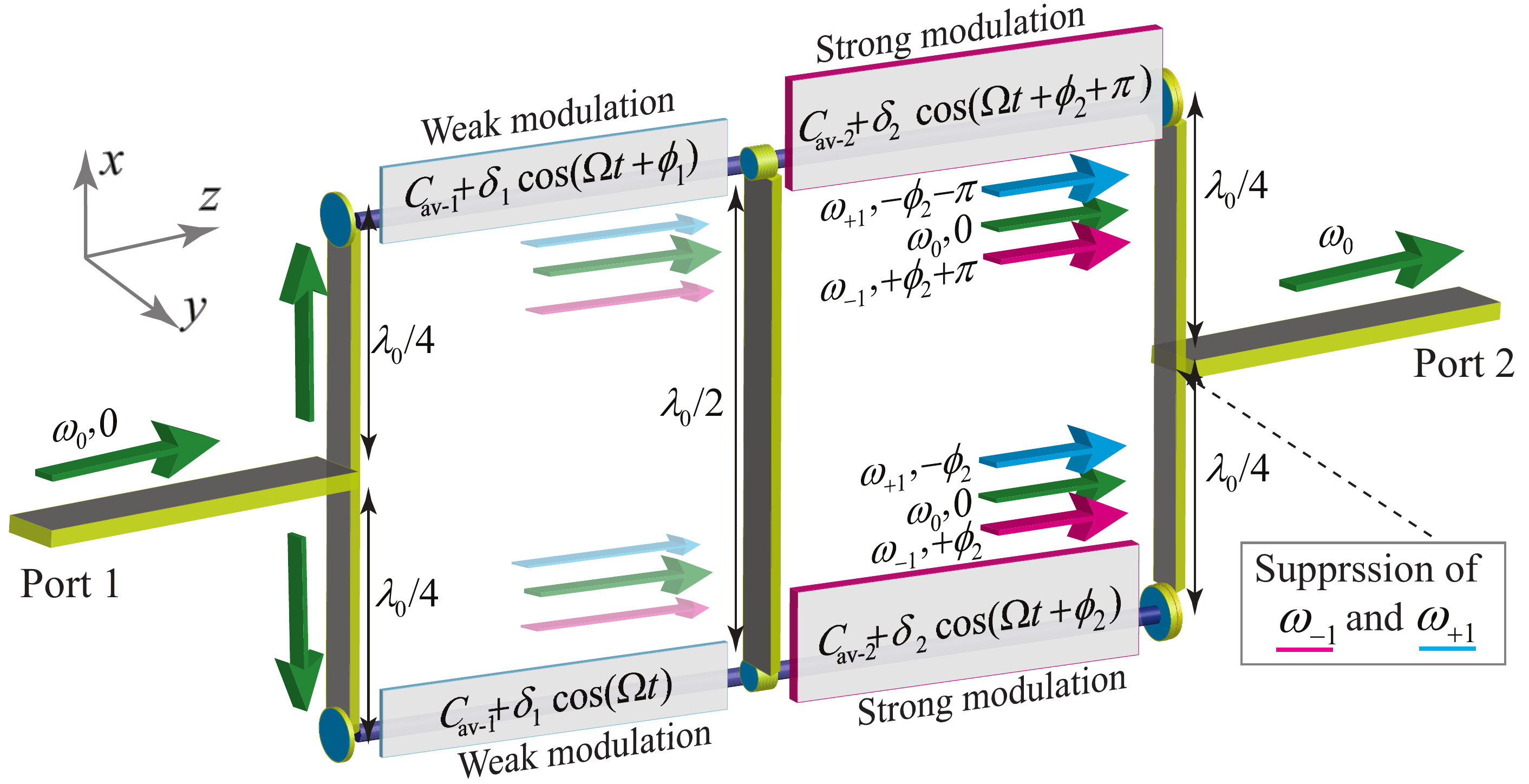}}
		\subfigure[]{ \label{Fig:loopIso_B}
			\includegraphics[width=0.47\columnwidth]{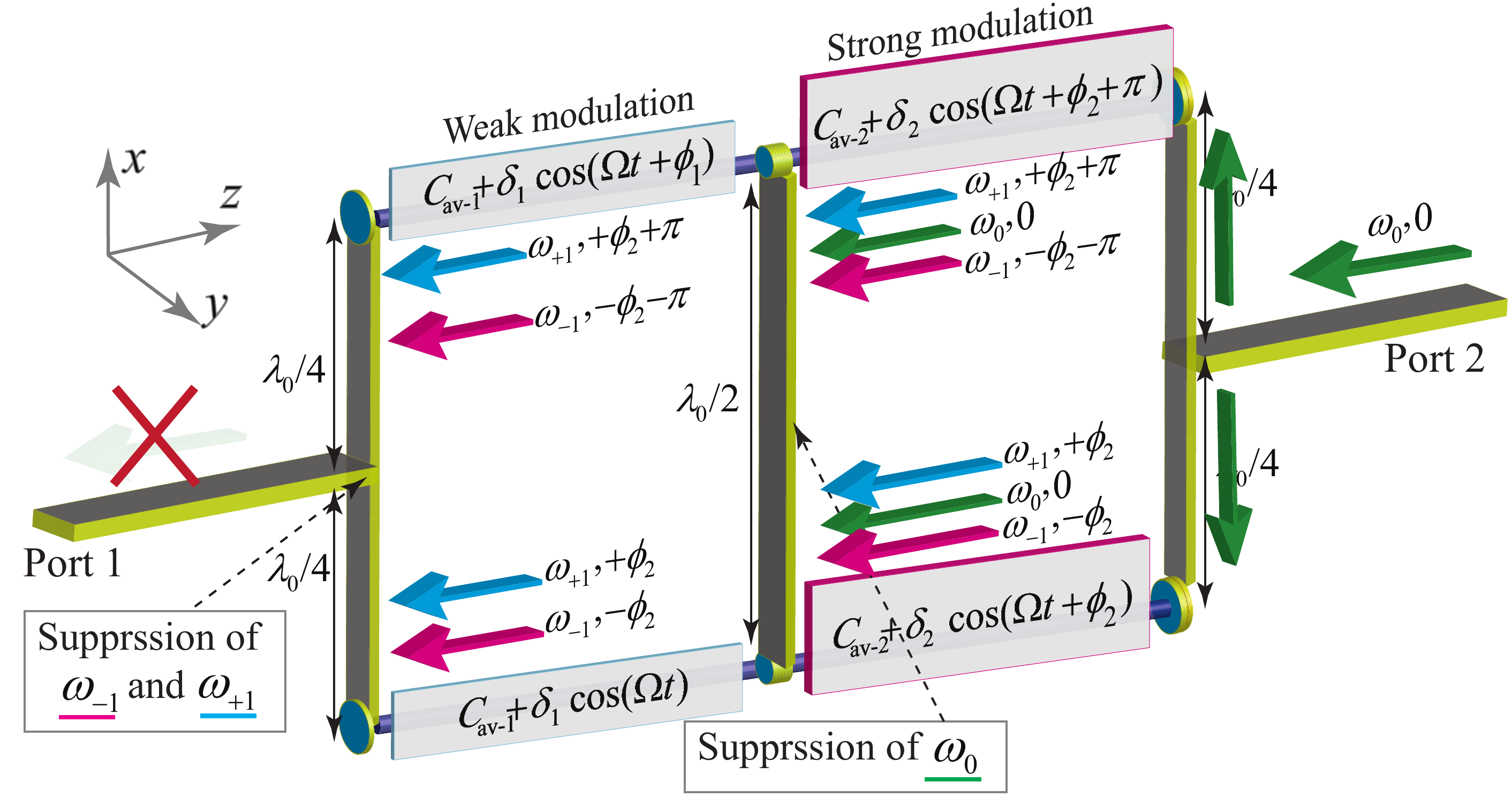}}
		\subfigure[]{ \label{Fig:loopIso_Photo}
			\includegraphics[width=0.6\columnwidth]{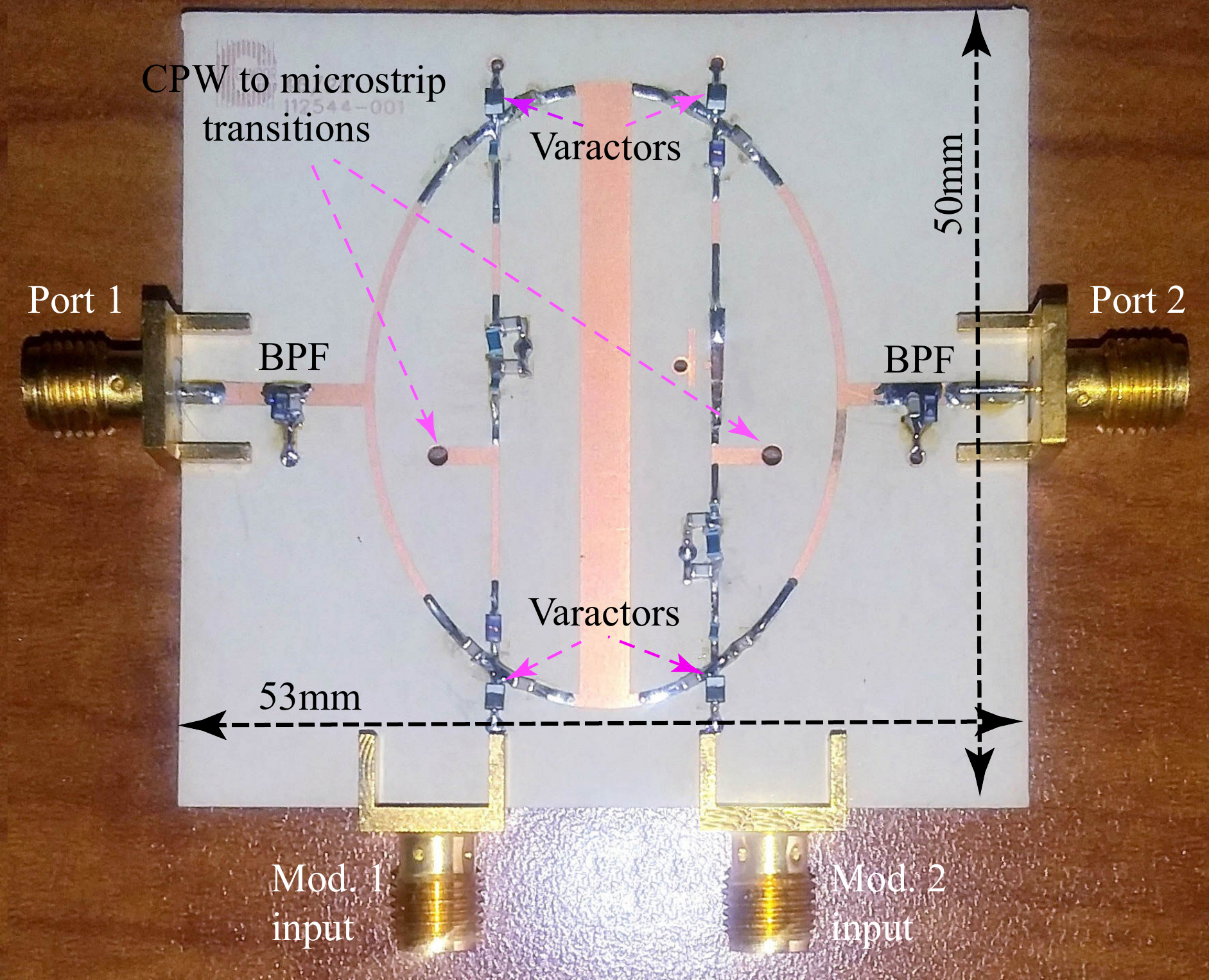}}
		\caption{Realization of an isolator based on time-modulated loops. (a)~Dispersion diagram showing  asymmetric phase shifts and frequency conversions. (b)~Schematic of the loop-based phase-engineered isolator, where one loop possesses a strong modulation amplitude and the other a weak one~\cite{Taravati_AMTech_2021}. (c, d)~Schematic visualization of forward and backward polychromatic signal transmission, demonstrating isolation~\cite{Taravati_AMTech_2021}. (e)~Photograph of the fabricated prototype~\cite{Taravati_AMTech_2021}.}
	\end{center}
\end{figure}

Figures~\ref{Fig:loopIso_disp} to~\ref{Fig:loopIso_Photo}  illustrate the realization of an isolator through the strategic control of frequency and phase transitions in time-modulated transmission lines, thereby achieving nonreciprocal signal propagation. Unlike space-time modulated systems that rely on progressive wave interaction over long distances, this design leverages the dispersion properties of compact, lumped temporal loops. The core principle is that a transmission line with a time-varying capacitance as in Eq.~\eqref{eqa:C}, possesses a dispersion diagram that is periodic along the frequency (\(\omega\)) axis. This periodicity means that the wavenumber is the same for frequencies separated by the modulation frequency, i.e., \(k(\omega_0) = k(\omega_0 + \Omega)\). This unique dispersion characteristic enables vertical transitions in the \(\omega-k\) diagram, allowing an input signal to be converted to a new frequency (\(\omega_0 \rightarrow \omega_0 \pm \Omega\)) while acquiring a specific, engineered phase shift (\(+\phi\) for up-conversion, \(-\phi\) for down-conversion) that is dictated by the phase of the modulation signal itself.

The isolator is realized by architecting a system of two such temporal loops, each comprising a pair of these phase-shifted, time-varying transmission lines. The overall nonreciprocal function is achieved by meticulously engineering the constructive and destructive interference of the fundamental and converted harmonic signals at strategic nodes within the circuit. For forward propagation, the dispersion is engineered so that the fundamental harmonic (\(\omega_0\)) constructively combines at the output port, while the first higher-order time harmonics (\(\omega_0 \pm \Omega\)) destructively interfere and are suppressed. This is facilitated by transmission line sections of specific electrical lengths (\(\lambda/4\) and \(\lambda/2\) arms) that act as interferometers, guiding the phase-shifted signals to recombine in a desired manner. Crucially, the asymmetry between the two loops, specifically a much larger modulation amplitude in the second loop, ensures that the harmonic generation process is nonreciprocal.

In the backward direction, the same dispersion principles and circuit layout lead to a completely different interference pattern. The fundamental harmonic is now engineered to experience destructive interference at the middle of the structure, effectively being blocked. While the higher-order harmonics may constructively interfere at this point, the weak modulation of the first loop prevents the efficient regeneration of the fundamental frequency needed for transmission to the output port. Therefore, the dispersion engineering of the phase and amplitude of frequency conversions within the asymmetric loops ensures that a signal can propagate freely in one direction but is blocked in the reverse, realizing isolation through synthesized, nonreciprocal dispersion in the time domain rather than through magnetic materials or nonlinearities.

\subsubsection{Nonreciprocal Phase Shifter}~\label{Sec:NRPh}

\begin{figure*}
	\begin{center}
		\subfigure[]{ \label{Fig:NRPhSh_Disp}
			\includegraphics[width=0.45\columnwidth]{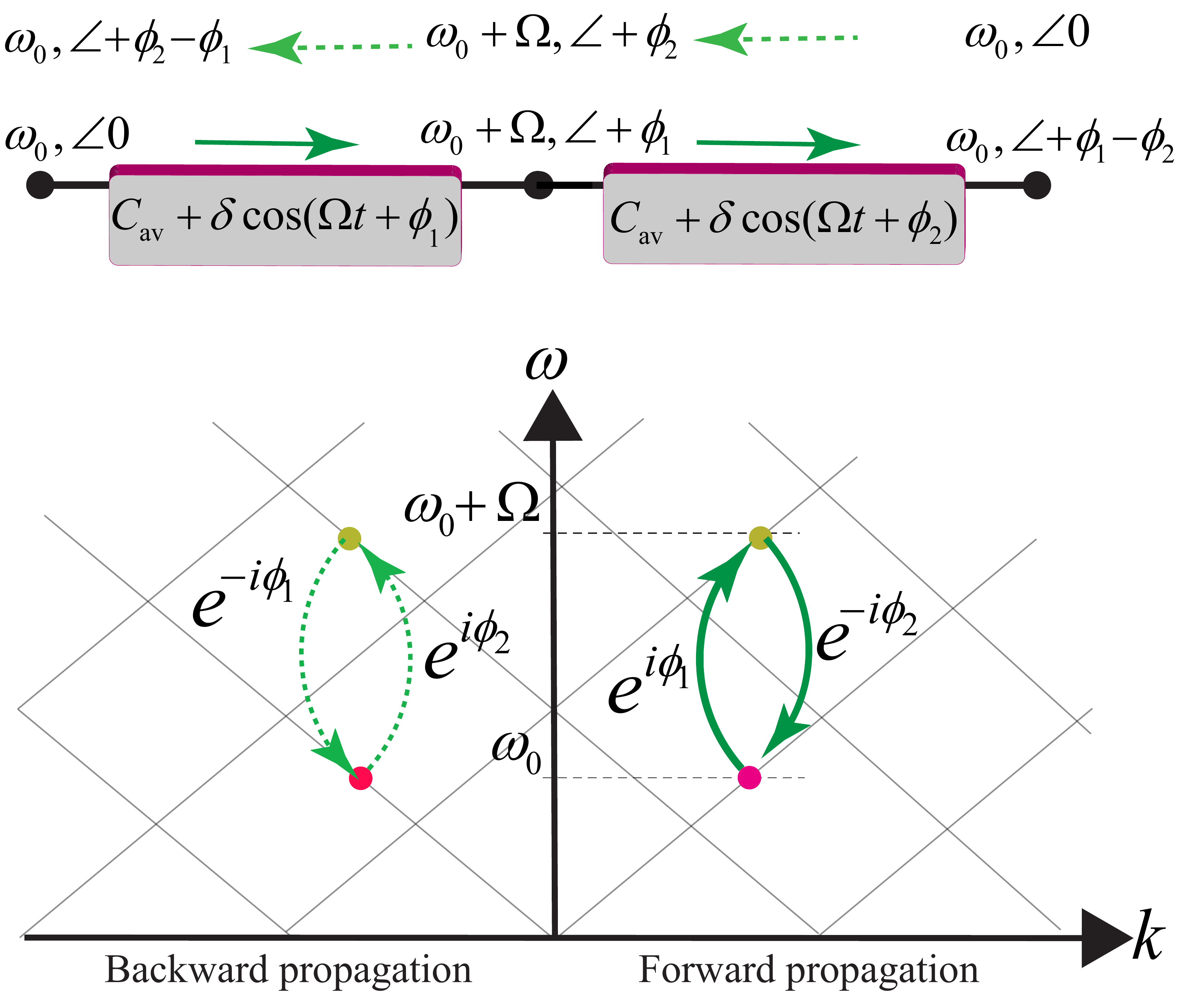}}
		\subfigure[]{ \label{Fig:NRPhSh_Oper}
			\includegraphics[width=0.54\columnwidth]{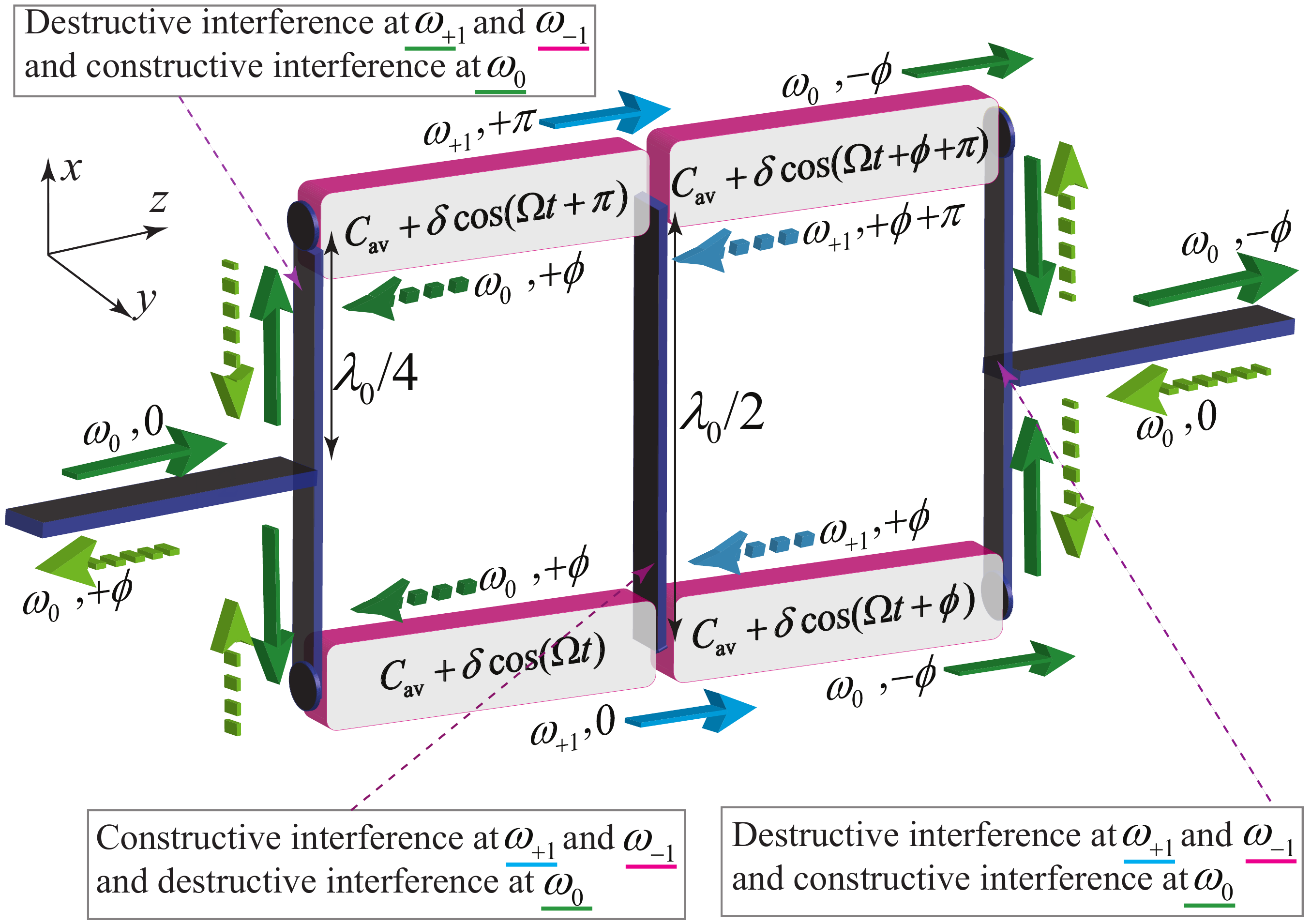}}
		\subfigure[]{ \label{Fig:NRPhSh_circ}
			\includegraphics[width=0.54\columnwidth]{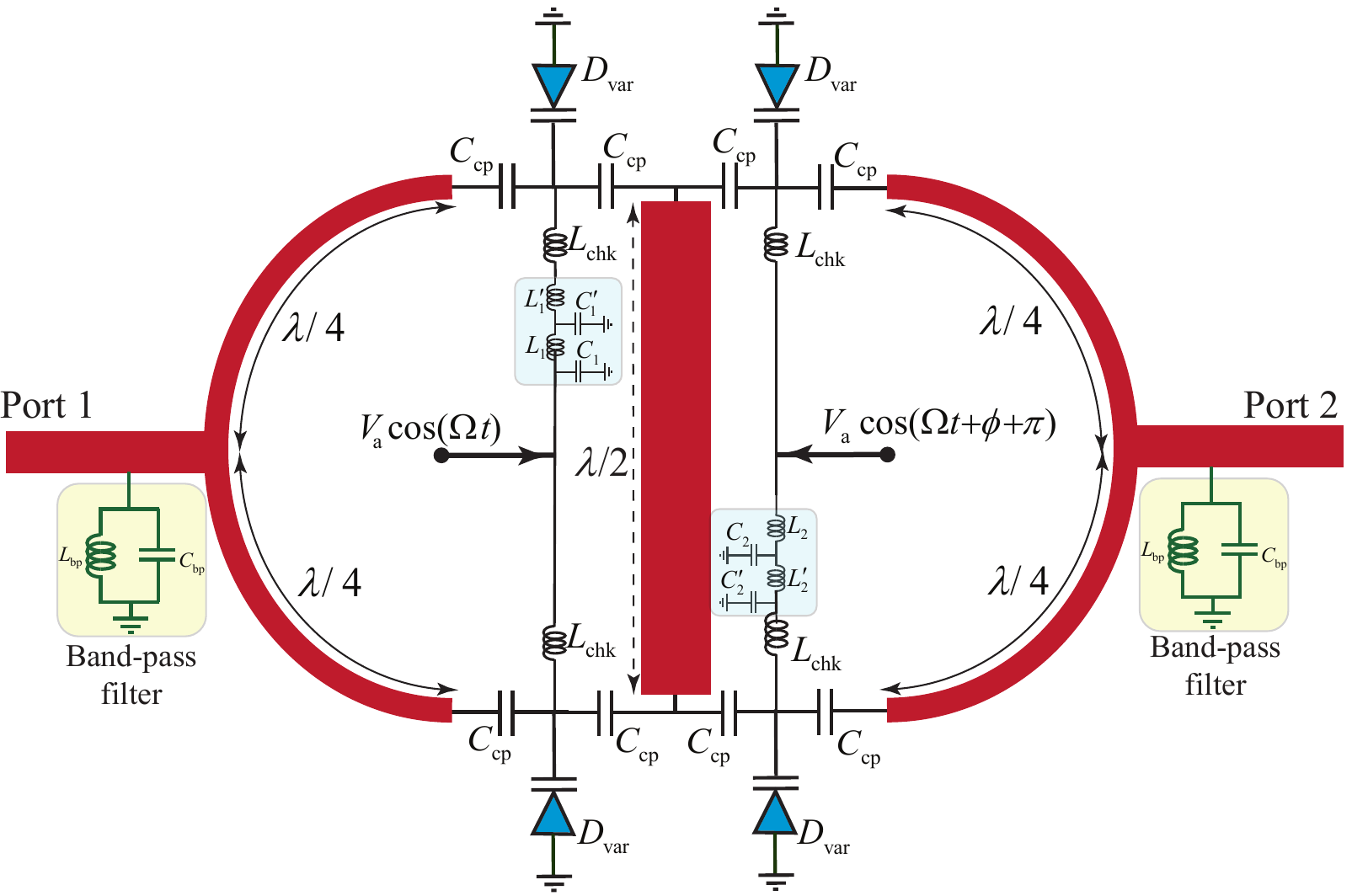}}
		\subfigure[]{ \label{Fig:NRPhSh_Photo}
			\includegraphics[width=0.42\columnwidth]{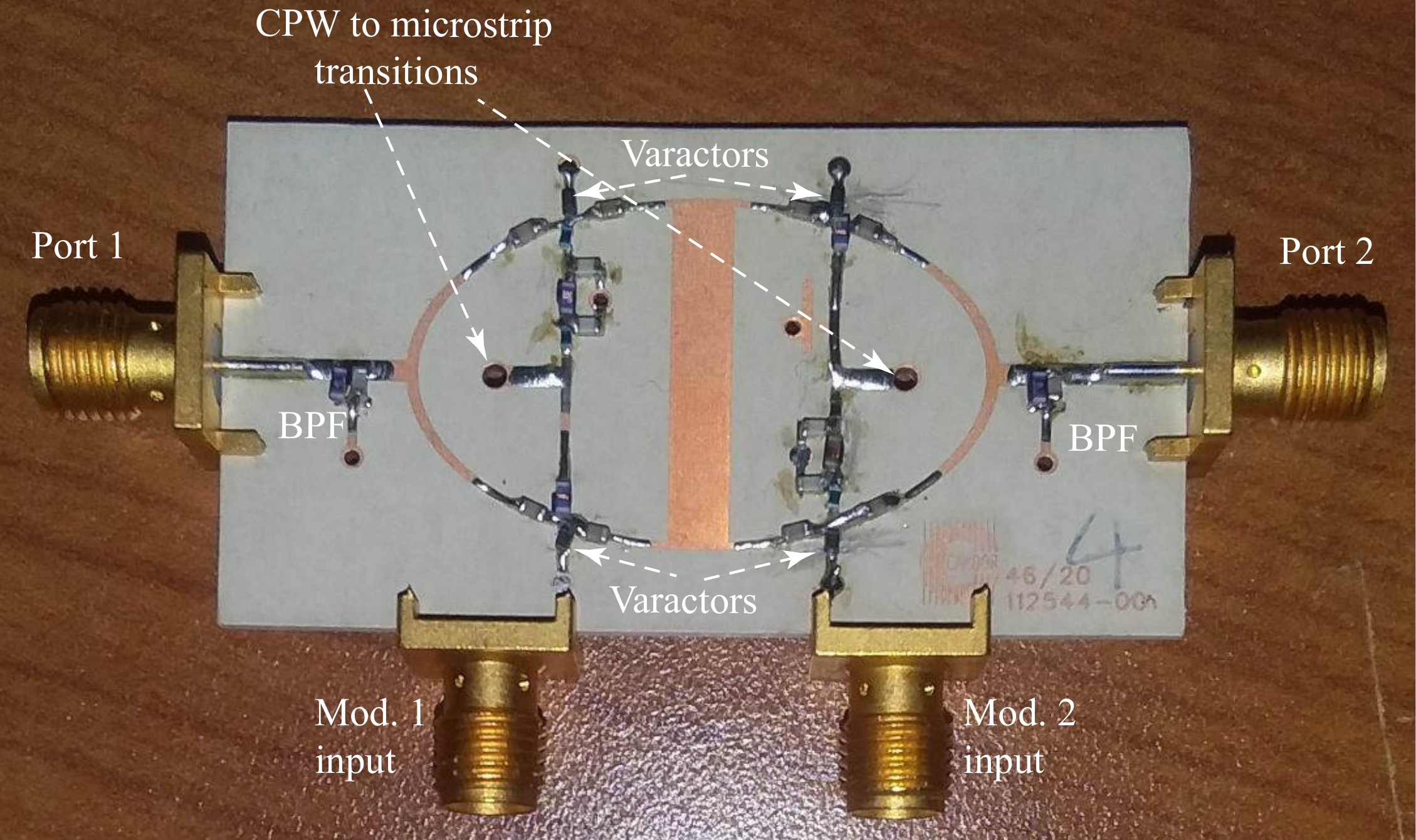}}
		\caption{Realization of an nonreciprocal phase shifter based on two dispersion-engineered time-modulated loops. (a)~Dispersion diagram showing nonreciprocal phase shifts~\cite{taravati2024nonreciprocal}. (b)~Schematic representation of the forward and backward polychromatic transmissions in the loop-based phase shifter~\cite{taravati2024nonreciprocal}. (c)~Circuit model~\cite{taravati2024nonreciprocal}. (e)~A photo of the fabricated nonreciprocal phase shifter~\cite{taravati2024nonreciprocal}.}
	\end{center}
\end{figure*}

Figures~\ref{Fig:NRPhSh_Disp} to~\ref{Fig:NRPhSh_Photo} provide the details for a nonreciprocal phase shifter composing two phase-engineered time-modulated loops~\cite{taravati2024nonreciprocal}. Unlike spatial modulation, which alters the structure along the direction of propagation, temporal modulation varies the electrical properties of the medium—specifically, the capacitance—over time. This creates a time-periodic system that supports an infinite number of time harmonics (THs), corresponding to frequency shifts of the input signal by integer multiples of the modulation frequency \(\Omega\).

The core principle lies in the asymmetric phase accumulation experienced by forward and backward propagating waves. When a signal passes through a time-modulated varactor-loaded TL, its frequency is converted—up-converted in one direction and down-converted in the other—while simultaneously acquiring a phase shift that depends on the modulation phase \(\phi\). For instance, a forward wave may undergo up-conversion with a phase shift of \(+\phi\), while a backward wave experiences down-conversion with a phase shift of \(-\phi\). This directional phase difference forms the basis of nonreciprocity.

To eliminate unwanted frequency conversion while preserving nonreciprocal phase shifting, the paper introduces a cascaded loop architecture with two temporal TLs modulated with different phases (\(\phi_1\) and \(\phi_2\)). In the forward direction, the signal undergoes up-conversion followed by down-conversion, resulting in no net frequency change but a net phase shift of \(+\phi_1 - \phi_2\). In the backward direction, the process is reversed, yielding a phase shift of \(+\phi_2 - \phi_1\). By ensuring \(\phi_1 \neq \phi_2\), a nonreciprocal differential phase shift of \(2(\phi_1 - \phi_2)\) is achieved.

Additionally, the loop-based design enforces constructive and destructive interference at specific time harmonics. The architecture suppresses spurious sidebands by ensuring that undesired harmonics interfere destructively at the output ports, while the desired fundamental harmonic is constructively reinforced. This is achieved through careful phase engineering of the modulation signals applied to the top and bottom branches of each loop, along with the use of power dividers and a half-wavelength interconnecting line. The result is a compact, magnet-free, and linear nonreciprocal phase shifter with low insertion loss and high power handling~\cite{taravati2024nonreciprocal}.

\subsubsection{Nonreciprocal-Beam Metasurface}~\label{Sec:NRBM}

\begin{figure*}
	\begin{center}
		\subfigure[]{ \label{Fig:sat}
			\includegraphics[width=0.45\columnwidth]{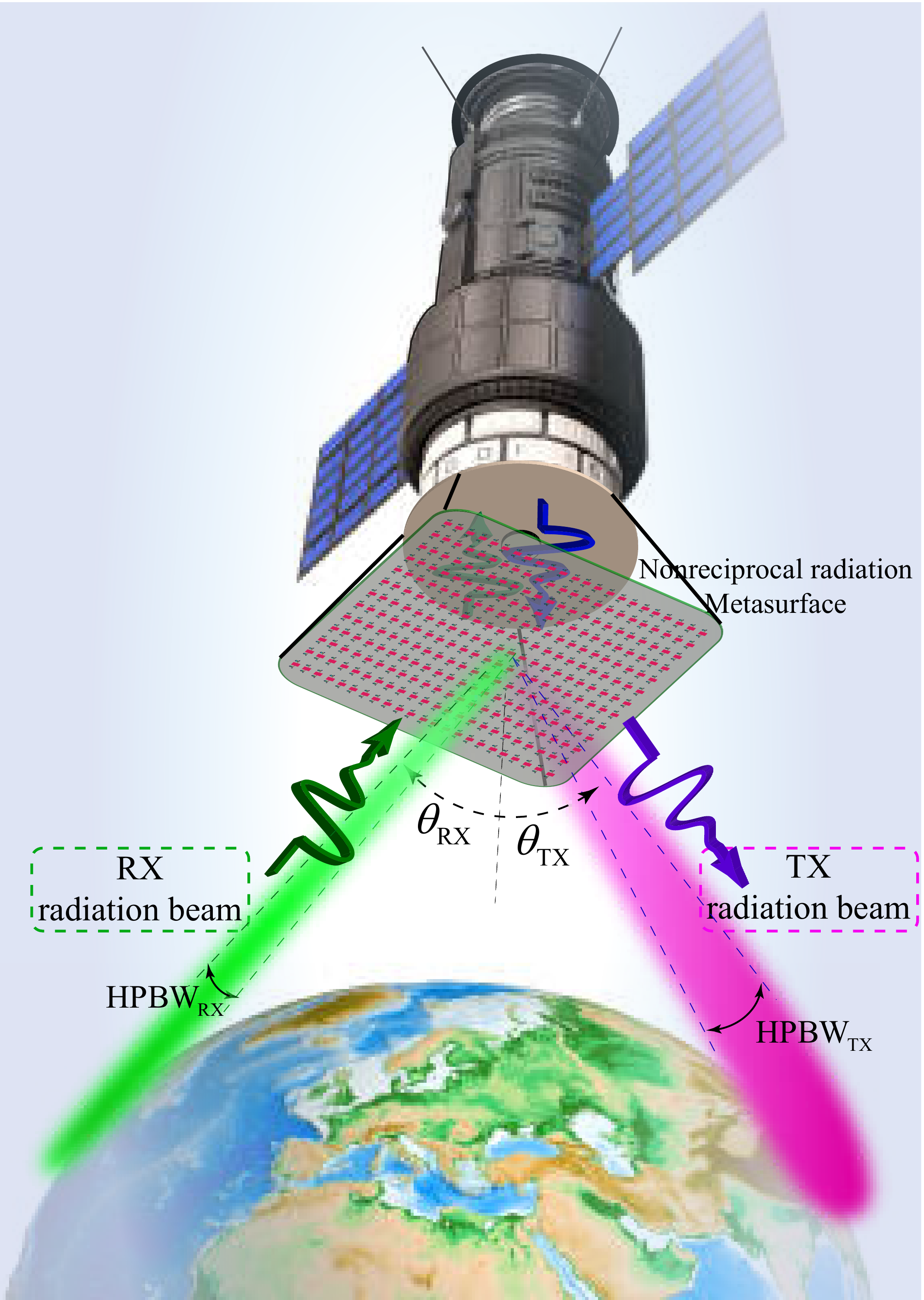}}
		\subfigure[]{ \label{Fig:twin}
			\includegraphics[width=0.49\columnwidth]{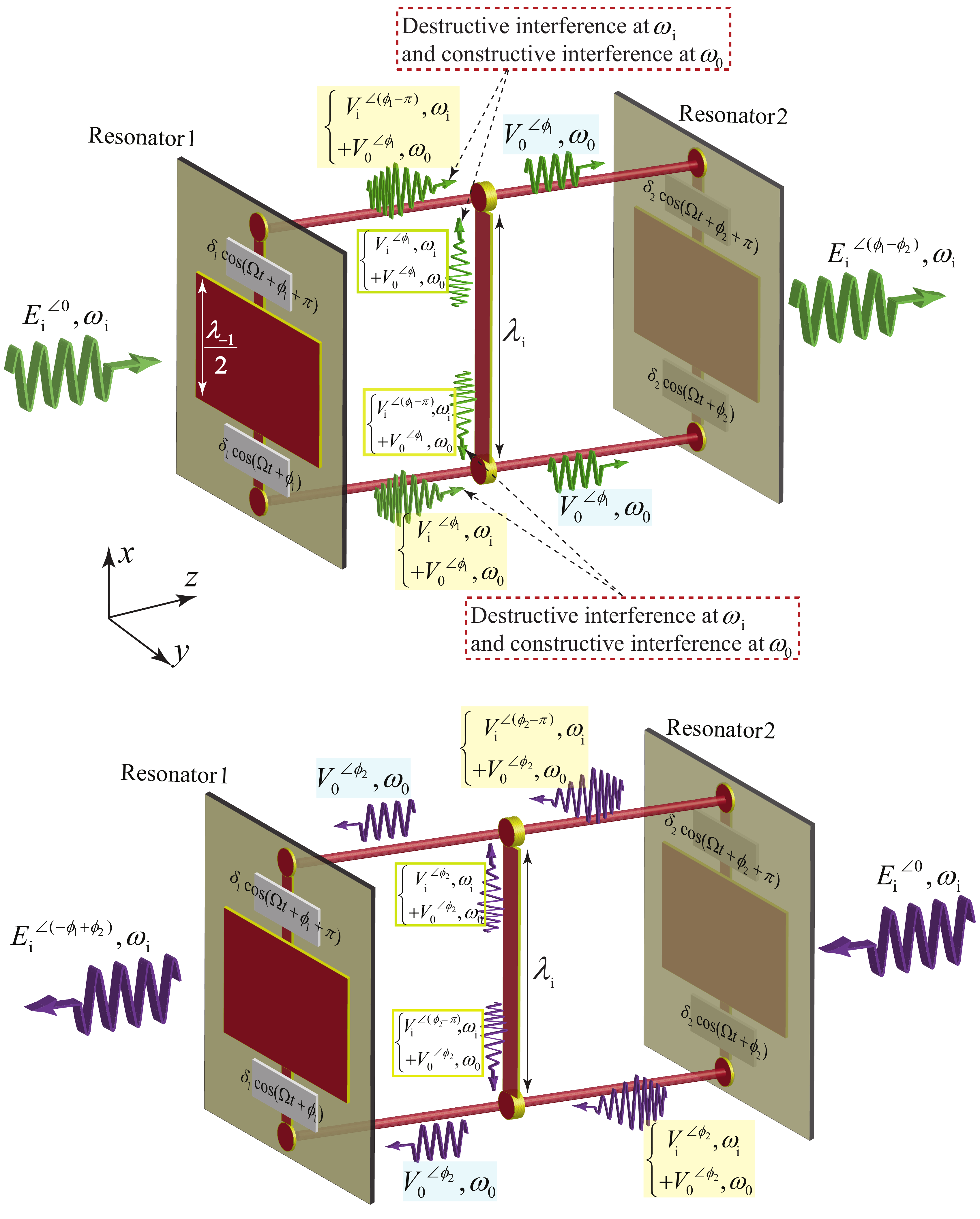}}
		\subfigure[]{ \label{Fig:NRBM}
			\includegraphics[width=0.4\columnwidth]{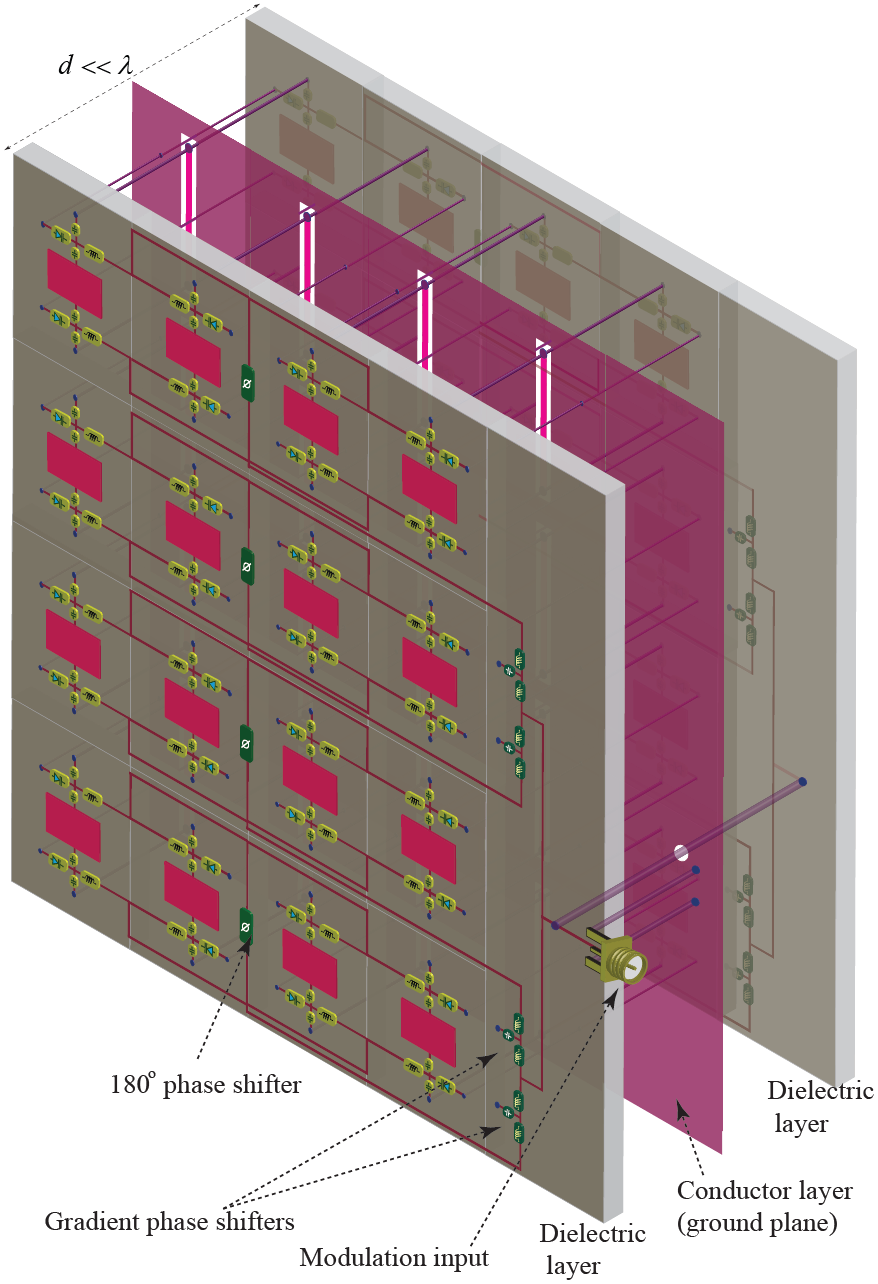}}
		\subfigure[]{ \label{Fig:Fab_NRBM}
			\includegraphics[width=0.6\columnwidth]{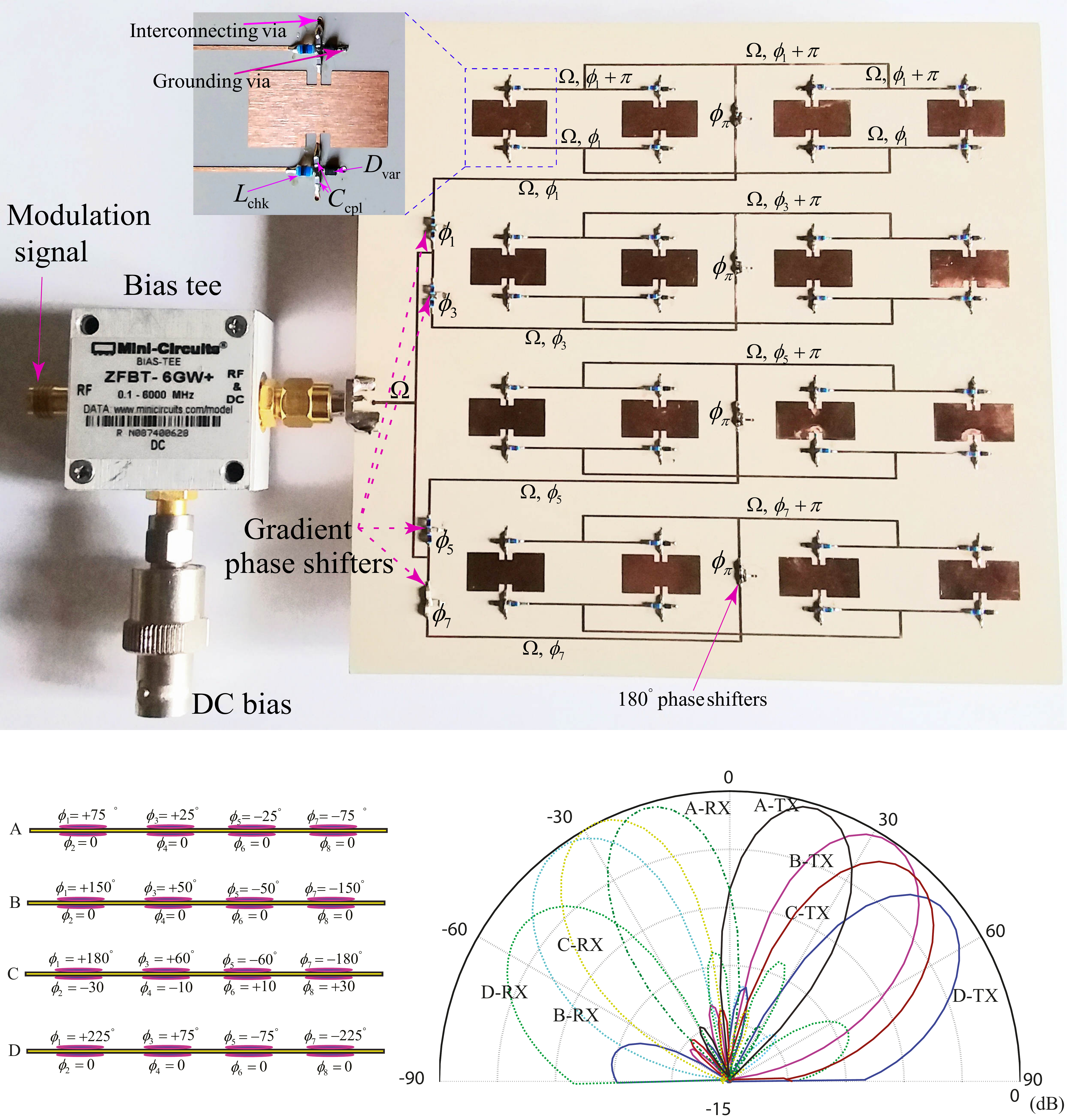}}
		\caption{Nonreciprocal-beam metasurface through asymmetric phase shifts with symmetric frequency conversions via a coupled-patches architecture. (a)~Schematic illustrating the operational principle~\cite{taravati2020full}. (b)~Forward and backward transmissions~\cite{taravati2020full}. (c)~Exploded view of the fabricated multi-layer structure~\cite{taravati2020full}. (d)~Photograph of the prototype and the measured nonreciprocal beam-steering patterns~\cite{taravati2020full}.}
	\end{center}
\end{figure*}

Next, we discuss the fundamental dispersion engineering mechanism enabling the nonreciprocal beam metasurface is the strategic induction of asymmetric frequency-phase transitions within "twin" time-modulated unit cells~\cite{taravati2020full,taravati2025nonreciprocal}, as shown in Figs.~\ref{Fig:sat} to~\ref{Fig:Fab_NRBM}. This approach allows the metasurface to impart a nonreciprocal phase gradient to a wave without altering its frequency, a key advancement over other time-modulated metasurfaces and antennas~\cite{zang2019nonreciprocal,zang2019nonreciprocal_metas,cardin2020surface}. The core principle leverages the unique dispersion properties of a medium with a time-periodic permittivity. The temporal modulation creates a dispersion diagram that is periodic in frequency, meaning a wave with wavenumber \(k_0\) can exist at frequencies \(\omega_i\) and \(\omega_i + \Omega\). Crucially, the system enables vertical transitions between these dispersion bands. The paper's theoretical model reveals that the phase of the modulation signal, \(\phi\), dictates the phase shift acquired during these transitions, but with a sign that depends on the transition direction: an up-conversion (\(\omega_i \rightarrow \omega_i + \Omega\)) imparts a phase shift of \(+\phi\), while a down-conversion (\(\omega_i + \Omega \rightarrow \omega_i\)) imparts a phase shift of \(-\phi\). This sign asymmetry is the foundational source of nonreciprocity.

The metasurface ingeniously cancels the frequency shift while preserving the nonreciprocal phase shift by cascading two such time-modulated unit cells into a "twin" configuration. In the forward direction (e.g., left-to-right), the first unit cell up-converts the signal from \((\omega_i, 0)\) to \((\omega_i + \Omega, \phi_1)\), and the second unit cell down-converts it back to the original frequency, resulting in a final phase of \(\phi_1 - \phi_2\). In the backward direction, the process is reversed: the signal traverses the second unit cell first, acquiring a phase of \(\phi_2 - \phi_1\). The net phase shift for the two directions is thus equal in magnitude but opposite in sign, realizing a frequency-invariant nonreciprocal phase shifter. By arranging an array of these twin cells, each programmed with a specific modulation phase (\(\phi_m\)), a linear phase gradient \(\partial\phi/\partial x\) is synthesized across the metasurface. This gradient steers the transmitted beam according to the generalized Snell's law, but crucially, the steering direction is nonreciprocal because the effective phase gradient is opposite for reception compared to transmission.

Furthermore, the design incorporates sophisticated dispersion engineering to suppress unwanted time harmonics, which is critical for efficiency and signal purity. The twin-unit-cell architecture is designed such that the two arms of each cell are modulated 180 degrees out of phase. This creates a destructive interference condition for all odd-order harmonics and a constructive one for even-order harmonics at a strategically placed interconnector. By designing the patch radiators to have a passband that only includes the fundamental (\(\omega_i\)) and the first even harmonic (\(\omega_i + \Omega\)), all other spurious harmonics are filtered out. This ensures that the system operates cleanly at the desired frequencies, enabling a highly efficient, full-duplex nonreciprocal beam-steering metasurface where the same physical aperture can transmit and receive signals at the same frequency but along independently steerable, highly directive beams.

\subsubsection{Circulator}~\label{Sec:Circ}

\begin{figure}
	\begin{center}
		\subfigure[]{ \label{Fig:cir}
			\includegraphics[width=0.7\columnwidth]{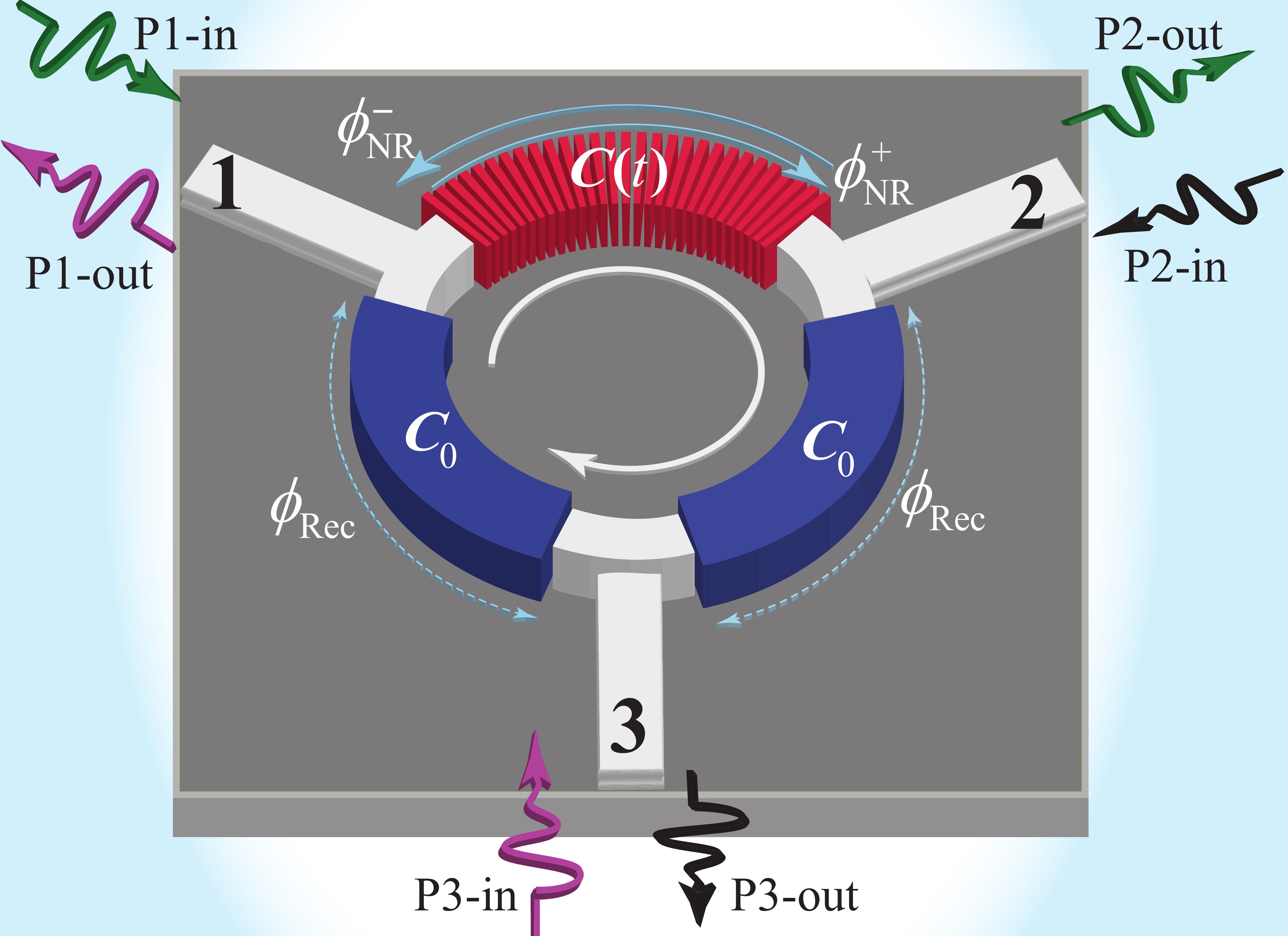}}
		\subfigure[]{ \label{Fig:circ_circ}
			\includegraphics[width=0.52\columnwidth]{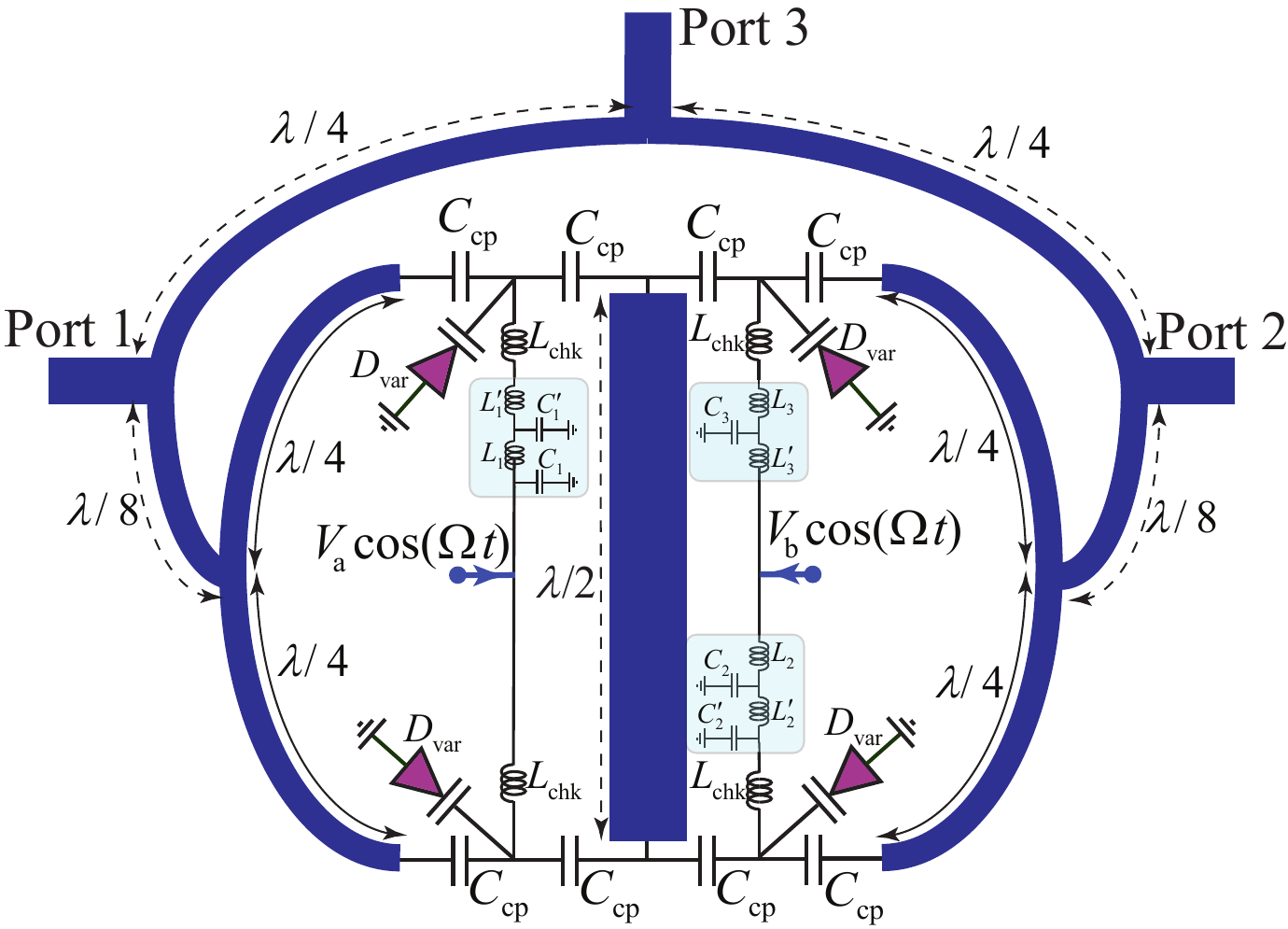}}
		\subfigure[]{ \label{Fig:photo_circ}
			\includegraphics[width=0.45\columnwidth]{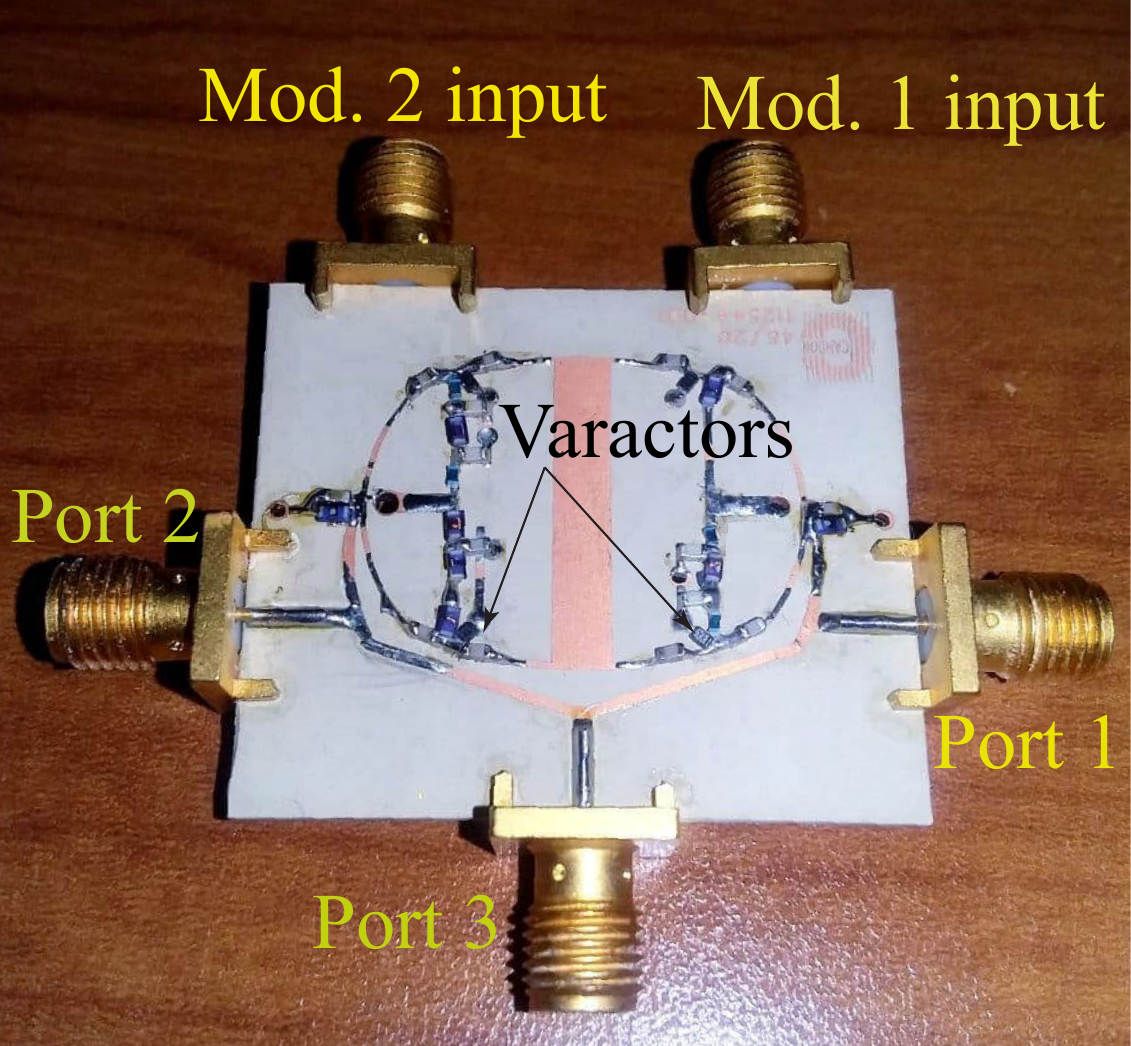}}
		\caption{Circulator based on nonreciprocal phase shifter. (a)~Operational principle of the device~\cite{taravati2022low}. (b)~Circuit schematic of the fabricated prototype~\cite{taravati2022low}. (c)~Photograph of the realized circulator~\cite{taravati2022low}.}
	\end{center}
\end{figure}

Figure~\ref{Fig:cir} to~\ref{Fig:photo_circ} This approach leverages time-varying capacitance, rather than magnetic materials or spatial gradients, to break Lorentz reciprocity and realize a nonreciprocal-phase-shifter-based circulator~\cite{taravati2022low}. The core principle is that a signal propagating through a time-modulated medium experiences asymmetric frequency and phase transitions depending on its direction of travel. The dispersion engineering is achieved by loading transmission lines with varactor diodes, whose capacitance is modulated in time as in Eq.~\eqref{eqa:C}. This temporal periodicity results in a dispersion diagram that is periodic in frequency, allowing for vertical transitions between time harmonics. Crucially, a forward-propagating wave (e.g., from port 1 to port 2) undergoes frequency up-conversion with a positive phase shift (\(+\phi\)), while a backward-propagating wave (from port 2 to port 1) experiences frequency down-conversion with a negative phase shift (\(-\phi\)). This direction-dependent phase accumulation is the genesis of nonreciprocity.

To construct a circulator, this nonreciprocal phase shifter is integrated into a three-port architecture alongside reciprocal phase shifters. The system is designed such that constructive interference occurs at the desired output port and destructive interference at the isolated port for each excitation scenario. For instance, when port 1 is excited, the nonreciprocal phase shifter and reciprocal delays are engineered to ensure signals constructively interfere at port 2 and destructively at port 3. This is achieved by satisfying specific phase conditions: \(\phi_{\text{rec}} = \pi/2\), \(\phi_{\text{NR}}^{+} = \pi\) (forward phase shift), and \(\phi_{\text{NR}}^{-} = 0\) (backward phase shift).

Furthermore, the loop-based architecture of the temporal phase shifter enforces controlled constructive and destructive interference for desired and spurious time harmonics. Using power dividers, combiners, and a half-wavelength interconnector, the structure ensures that the fundamental harmonic is transmitted with the desired nonreciprocal phase shift, while undesired sidebands are suppressed. This sophisticated dispersion engineering through phased time modulation enables a compact, linear, low-noise, and magnet-free circulator that is compatible with integrated circuit technology~\cite{taravati2022low}.

\subsection{Space-Time Coherency of Identical Dispersion Modes Offering Unidirectional Beamsplitting and Amplification}

A central feature of the work in~\cite{Taravati_PRB_SB_2017,Taravati_Kishk_PRB_2018} is the deliberate engineering of the three-dimensional dispersion landscape of space–time metamaterials to enforce synchronized interactions between \emph{identical dispersion modes}. This work leverages the unique coherent space-time-modulated metamaterials, where $\omega_\text{m} = 2\omega_0$. This spatiotemporal periodicity creates an infinite set of space-time harmonics (STHs), each with distinct temporal frequencies, $\omega_m = (1 + 2m)\omega_0$, and spatial frequencies, $k_{z,m} = \beta_0 + m \beta_{m}$. A key achievement of this dispersion engineering is the creation of a nonreciprocal synchronization condition between specific harmonics. For an obliquely incident wave at $45^\circ$, the design ensures that the fundamental harmonic ($m=0$) and the first lower harmonic ($m=-1$) share the same temporal frequency ($\omega_0$) but possess opposite spatial frequencies in the $x$-direction ($k_{x,-1} = -k_{x,0}$). This is visualized in the three-dimensional dispersion diagram (Fig. 3), where the isofrequency contours for the \textit{forward} propagating modes ($m=0$ and $m=-1$) are brought into close proximity, enabling strong, coherent energy exchange. In contrast, the corresponding \textit{backward} harmonics are widely separated, preventing significant interaction. This asymmetry, controlled by the ST velocity ratio, is the foundation of the one-way response.

Furthermore, the dispersion relation is engineered to yield specific desired transmission angles, with the $m$-th harmonic transmuted under the angle $\theta_{\text{T},m} = \sin^{-1}[\sin(\theta_{\text{I}})/(1 + 2m)]$. This results in the $m=0$ and $m=-1$ beams being transmitted at $+45^\circ$ and $-45^\circ$ respectively, achieving a $90^\circ$ beam split. The modulation parameters ($\gamma$, $\delta_\epsilon$, $\epsilon_{\text{av}}$) provide a lever to tune this splitting angle, the power division ratio between the two output beams, and even achieve transmission gain, particularly in the luminal ($\gamma=1$) regime where the closed-form solution reveals an exponential growth of the field amplitudes.

\begin{figure}
	\begin{center}
			\subfigure[]{ \label{Fig:3D_disp}
			\includegraphics[width=0.45\linewidth]{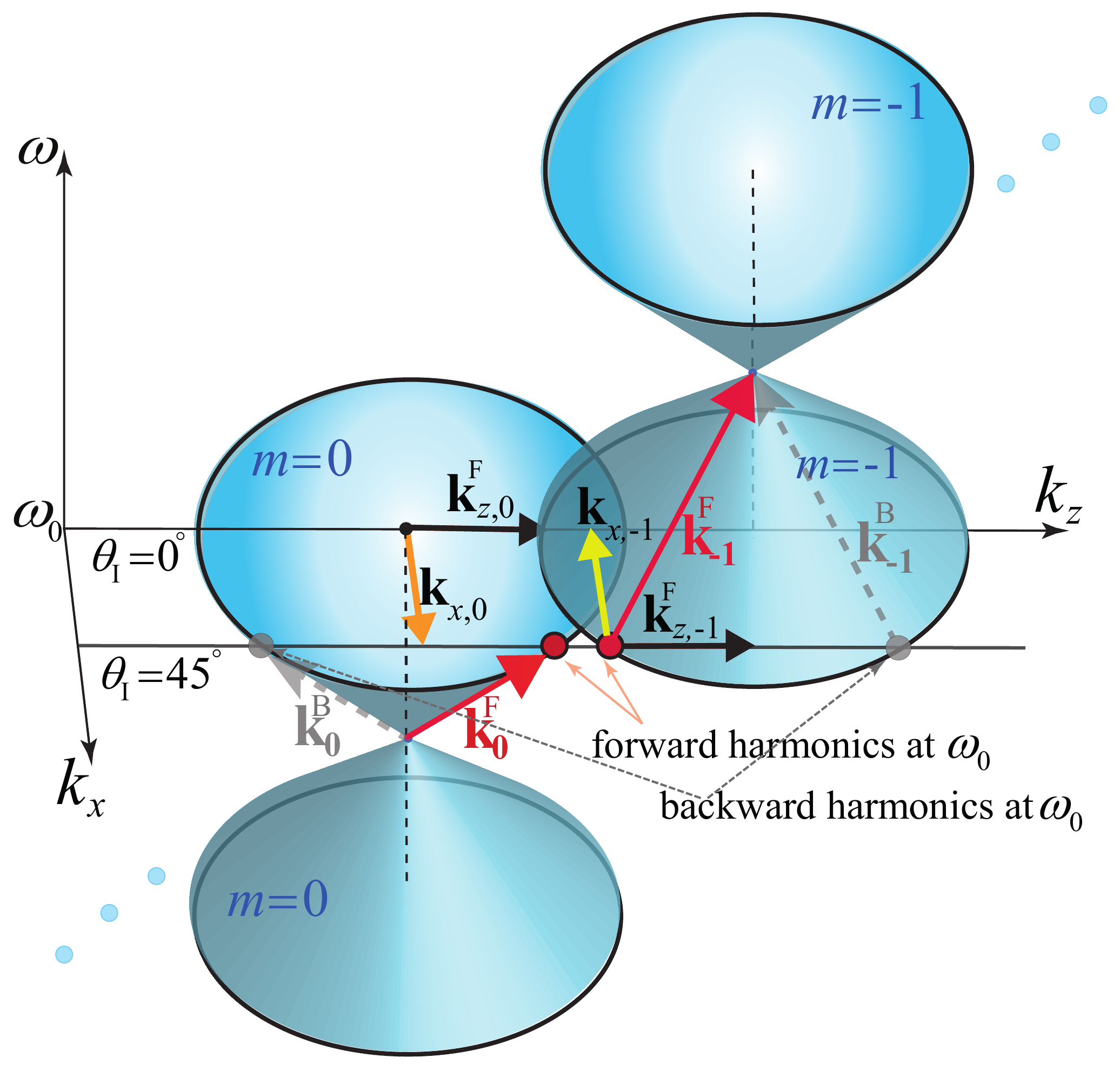}}
		\subfigure[]{ \label{Fig:BS}
			\includegraphics[width=0.45\linewidth]{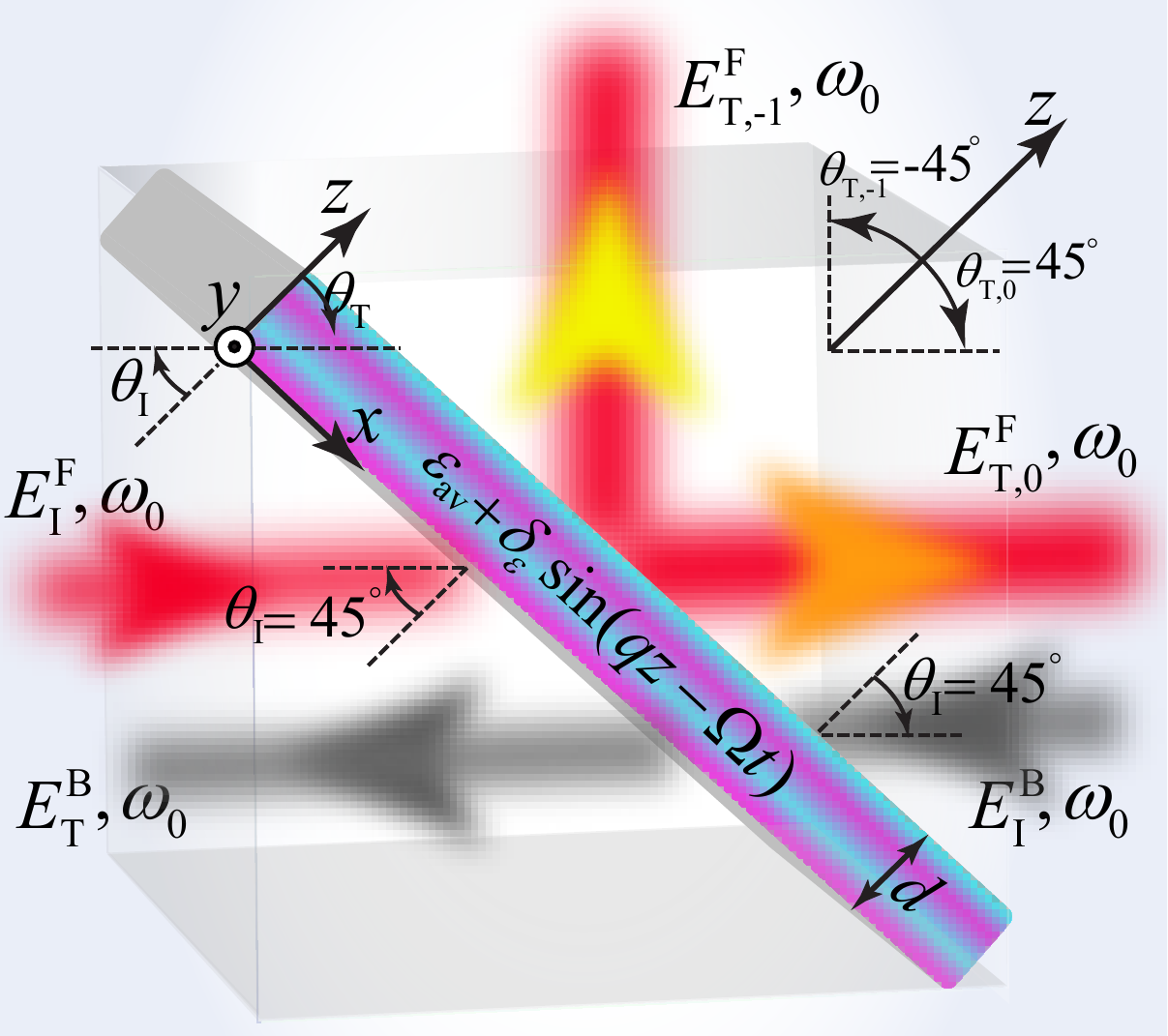}}
				\subfigure[]{\label{Fig:disp-a}
			\includegraphics[width=0.45\linewidth]{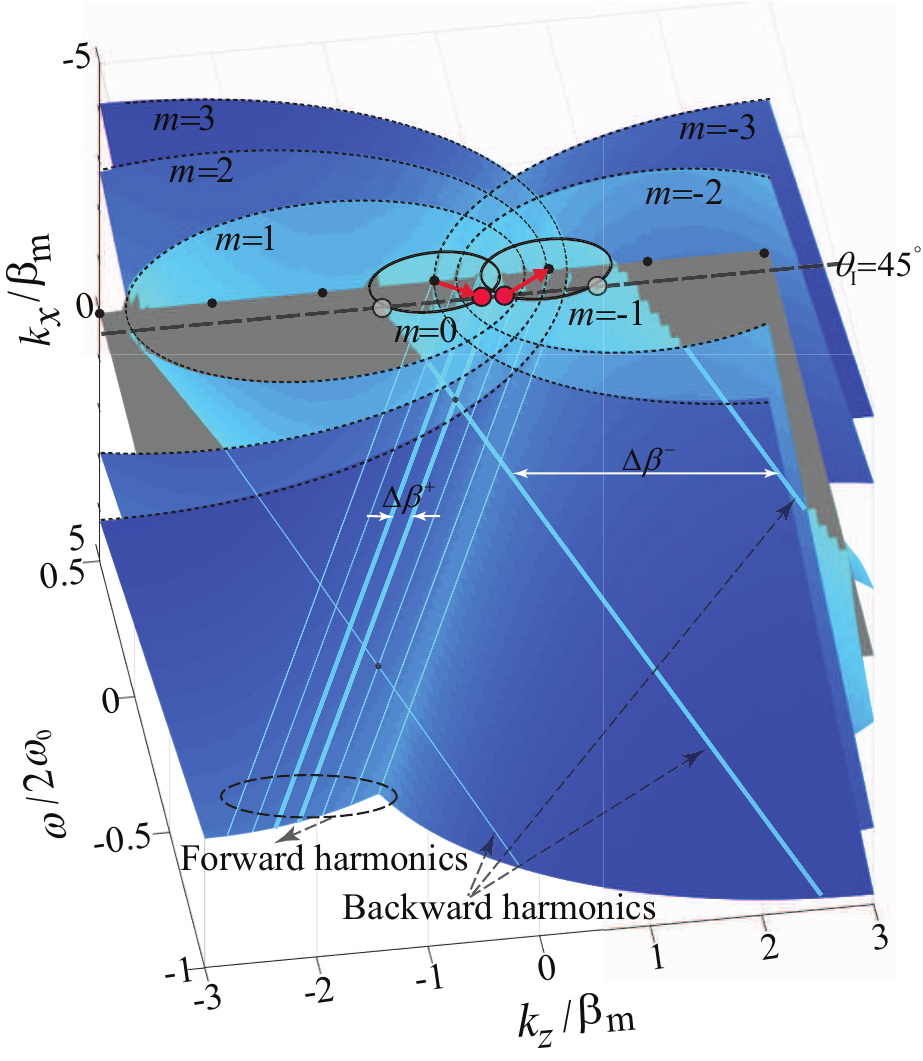}  }
		\subfigure[]{\label{Fig:disp-b}
			\includegraphics[width=0.45\linewidth]{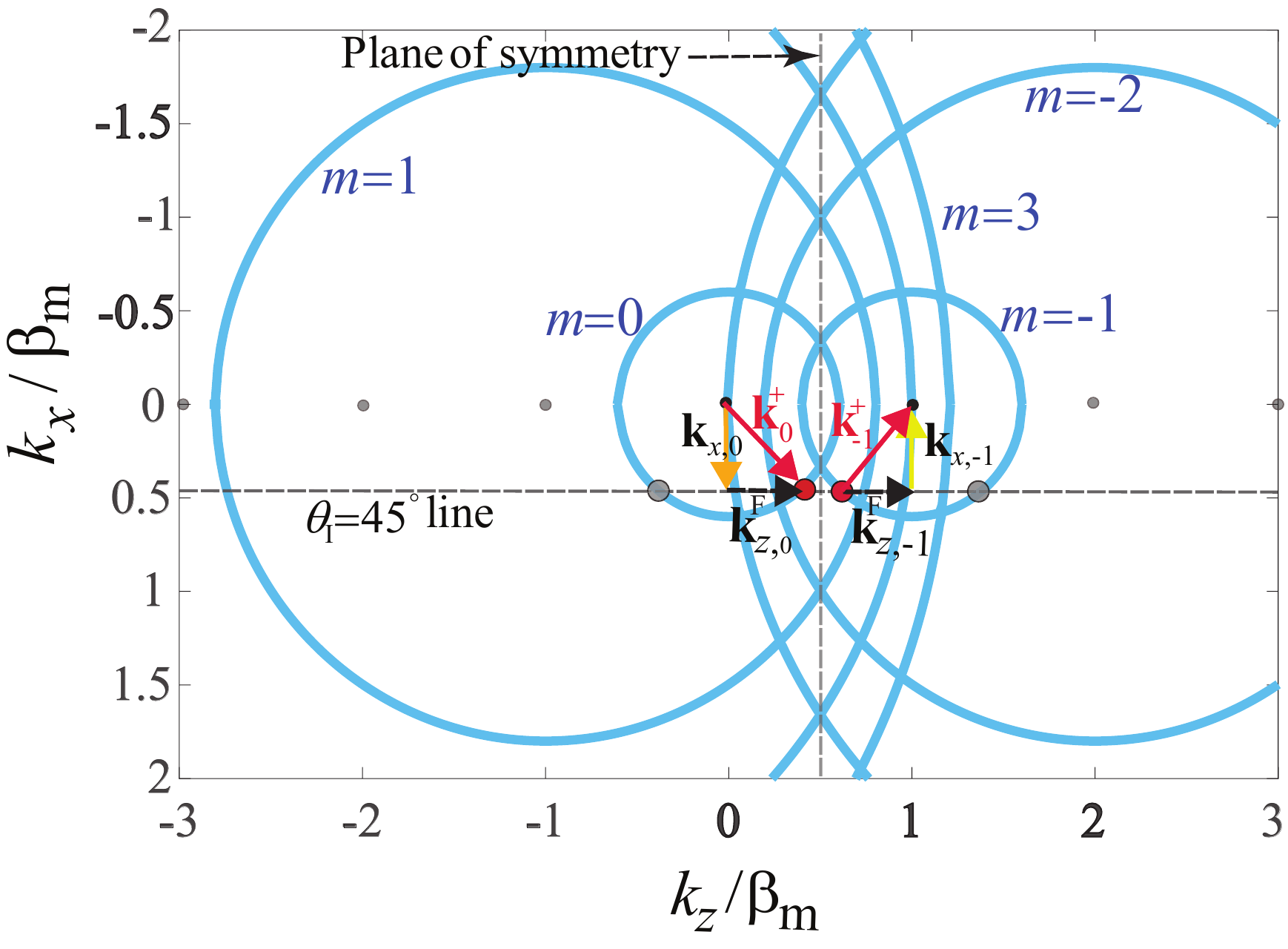}  }
		\caption{Realization of one-way beam splitting through creation of identical isofrequency circles for the $m=0$ and $m=-1$ space-time harmonics. (a)~Illustrative 3D dispersion diagram showing the operating principle~\cite{Taravati_Kishk_PRB_2018}. (b)~Schematic of the space-time-modulated beam splitter~\cite{Taravati_Kishk_PRB_2018}. (c)~Analytical 3D dispersion diagram~\cite{Taravati_Kishk_PRB_2018}. (d)~Analytical isofrequency diagram showing strong coupling between the $m=0$ and $m=-1$ time harmonics~\cite{Taravati_Kishk_PRB_2018}.}
		\label{Fig:3D_disp_sch}
	\end{center}
\end{figure}

\subsection{Space-Time Surface-Waves Engineering}

A sophisticated form of dispersion engineering that leverages space-time surface waves in a subluminal space-time modulated medium to create a multifunctional transceiver is proposed in~\cite{Taravati_PhNt_BS_2018,taravati2019_mix_ant,taravati2018space,Taravati_AMA_PRApp_2020}. The fundamental principle involves engineering the dispersion characteristics such that specific space-time harmonics transition between propagating waves and surface waves. The medium is designed with a spatiotemporally periodic electric permittivity, where the critical parameter is the space-time velocity ratio. In the subluminal regime, the dispersion diagram consists of an infinite periodic set of double cones with apexes at $\beta_{z,0} = -n\beta_\text{m}$, creating a complex landscape where specific harmonics can be engineered to exhibit purely imaginary wave vector components, thereby becoming surface waves confined to the boundaries of the medium.

The key innovation lies in the precise dispersion engineering that enables pure transitions between specific harmonics. By carefully selecting the modulation parameters and angle of incidence $\theta_i$, the system is designed such that the first lower space-time harmonic ($n = -1$) acquires a purely imaginary wave vector component $\gamma_{z,-1} = i\alpha_{z,-1}$ while maintaining a real tangential component. This creates a space-time surface wave that propagates along the interface while being evanescent in the normal direction. Simultaneously, the fundamental harmonic ($n = 0$) remains a propagating wave, allowing for efficient conversion between space waves and surface waves. The condition $\theta_{r,-1} = \theta_{t,-1} = 90^\circ$ ensures optimal coupling to surface waves, achieved when $\theta_i = \sin^{-1}(1 - \omega_\text{m}/\omega_0)$.

This engineered dispersion enables remarkable multifunctionality: in reception (down-link), an incident propagating wave at $\omega_0$ undergoes transition to a surface wave at $\omega_{\text{IF}} = \omega_0 - \omega_\text{m}$, while in transmission (up-link), a surface wave at $\omega_{\text{IF}}$ converts to a radiating wave at $\omega_0$. The complex dispersion diagram reveals electromagnetic band gaps at harmonic intersections that facilitate strong coupling between specific harmonics while suppressing others. The analytical solution shows that this approach can achieve power conversion gain in the down-link direction due to energy transfer from the modulation to the wave. This represents a paradigm where dispersion engineering enables a single medium to function simultaneously as antenna, mixer, amplifier, and filter.

\begin{figure}
	\begin{center} 
			\subfigure[]{\label{Fig:AMA_disp}
		\includegraphics[width=0.8\columnwidth]{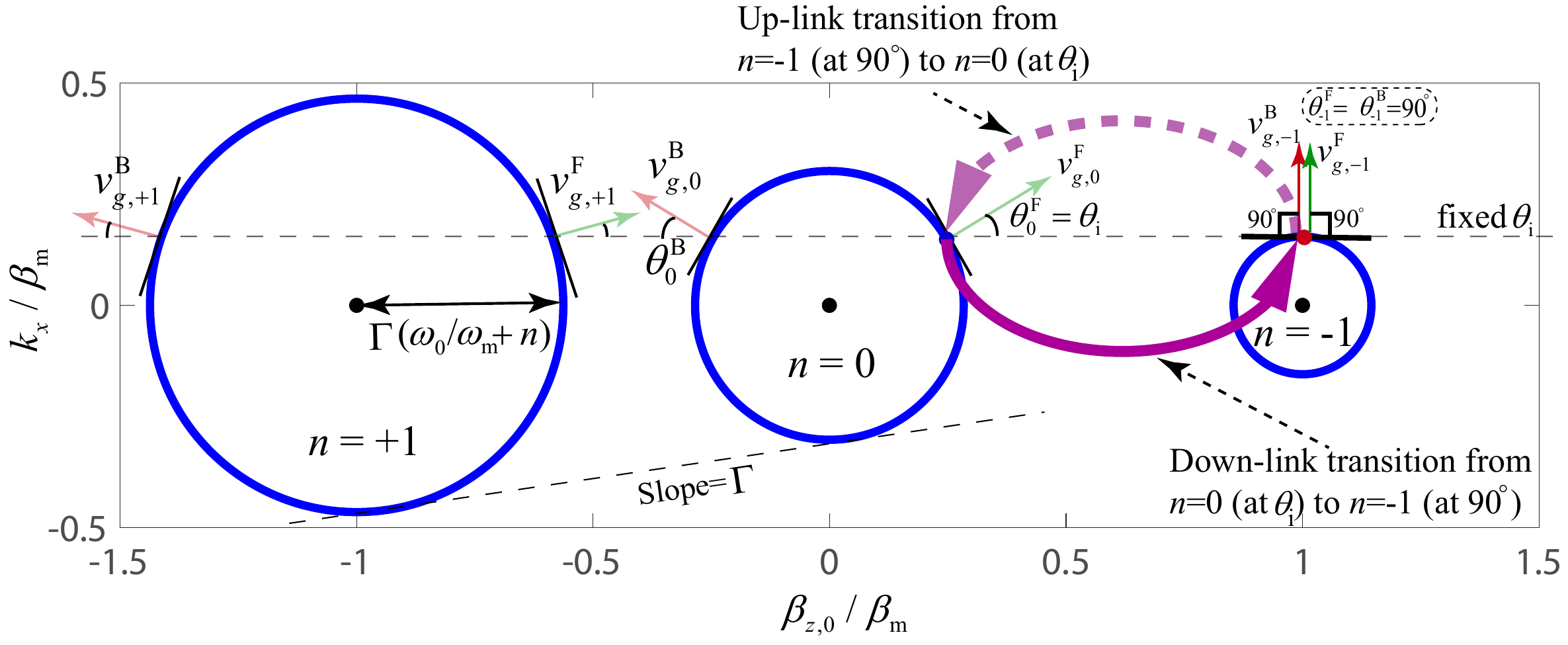}} 
		\subfigure[]{\label{Fig:AMA_sc}
			\includegraphics[width=0.7\columnwidth]{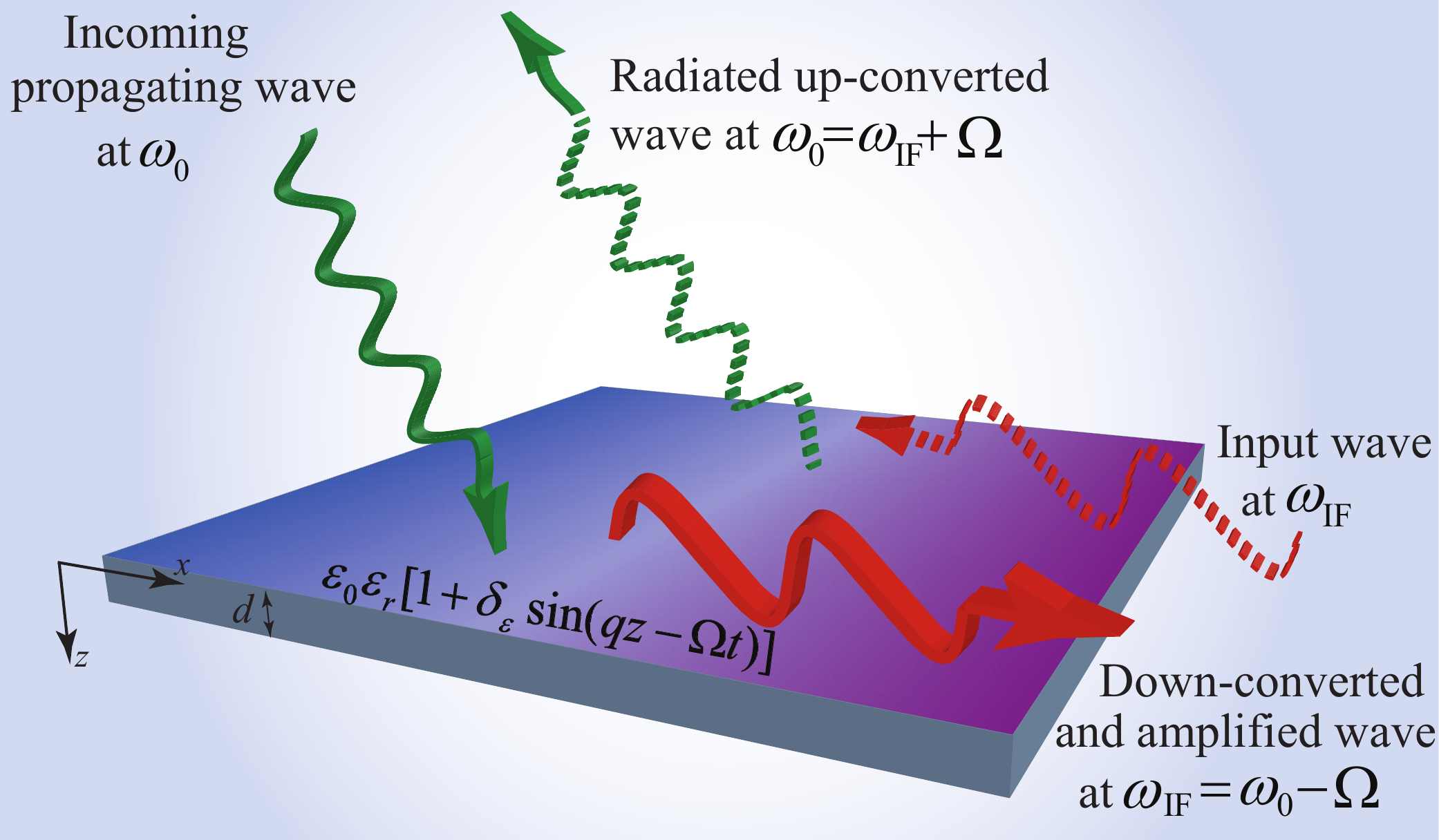} } 
	\subfigure[]{\label{fig:dispiso_4}	
	\includegraphics[width=0.36\columnwidth]{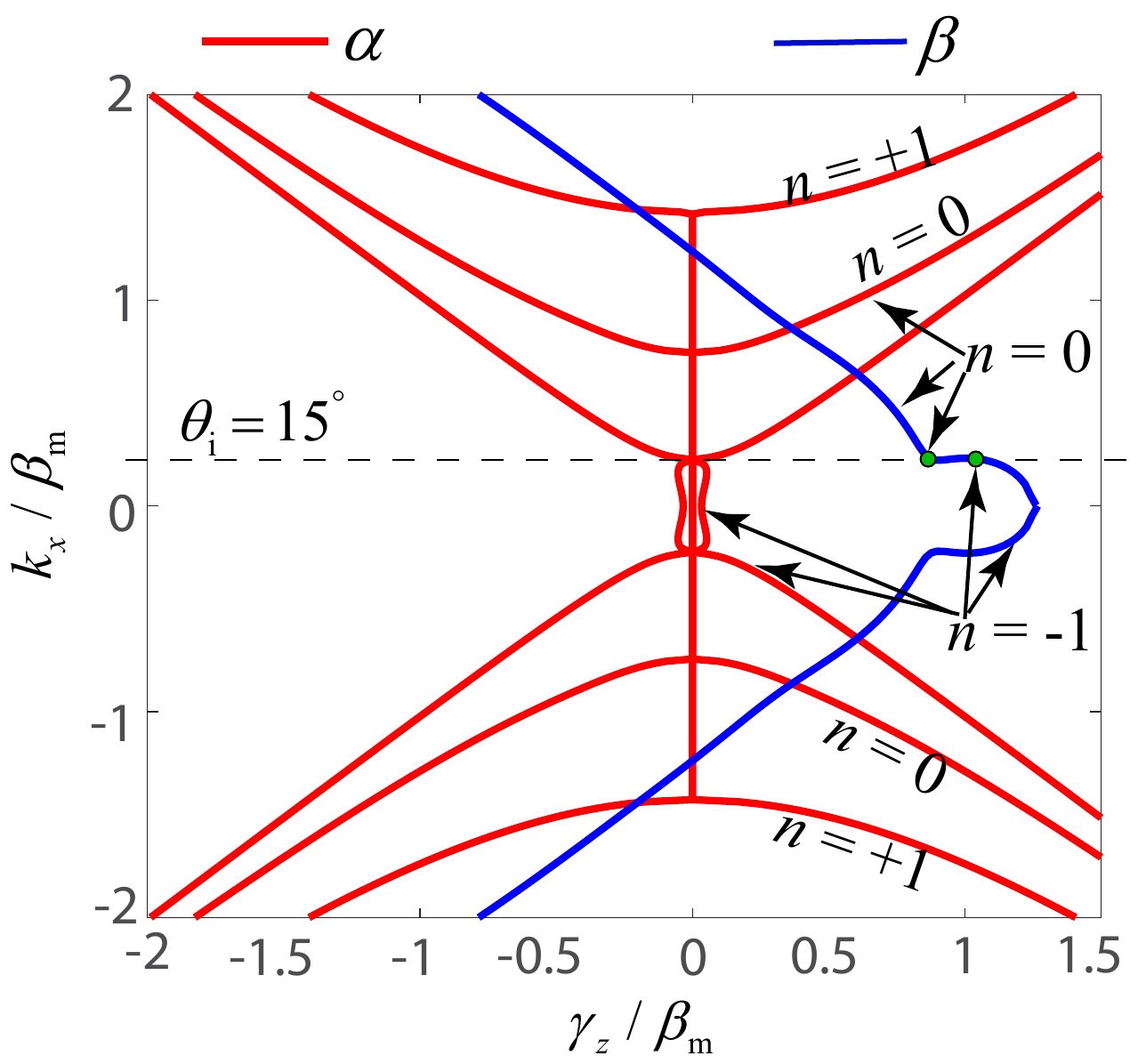}} 	
\subfigure[]{\label{fig:disp_5}	
	\includegraphics[width=0.5\columnwidth]{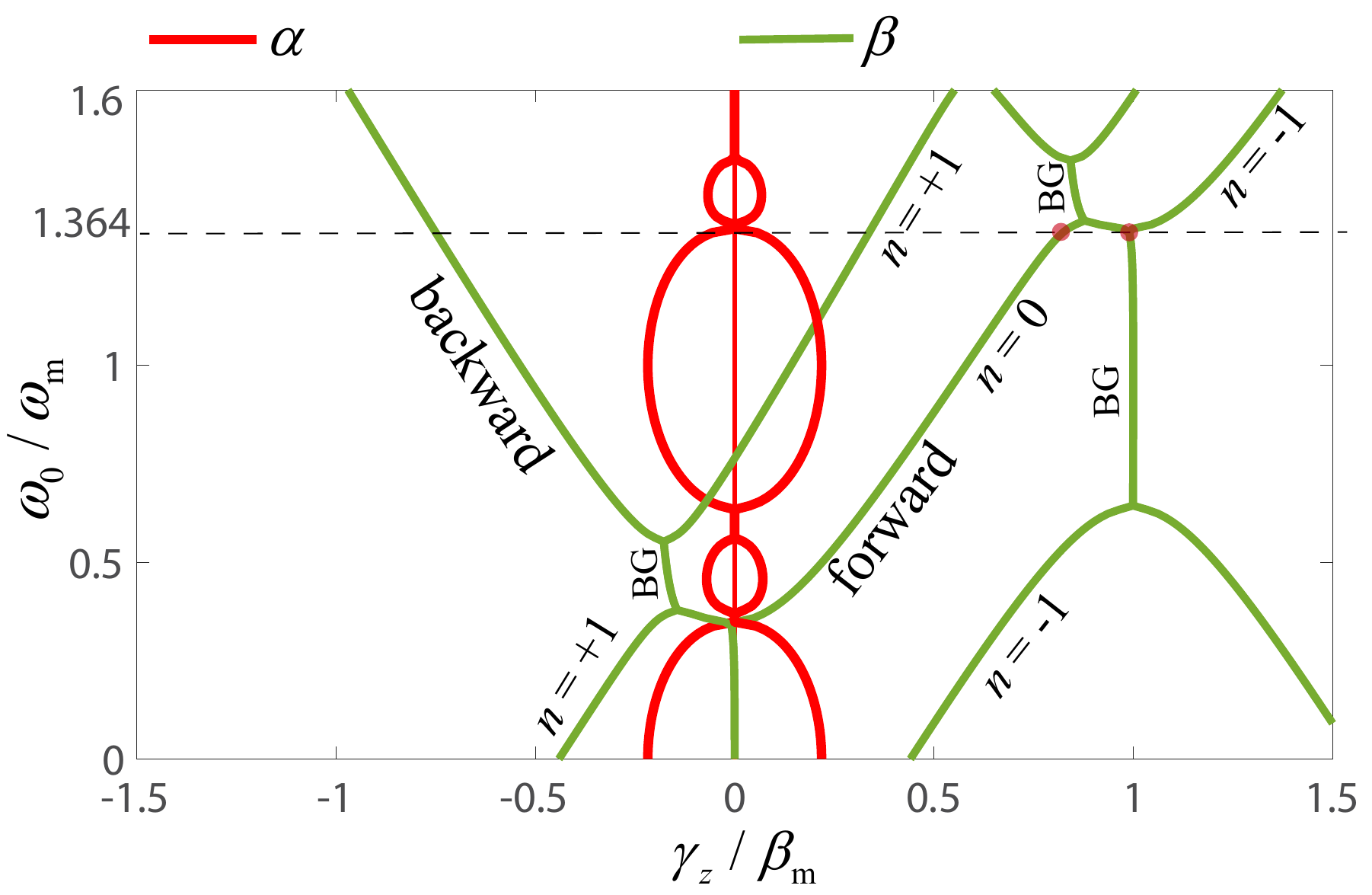}}	
		\caption{Excitation of space-time surface waves to realize an integrated Antenna-Mixer-Amplifier Transceiver. (a)~Illustrative isofrequency dispersion diagram demonstrating the operation principle~\cite{Taravati_AMA_PRApp_2020}. (b)~Schematic representation of the space-time transceiver. (c,d)~Analytically derived isofrequency and dispersion diagrams~\cite{Taravati_AMA_PRApp_2020}.}
		\label{Fig:AMA}
	\end{center}
\end{figure}

\subsection{Aperiodic Linear Space-Time Metamaterials}	

\begin{figure}
	\begin{center}
		\subfigure[]{\label{Fig:Mix_a}
			\includegraphics[width=0.7\columnwidth]{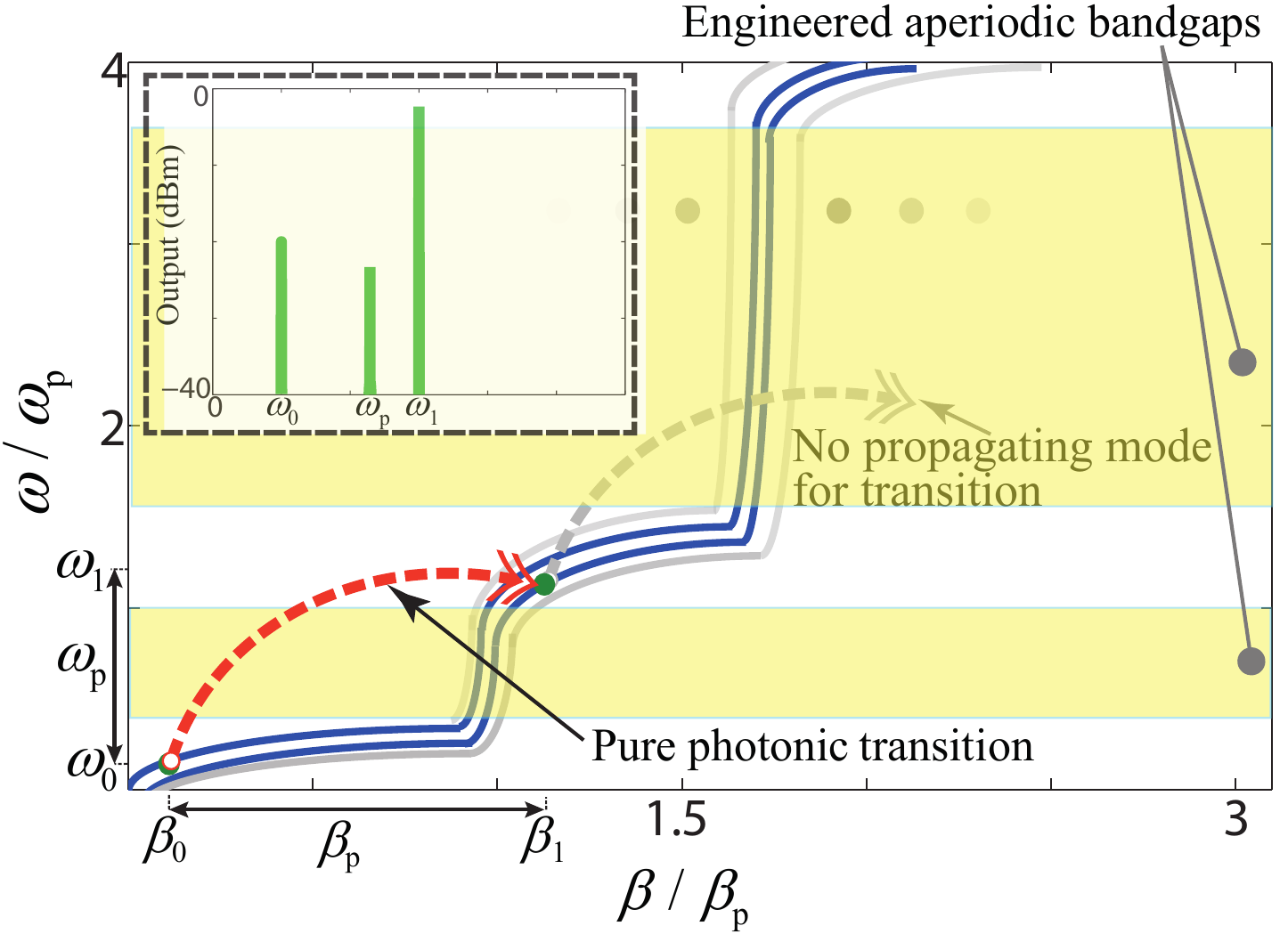}} 	
		\subfigure[]{\label{Fig:Mix_b}
			\includegraphics[width=0.7\columnwidth]{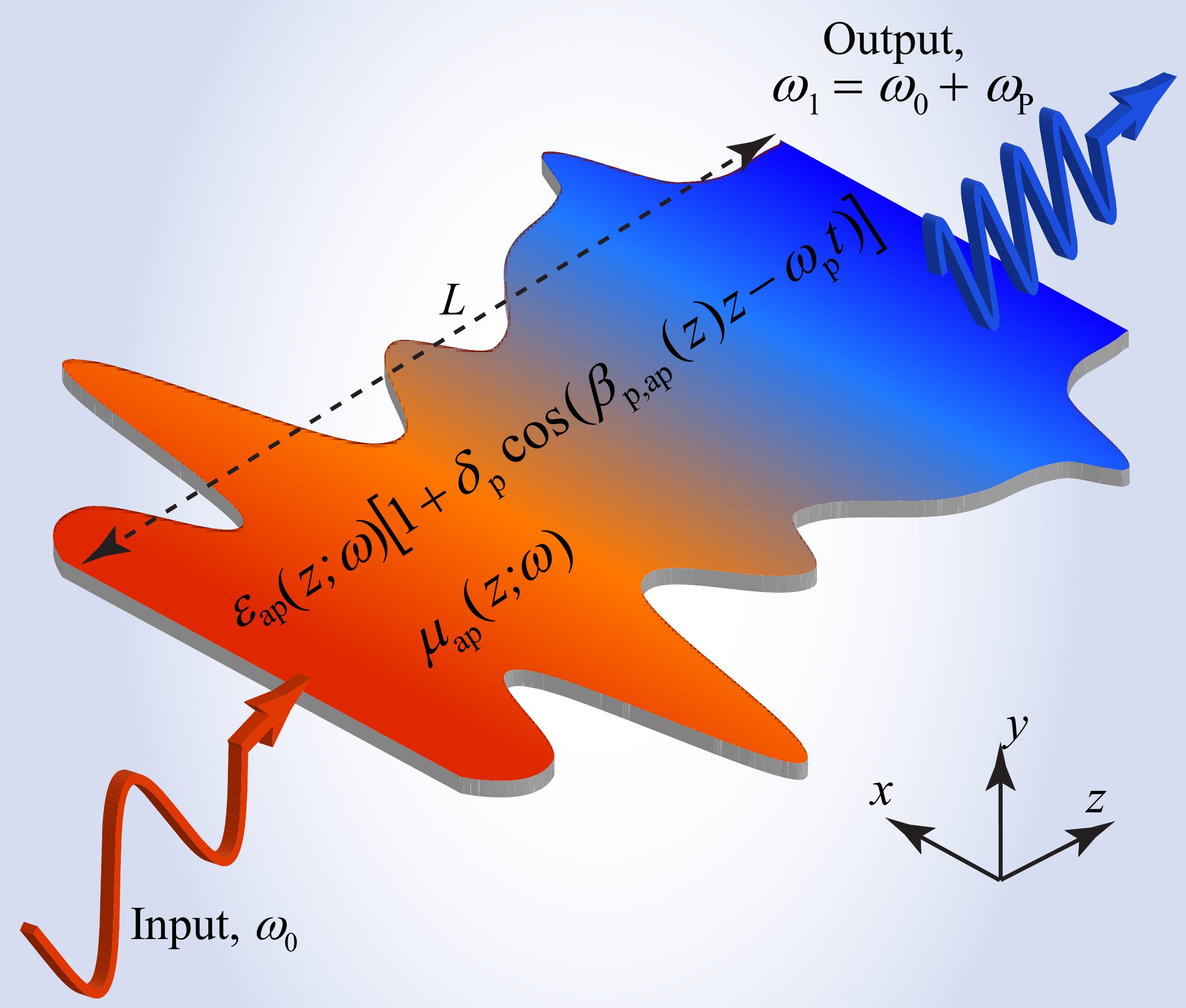}} 
		\subfigure[]{\label{Fig:Mix_c}
			\includegraphics[width=0.8\columnwidth]{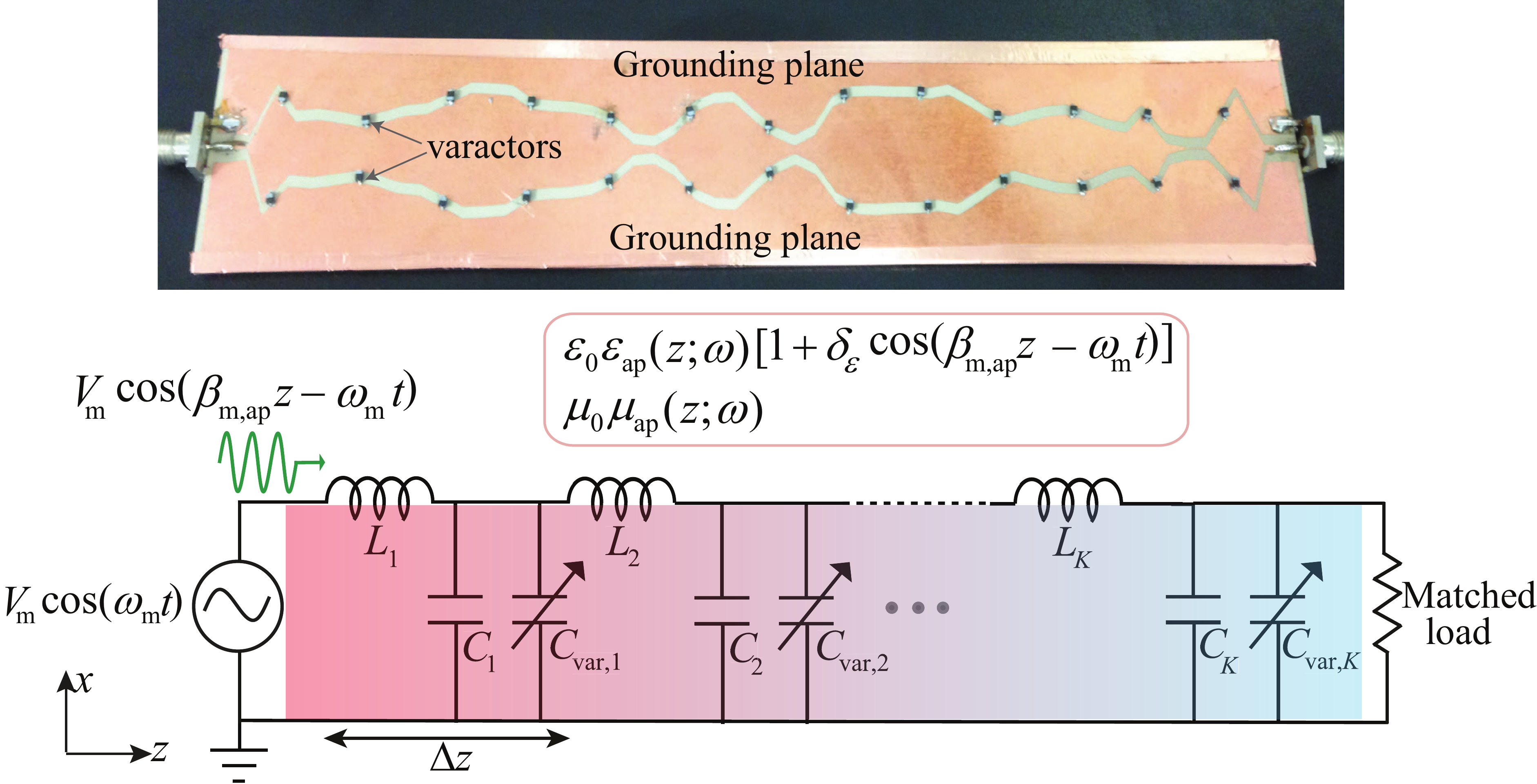}} 
		\caption{Pure frequency conversion via aperiodic space-time transitions in metamaterials. (a)~The underlying concept of tailored aperiodic transitions~\cite{Taravati_PRB_Mixer_2018}. (b,c)~Schematic and a photo of the realized aperiodic space-time frequency converter~\cite{Taravati_PRB_Mixer_2018}.} 
		\label{fig:ap}
	\end{center}
\end{figure}

This section derives a general analytical solution which may be used for aperiodic/periodic space-time modulated media, where electromagnetic fields can be expressed as $\textbf{E} (z,t;\omega_n)$ and $\textbf{H} (z,t;\omega_n)$~\cite{Taravati_PRB_Mixer_2018}. The core dispersion engineering method in this work involves the strategic design of aperiodic spatial modulation in a time-periodic medium to create tailored photonic band gaps that suppress unwanted harmonic generation, enabling pure frequency conversion, as shown in Figs~\ref{Fig:Mix_a} to~\ref{Fig:Mix_c}. Key principles of the method are as follows. i)~Fundamental Concept: The medium is engineered to be spatially aperiodic while remaining temporally periodic. This contrasts with conventional space-time metamaterials that are periodic in both dimensions. The temporal periodicity inherently generates frequency harmonics at $\omega_0 + n\omega_p$, but the spatial aperiodicity shapes the dispersion diagram to control which of these harmonics can propagate. ii)~Engineered Band Gaps: The primary mechanism is the creation of aperiodic photonic band gaps in the dispersion diagram. The spatial variation of the constitutive parameters (permittivity $\epsilon_{\text{ap}}(z;\omega)$ and permeability $\mu_{\text{ap}}(z;\omega)$) is synthesized such that the dispersion relation $\beta(\omega)$ features stop bands at the frequencies of all unwanted space-time harmonics (e.g., $\omega_0 - \omega_p$, $\omega_0 + 2\omega_p$). This prohibits the progressive growth of these harmonics as the wave propagates. iii)~Synthesis Procedure: The method employs a practical synthesis technique to achieve the desired dispersion, i.e., the spatially variant transmission line is discretized into deep subwavelength sections. 

The characteristic impedance profile $Z_{\text{ap}}(z;\omega)$ is iteratively optimized to satisfy stop bands (band gaps) at all other undesirable harmonic frequencies, and passbands at the desired frequencies (the input ($\omega_0$), the pump ($\omega_p$), and the target output $\omega_1 = \omega_0 + \omega_p$). This impedance profile is then translated into the required spatial profiles for the effective permittivity and permeability of the metamaterial. This dispersion engineering results in a "pure" and distinct photonic transition. While the time-periodic modulation generates all harmonics, the engineered band gaps ensure that only the desired harmonic ($\omega_1$) experiences a passband and thus builds up power progressively. Unwanted harmonics, being in a stop band, remain as weak, non-propagating waves and are effectively suppressed. This demonstrates a shift from passive analysis of dispersion in periodic structures to active dispersion engineering in aperiodic structures. It uses spatial tailoring as a powerful design tool to sculpt the band structure and achieve a specific, high-purity functionality—pure frequency mixing, that is unattainable with conventional periodic space-time modulations.

\begin{figure}
	\begin{center}
		\subfigure[]{\label{Fig:MetMix_a}
			\includegraphics[width=0.9\columnwidth]{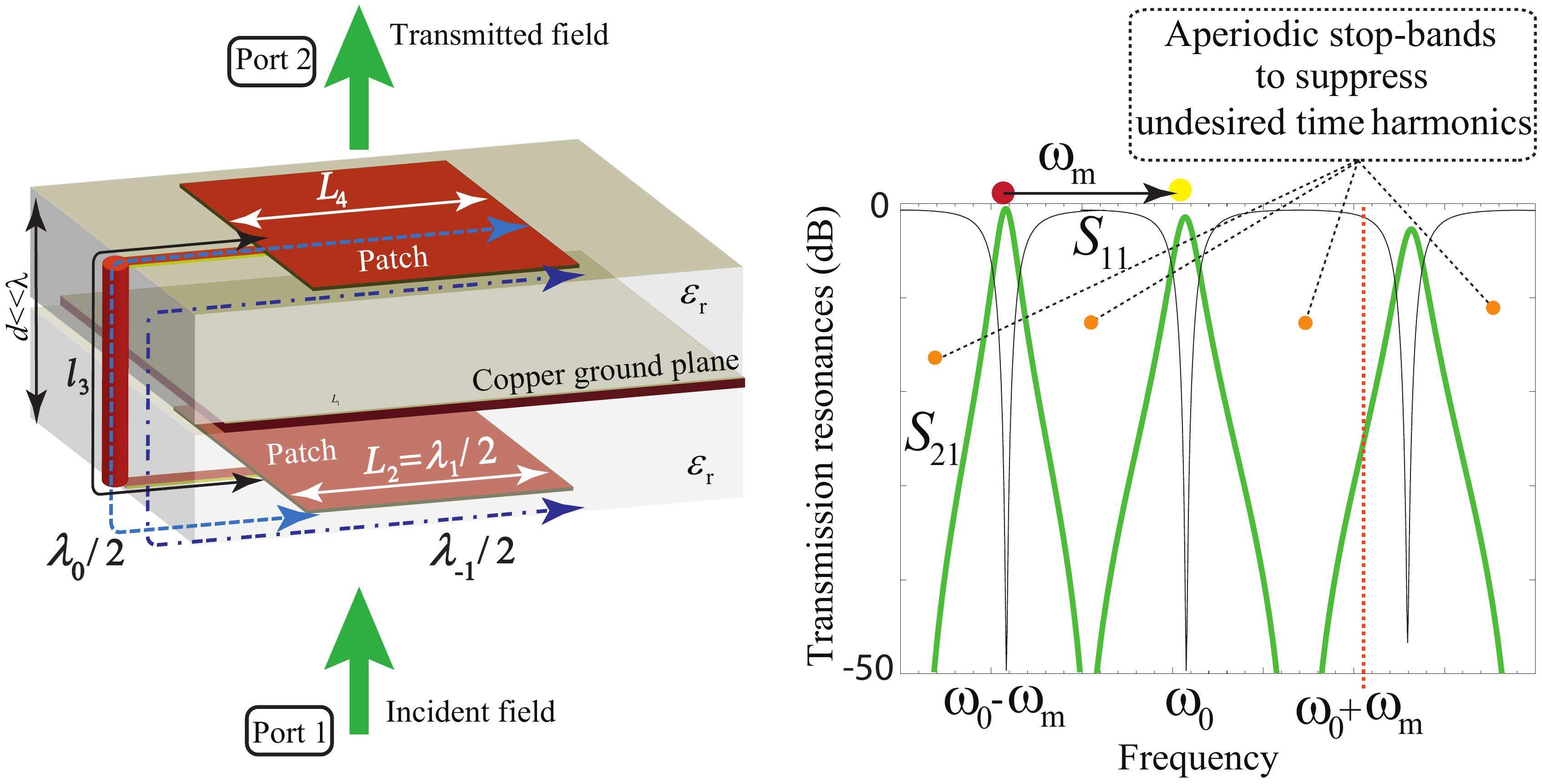}} 	
		\subfigure[]{\label{Fig:MetMix_b}
			\includegraphics[width=1\columnwidth]{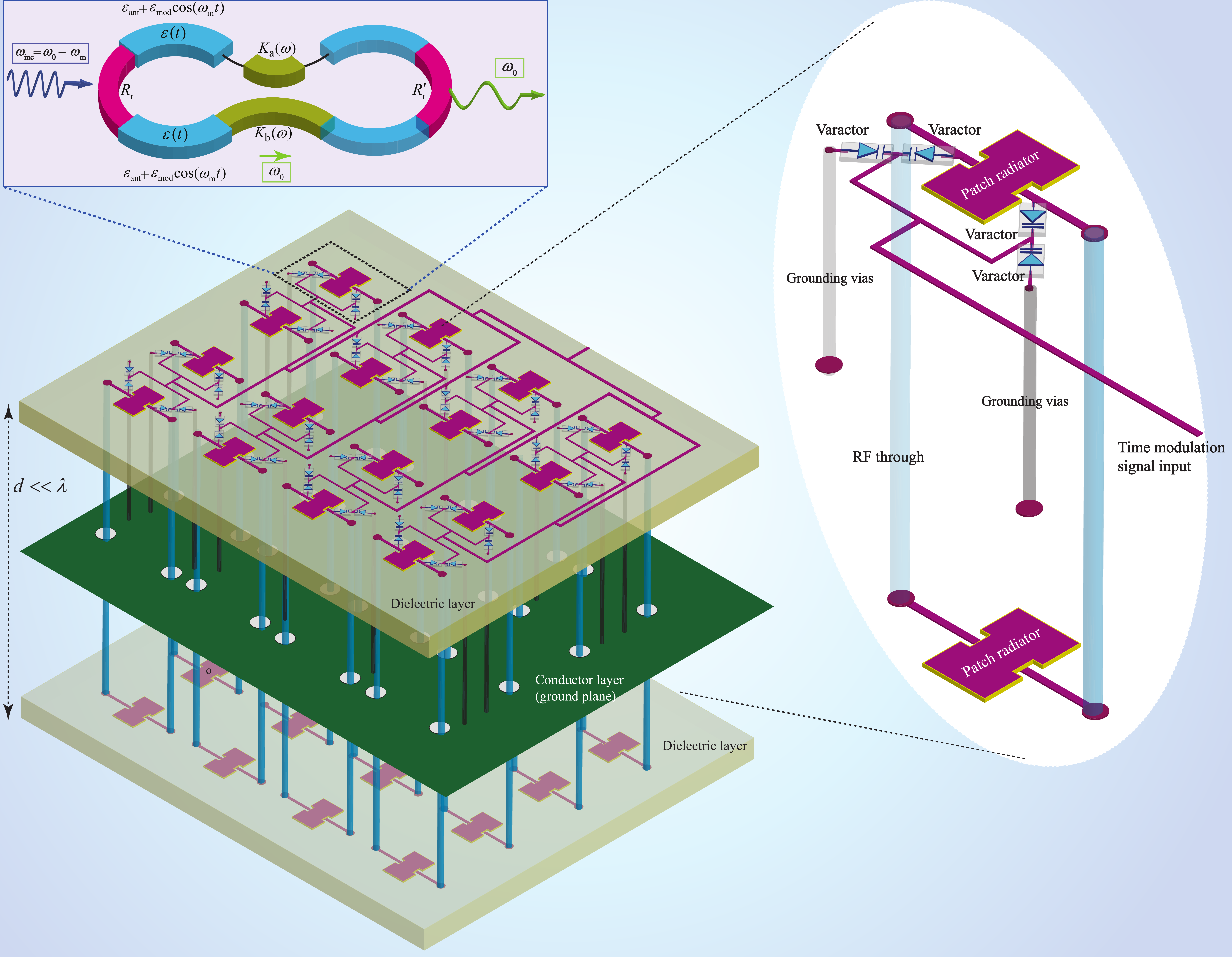}} 		
		\subfigure[]{\label{Fig:MetMix_c}
			\includegraphics[width=1\columnwidth]{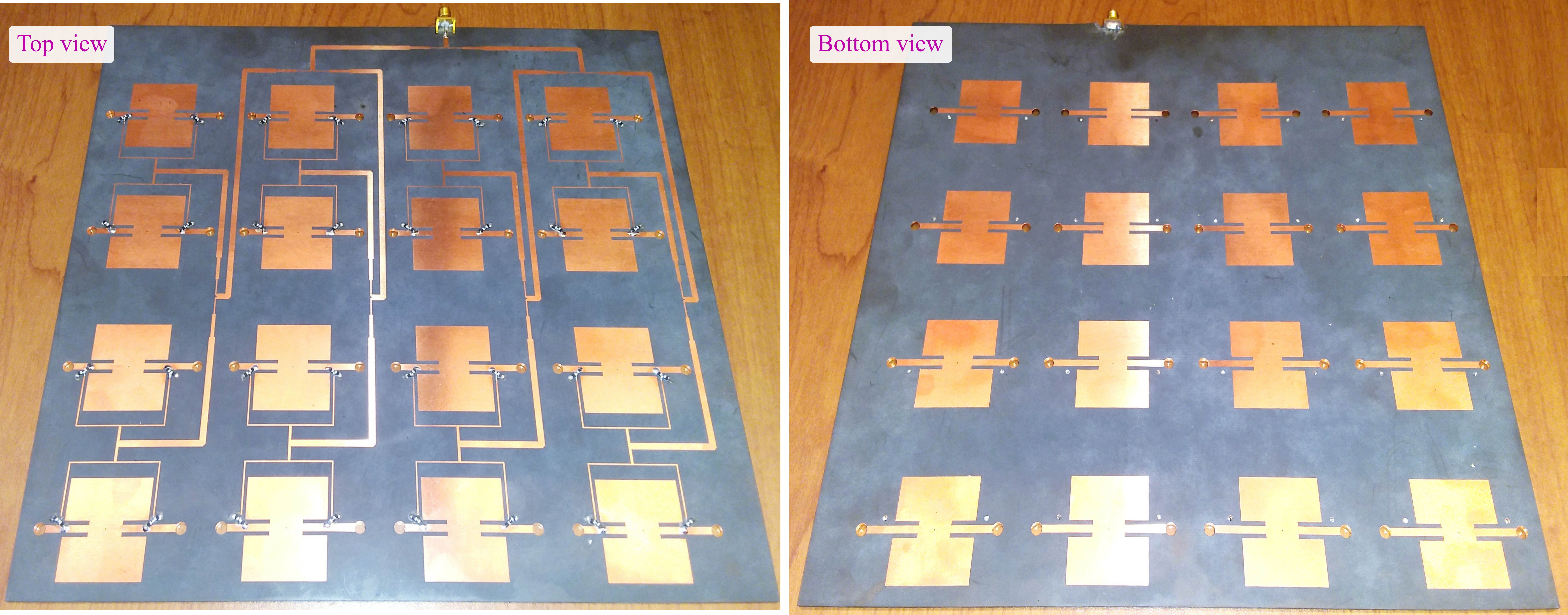}} 
		\caption{Design of aperiodic dispersion stop bands for realizing a pure frequency-converting metasurface~\cite{Taravati_MixMet_APS_2020,taravati2021pure}.
			(a)~Conceptual illustration of creating aperiodic stop bands using a coupled-patch architecture~\cite{taravati2021pure}.
			(b)~Schematic of the designed metasurface frequency converter~\cite{taravati2021pure}. (c)~Photograph of the fabricated metasurface~\cite{taravati2021pure}.} 
		\label{fig:apMetMix}
	\end{center}
\end{figure}

This dispersion engineering methodology in~\cite{Taravati_PRB_Mixer_2018} is extended to two-dimensional metasurfaces through a Surface-Interconnector-Phaser-Surface (SIPS) architecture~\cite{Taravati_MixMet_APS_2020,taravati2021pure}. Here, the spatial periodicity is replaced by a carefully designed aperiodic supercell structure that functions as an inherent multi-band bandpass filter. Each supercell consists of interconnected patch resonators and phase shifters, engineered to provide three primary full-transmission passbands at the desired frequencies, the input frequency $\omega_0$, the up-conversion frequency $\omega_0 + \omega_\text{m}$, and the down-conversion frequency $\omega_0 - \omega_\text{m}$, while simultaneously creating strong stopbands at all other unwanted harmonic frequencies. This spatial filtering effect is crucial, as it locally suppresses the progressive growth of spurious time harmonics generated by the temporal modulation, thereby enabling a pure frequency conversion process.

The operational principle leverages the synergy between the global time modulation and the local spatial dispersion control. While the time-varying permittivity $\epsilon(t) = \epsilon_{\text{ant}} + \epsilon_{\text{mod}} \cos(\omega_\text{m} t)$ provides the energy for photonic transitions, generating an infinite set of time harmonics at $\omega_0 \pm n\omega_\text{m}$, the tailored SIPS architecture ensures that only the desired harmonic (e.g., $\omega_0 + \omega_\text{m}$ for up-conversion) lies within a transmission passband. All other unwanted harmonics, such as $\omega_0 + 2\omega_\text{m}$ or $\omega_0 - \omega_\text{m}$, fall within the engineered stopbands and are strongly attenuated. This combined space-time strategy achieves a spurious-free and linear frequency converter, demonstrating that dispersion engineering via aperiodic spatial design is a powerful and generalizable approach applicable beyond bulk media to planar metasurface systems.

\subsection{Nonlinear Space-Time Metamaterials}	

\begin{figure}
	\begin{center}
		\includegraphics[width=0.8\columnwidth]{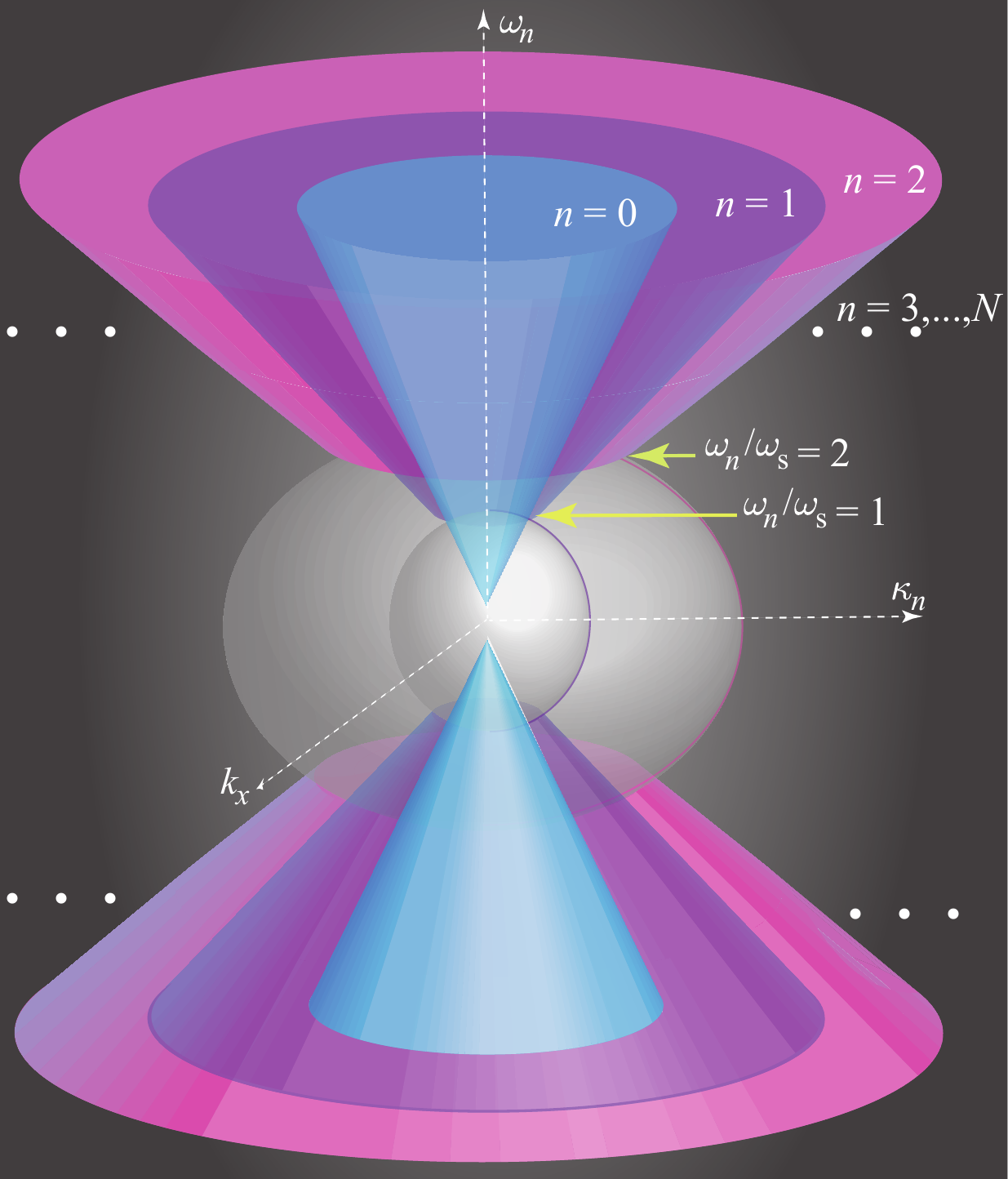}
		\caption{Three-dimensional dispersion diagram of a periodic nonlinear space–time metamaterial. The array of space–time cones represents the dispersion surfaces of multiple temporal harmonics ($n = 0, 1, 2, \ldots, N$), illustrating harmonic coupling and nonreciprocal dispersion effects~\cite{taravati2025light}.}
		\label{Fig:NL_3D_disp}
	\end{center}
\end{figure}

Figure~\ref{Fig:NL_3D_disp} depicts the three-dimensional dispersion diagram of a periodic nonlinear space–time metamaterial~\cite{taravati2024nonlinear,taravati2024one}. The diagram illustrates the formation of multiple space–time cones representing the dispersion surfaces of different temporal harmonics ($n = 0, 1, 2, \ldots, N$) in the modulated medium. Each cone corresponds to a distinct space–time harmonic generated by the periodic modulation, with their apices aligned along the temporal frequency axis $\omega_n$. The coupling between adjacent harmonics leads to the emergence of complex nonreciprocal dispersion characteristics. The labeled ratios $\omega_n / \omega_s = 1$ and $\omega_n / \omega_s = 2$ denote fundamental and higher-order harmonic interactions, respectively.

\subsubsection{Angular-Frequency Beam Multiplexing}	~\label{Sec:AngFreqMX}	

\begin{figure}
	\begin{center}
		\includegraphics[width=1\columnwidth]{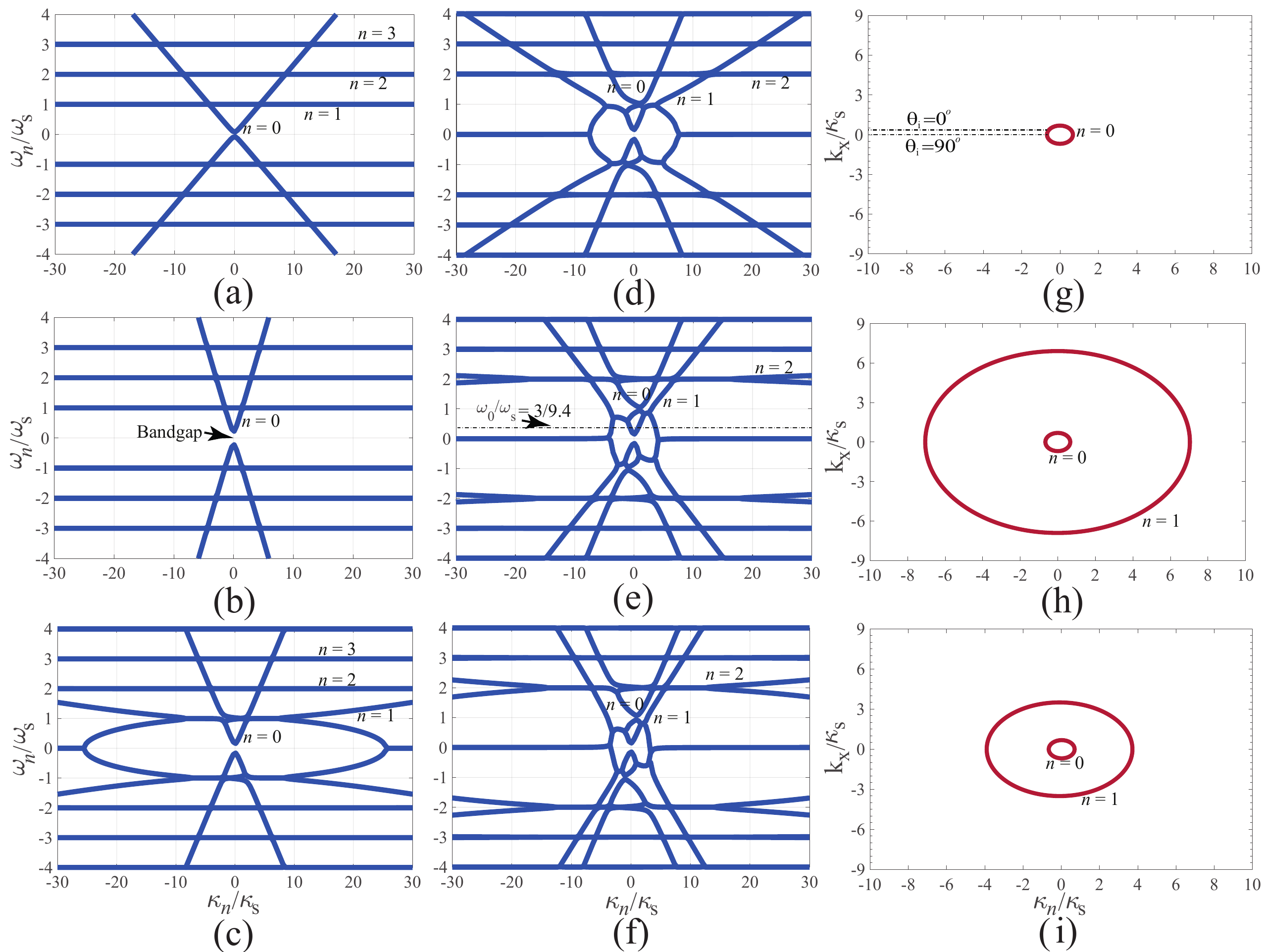}
		\caption{Dispersion and isofrequency diagrams (at $\omega_0 / \omega_\text{s} = 0.333$) for a space-time-modulated Josephson array, demonstrating its operational versatility. (a–f) Transition from linear to highly nonlinear regimes: (a) shows the linear limit ($\widetilde{\Phi}\text{dc} \to 0$, $\widetilde{\Phi}\text{rf} \to 0$), while (b–f) feature a fixed $\widetilde{\Phi}\text{dc} = 0.7$ with increasing rf modulation $\widetilde{\Phi}\text{rf} = 0, 0.1, 0.35, 0.7, 0.9$. (g–i) Corresponding isofrequency contours for select modulation strengths from (b), (d), and (e). These results reveal the profound potential of dynamic nonlinear metamaterials for advanced wave engineering~\cite{taravati2025light}.}
		\label{Fig:sch_TM}
	\end{center}
\end{figure}
\begin{figure*}
	\begin{center}
		\subfigure[]{\label{Fig:Multiplex}
			\includegraphics[width=0.65\columnwidth]{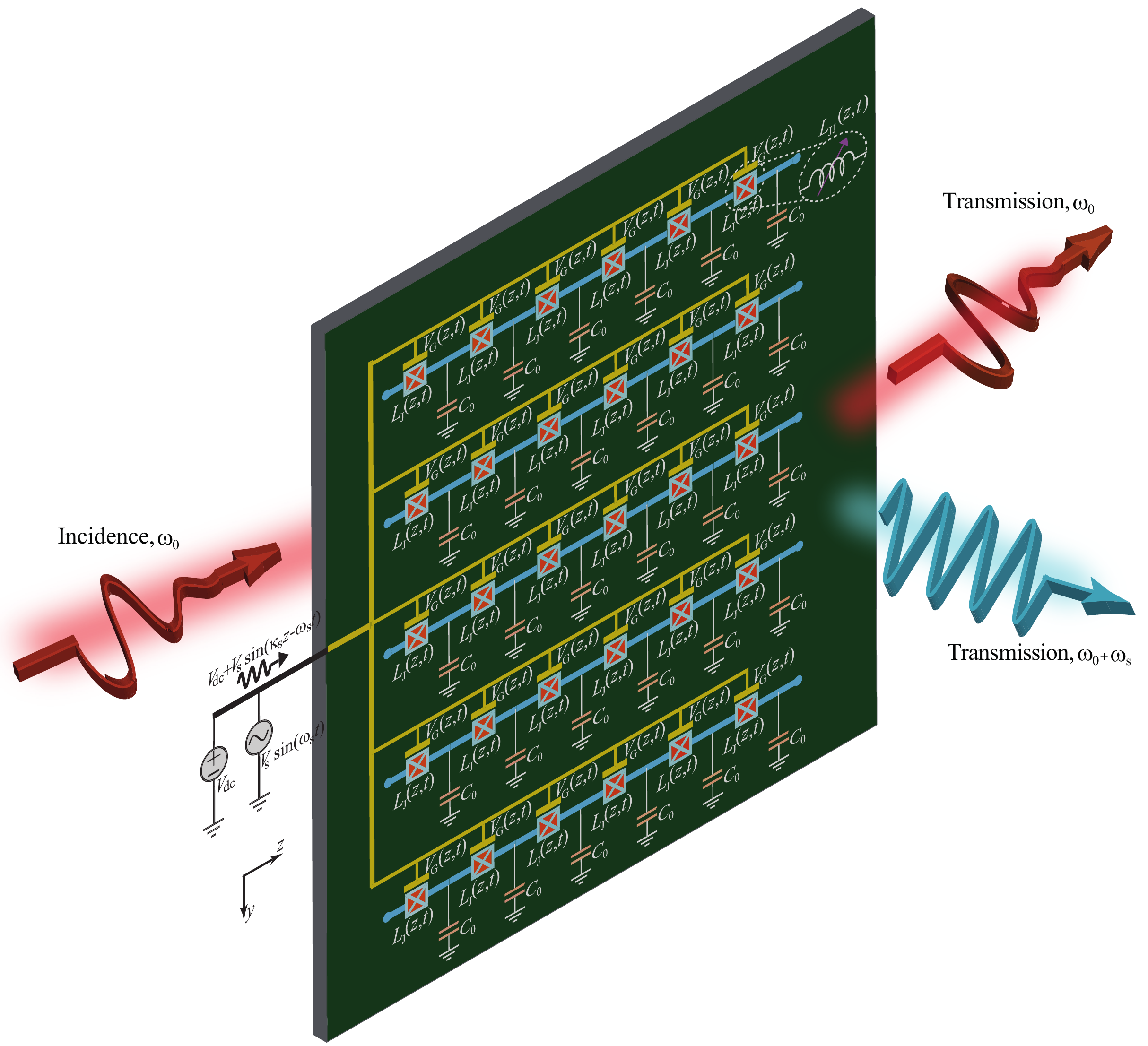}}
		\hfil
		\subfigure[]{\label{Fig:MX_a}
			\includegraphics[width=0.65\columnwidth]{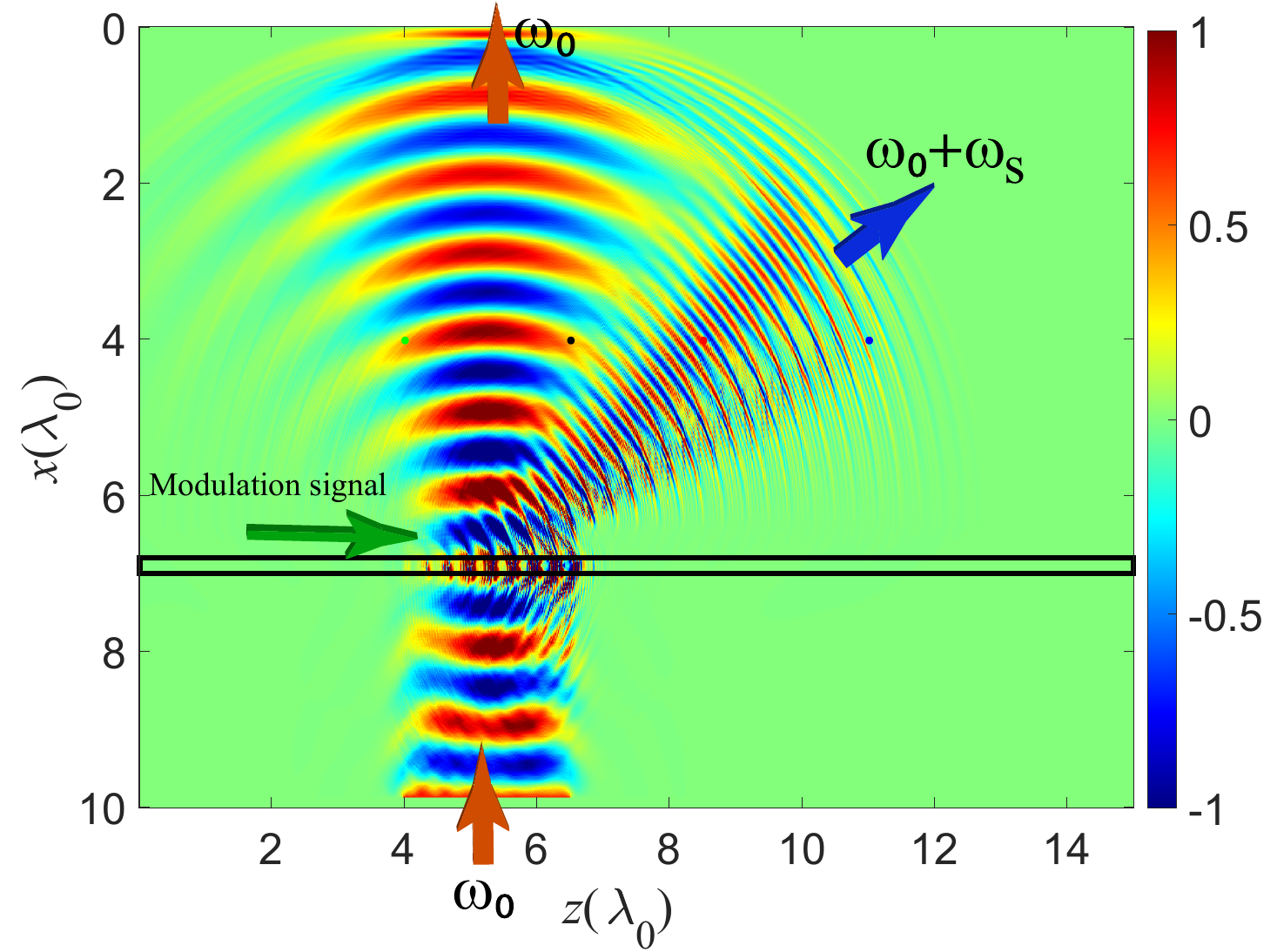}}
		\subfigure[]{\label{Fig:MX_b}
			\includegraphics[width=0.65\columnwidth]{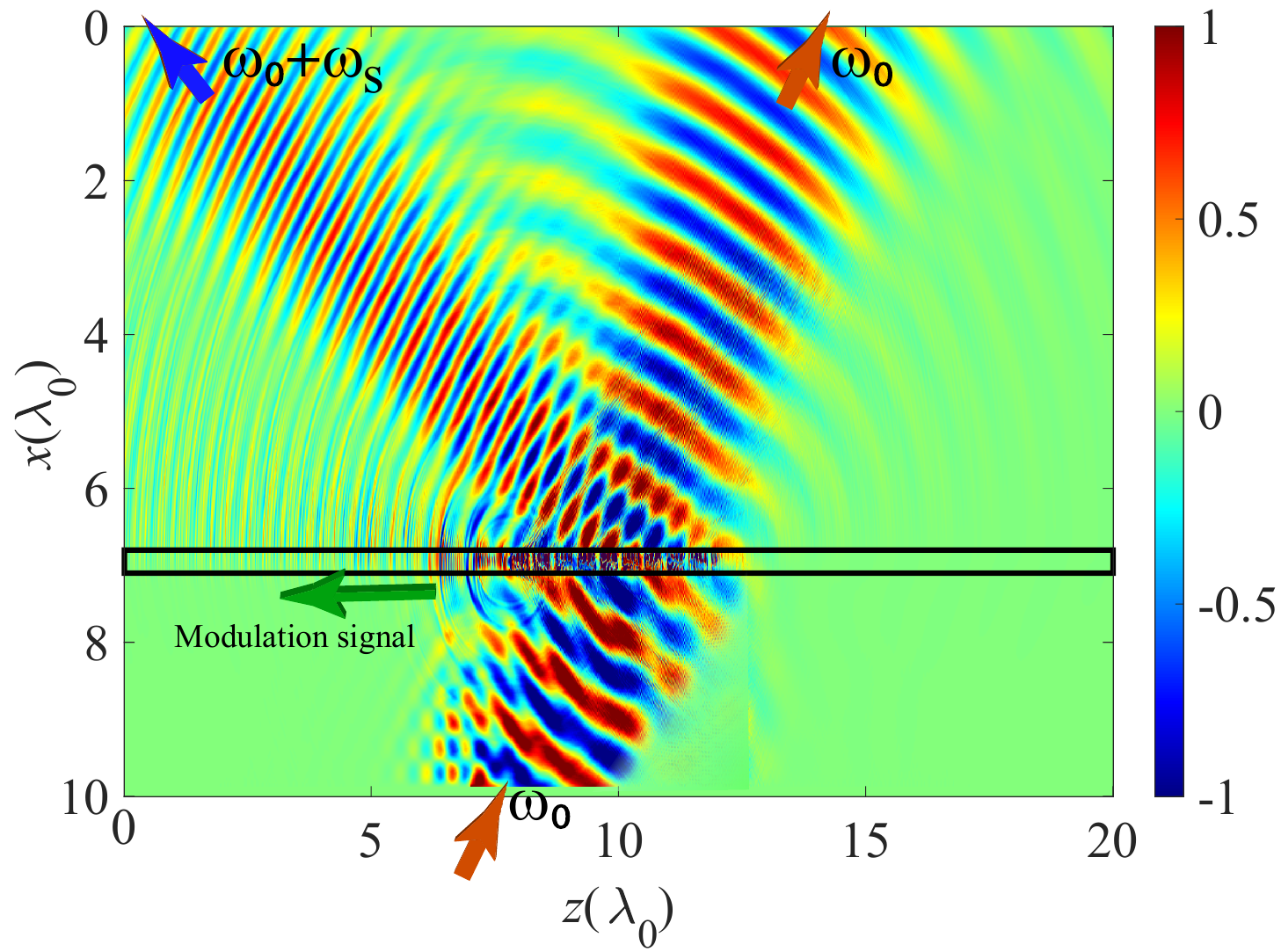} }
		\caption{Angular-frequency beam multiplexing with a space-time-modulated Josephson junction array. (a)~Conceptual schematic of the array, where a traveling-wave voltage \(V(z,t)\) dynamically modulates the junctions, enabling control over light propagation. (b-c)~Full-wave simulation results demonstrating the array's wavefront shaping capabilities. (b)~Under normal incidence (\(\theta_i = 90^\circ\)) and a modulation frequency of \(\omega_s = 3\omega_0\), the array performs efficient frequency up-conversion, generating a transmitted beam at \(4\omega_0\) that is steered to a specific angle \(\theta_1^T\). (c)~With oblique incidence (\(\theta_i = 65^\circ\)) and a reversed modulation velocity (\(\gamma = -0.6\)), the array dynamically alters the beam-steering direction of the up-converted harmonic~\cite{taravati2025light}.}
		\label{Fig:MX}
	\end{center}
\end{figure*}

Figures~\ref{Fig:sch_TM}(a) to~(i) plot dispersion and isofrequency diagrams for space-time-modulated Josephson arrays, demonstrating their operational versatility~\cite{taravati2025light}. We consider a space-time permeability-modulated nonlinear grating, as
\begin{equation}\label{eqa:ap_permeab}
	\mu_\text{m}(I,z,t) = f_\text{per,nl}(I,\kappa_\text{s}z,\omega_\text{s} t),
\end{equation}
where $f_\text{per,nl}$ represents a periodic, nonlinear function that is explicitly dependent on the induced current $I$, as well as the spatial coordinate $z$ and time $t$, while $\kappa_\text{s}$ and $\omega_\text{s}$ denote the spatial and temporal frequencies of the modulation, respectively. Since the nonlinear permeability of the grating is periodic in both space and time, it may be decomposed to Fourier series terms, namely 
\begin{equation}\label{eqa:perm}
	\mu_\text{m}(I,z,t)=  \sum_{m=-\infty}^{+\infty} \tilde{\mu}_m e^{-jm(\kappa_\text{s} z-\omega_\text{s} t+\phi)}.
\end{equation}

For non-trivial solutions (i.e., $[\overrightarrow{\varPsi}] \neq 0$), the matrix $[\textbf{A}]$ must be singular, meaning its determinant is zero, that is,
\begin{equation}
	\det\left[\textbf{K}\right]=0,
	\label{eqa:det_gen}
\end{equation}
where the matrix $[\textbf{K}]$ is a square matrix of size $(2N+1) \times (2N+1)$, with elements defined as
\begin{equation}
	\begin{split}
		K_{nn} &= \frac{ k_x^2+ \kappa_n ^2  }{k_n^2 }-  \tilde{\mu}_0,   \\
		K_{nm} &= -  \tilde{\mu}_{m+n},\quad\text{for }n\neq m.
	\end{split}
	\label{eqa:A_matrix}
\end{equation}

The core of the dispersion engineering lies in the creation of a dynamic, tunable band structure through the spatiotemporal modulation (STM) of an array of Josephson junctions. Unlike static materials or even linear STM systems, this platform uses the intrinsic nonlinearity of the Josephson effect, governed by the sinusoidal current-phase relation. The key modulated parameter is the effective permeability of the array, which is controlled by applying a traveling wave of magnetic flux, described by the parameters \(\widetilde{\Phi}_{\text{dc}}\) (the DC bias) and \(\widetilde{\Phi}_{\text{rf}}\) (the RF modulation amplitude). This unidirectional modulation fundamentally breaks Lorentz reciprocity and creates a medium that is periodic in both space and time.

The engineered dispersion relation reveals a complex landscape of "space-time harmonics." These are not simple replicas of the fundamental mode. Instead, the dispersion diagram consists of a series of "double semi-cones" for each harmonic (\(n=0, \pm1, \pm2,...\)), where the slopes represent the group velocity. The modulation parameters \(\widetilde{\Phi}_{\text{dc}}\) and \(\widetilde{\Phi}_{\text{rf}}\) act as independent knobs to sculpt this landscape: \(\widetilde{\Phi}_{\text{dc}}\) primarily controls the tilt and bandgap of the fundamental (\(n=0\)) harmonic, while \(\widetilde{\Phi}_{\text{rf}}\) controls the emergence and slope of the higher-order harmonics (\(n \neq 0\)). This allows for precise control over harmonic coupling, nonreciprocity, and bandgap width. For instance, increasing \(\widetilde{\Phi}_{\text{rf}}\) enhances the coupling between harmonics, enabling efficient energy transfer from the fundamental frequency \(\omega_0\) to sidebands like \(\omega_0 + \omega_s\) and \(\omega_0 + 2\omega_s\).

The isofrequency contours in Figs.~\ref{Fig:sch_TM}(g) to~(i) demonstrate that different space-time harmonics propagate at different angles for the same input frequency. This is a direct consequence of momentum conservation (\(\kappa_n = \kappa_0 + n\kappa_s\)) from the spatial modulation. Therefore, an incident beam at a single frequency \(\omega_0\) and angle \(\theta_i\) is simultaneously converted into multiple output beams, each at a distinct frequency (\(\omega_0 + n\omega_s\)) and a distinct angle (\(\theta_n^T\)). This seamlessly integrates the functionalities of a frequency mixer and a beam splitter into a single, reconfigurable superconducting device.

Figure~\ref{Fig:Multiplex} depicts a mechanism for parametric wave generation within an active, traveling-wave modulator~\cite{taravati2025light}. By precisely engineering the modulation parameters—specifically its temporal frequency (\(\omega_s\)) and phase velocity (controlled by \(\gamma\)), the space-time Josephson array acts as a wave-mixing engine that can simultaneously convert the frequency and control the exit angle of a transmitted beam, thereby realizing the core function of angular-frequency beam multiplexing. The core principle is that the ST modulation, achieved by applying a traveling-wave voltage \(V(z,t)=V_{\mathrm{dc}}+V_{\mathrm{rf}}\sin[\kappa_{\mathrm{s}}z-\omega_{\mathrm{s}}t+\phi]\), actively imparts momentum and energy to the incident photons. This dynamic interaction can be understood through the Floquet-Bloch theory applied to the time-varying medium, which dictates that the incident wave does not simply transmit at its original frequency \(\omega_0\) but couples to an infinite set of space-time harmonics with frequencies \(\omega_n = \omega_0 + n\omega_s\) and wavenumbers \(\kappa_n = \kappa_0 + n\kappa_s\). The specific outcome—namely, the frequency and direction of the transmitted beams—is governed by the conservation laws embedded in the system's dispersion relation, \(\omega_n(k_x, \kappa_n)\), where \(k_x = k_0 \cos(\theta_i)\) is the conserved transverse momentum.

In Fig.~\ref{Fig:MX_a}, the condition of increased modulation frequency (\(\omega_s = 3\omega_0\)) and a modulation phase velocity equal to the background wave velocity (\(\gamma = 1\)) leads to a highly efficient frequency up-conversion. The incident wave at \(\omega_0\) and normal incidence (\(\theta_i = 90^\circ\)) strongly couples to the \(n=+1\) harmonic, generating a transmitted beam at \(\omega_0 + \omega_s = 4\omega_0\). Crucially, the conservation of transverse momentum (\(k_x\)) and the need to satisfy the dispersion relation for the \(n=+1\) harmonic force this new beam to be radiated at a specific, non-zero angle \(\theta_1^T\). The higher the modulation frequency \(\omega_s\), the greater the required momentum compensation, resulting in a larger transmission angle, as vividly shown in the figure. Conversely, Fig.~\ref{Fig:MX_b} demonstrates the ability to dynamically steer the transmitted beam by controlling the modulation velocity. Here, with a fixed oblique angle of incidence (\(\theta_i = 65^\circ\)) and modulation frequency (\(\omega_s = 1.333\omega_0\)), the key parameter is the modulation velocity ratio, set to \(\gamma = -0.6\). A negative \(\gamma\) signifies that the modulation wave propagates in the opposite direction to the phase velocity of the incident wave in the medium. This reversal dramatically alters the phase-matching condition for the \(n=+1\) harmonic. As a result, the conservation of momentum is satisfied at a different radiative wavenumber, which manifests as a significant change in the transmission angle \(\theta_1^T\) of the up-converted beam compared to other cases.

\subsubsection{Nonreciprocal Quantum Absorption}~\label{Sec:NRQuantAbs}	

\begin{figure}
	\begin{center}
		\subfigure[]{\label{Fig:dispersion}
			\includegraphics[width=0.7\columnwidth]{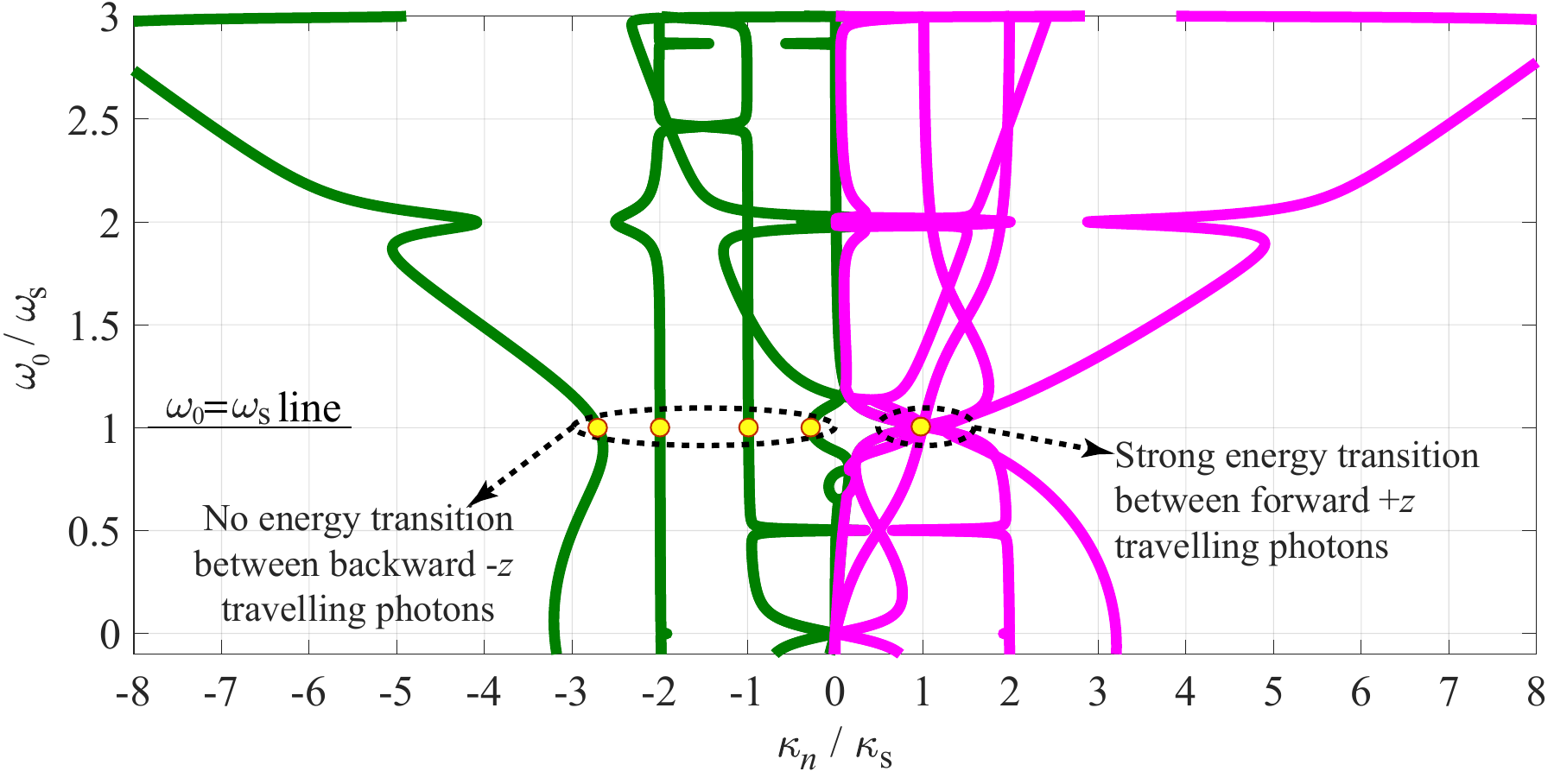}}
		\hfil
		\subfigure[]{\label{Fig:isof}
			\includegraphics[width=0.7\columnwidth]{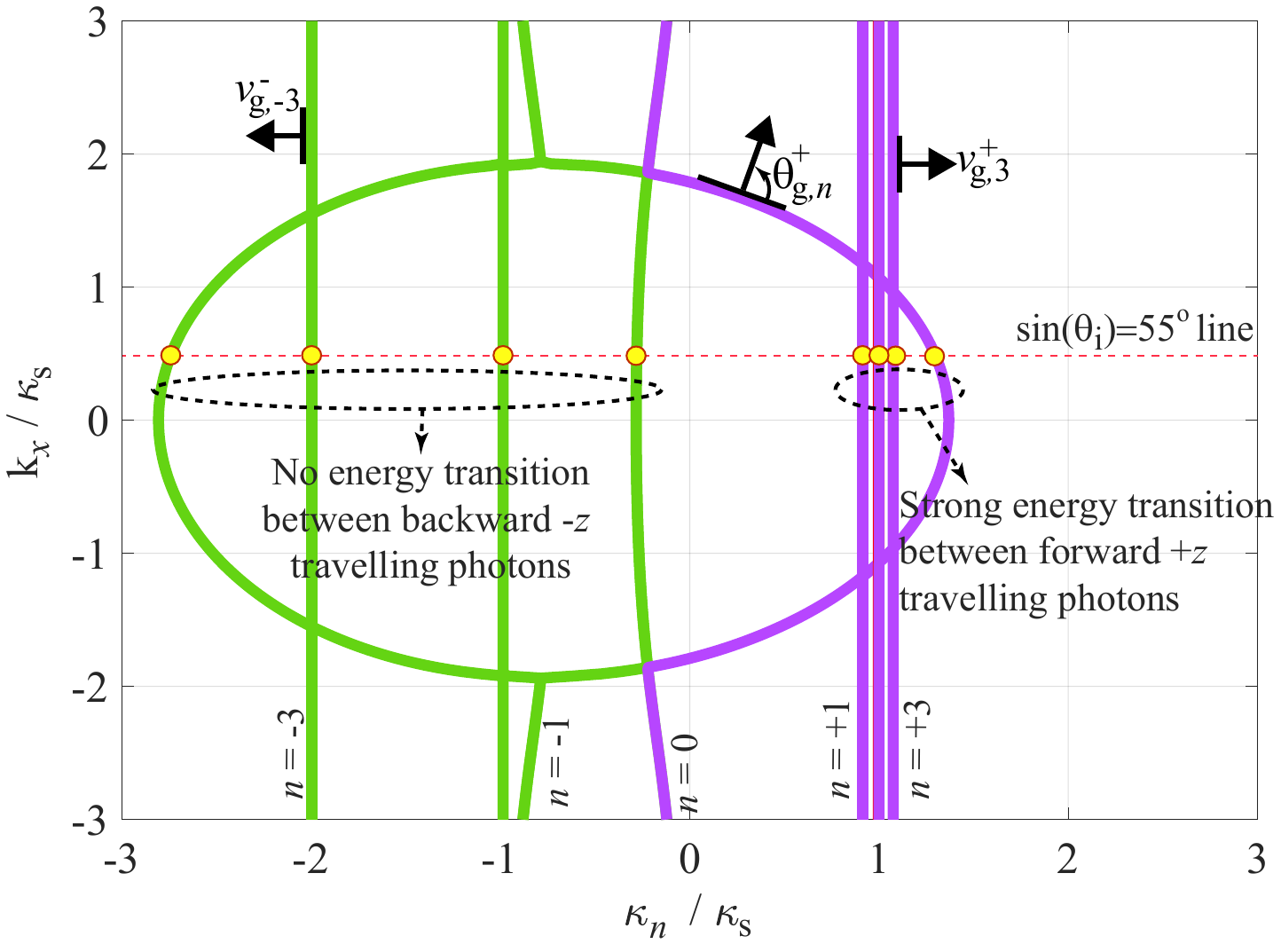}}
		\subfigure[]{\label{Fig:sch}
			\includegraphics[width=0.8\linewidth]{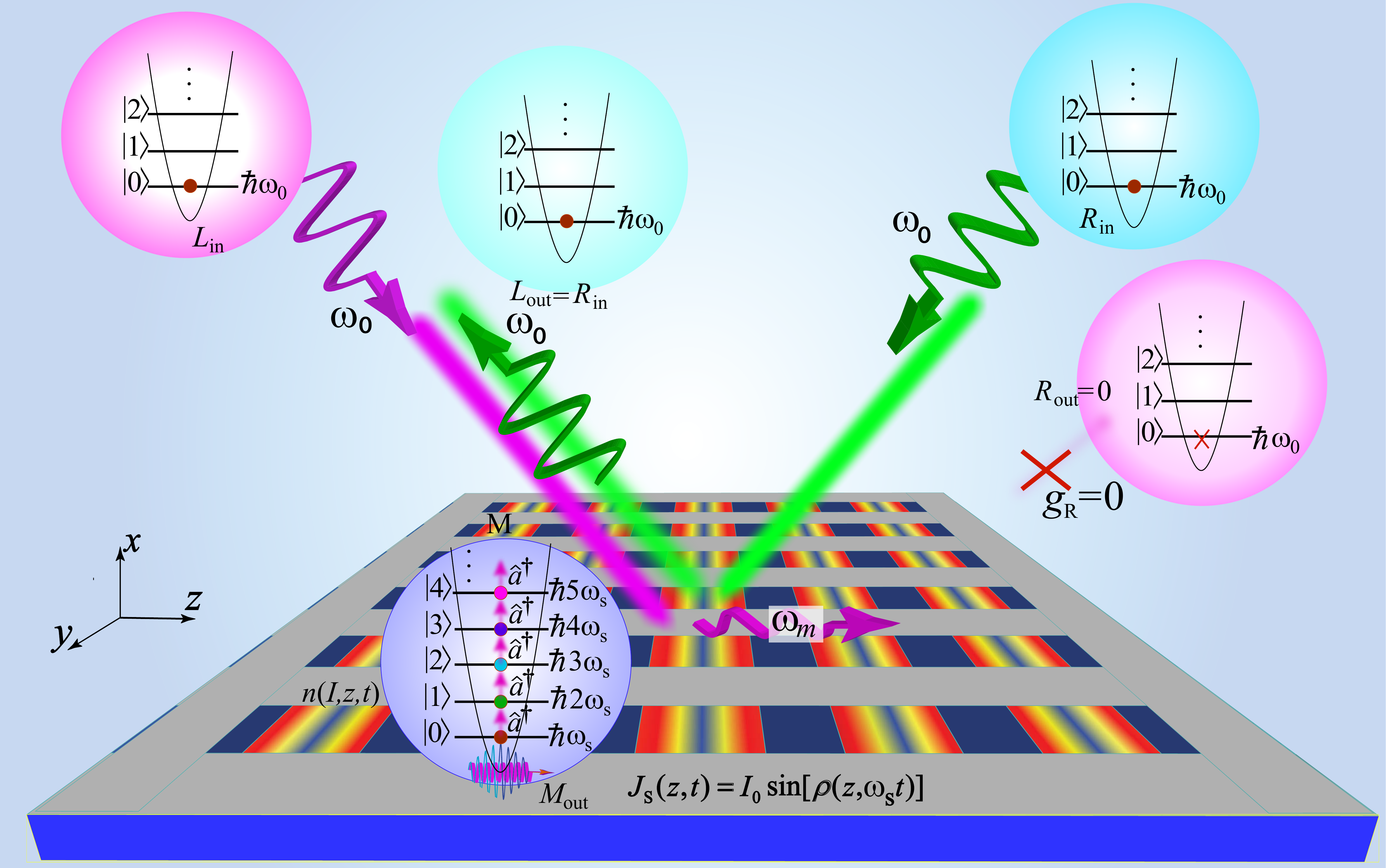} }
		\caption{Engineering dispersion and isofrequency characteristics of a nonlinear space–time Josephson metamaterial for one-way quantum absorption.
			(a)~The dispersion diagram reveals pronounced nonreciprocity at $\omega_0 / \omega_\text{s} = 1$. In this regime, all higher-order forward-propagating space–time harmonics coalesce at $\kappa_n / \kappa_\text{s} = 1$, leading to energy transfer from the fundamental harmonic to its higher-frequency counterparts. In contrast, the backward-propagating harmonics remain spatially separated, suppressing any reverse energy conversion~\cite{taravati2024spatiotemporal}.
			(b)~The corresponding isofrequency diagram at $\omega_0 / \omega_\text{s} = 1$ further demonstrates this unidirectional behavior, where forward harmonics cluster at $\kappa_n / \kappa_\text{s} = 1$, efficiently absorbing energy from the fundamental mode. The group velocity vectors $v_{\text{g},n}$ indicate that these harmonics propagate along the $+z$ direction—parallel to the grating boundary—without transmission beyond the structure~\cite{taravati2024spatiotemporal}.
			(c)~Schematic representation of the one-way photon blockade and absorption functionality~\cite{taravati2024spatiotemporal}.}
		\label{Fig:res}
	\end{center}
\end{figure}

A groundbreaking form of dispersion engineering that operates at the intersection of quantum optics and superconducting circuits is reported in~\cite{taravati2024one,taravati2024spatiotemporal}. A periodic space-time-varying superconducting array is proposed in~\cite{taravati2024spatiotemporal}, where the inductance of the dynamic superconducting structure is expressed as
\begin{equation}\label{eqa:L}
	\begin{split}
		L_\text{s}(I,z,t)
		= \dfrac{\Phi_0}{ 2\pi I_0}  \sec\left(\widetilde{\Phi}_\text{dc}+ \widetilde{\Phi}_\text{rf} \sin[\kappa_\text{s} z-\omega_\text{s} t+\phi]\right),
	\end{split}
\end{equation} 

The core achievement is the design of a medium whose dispersion diagram is engineered not just to control the path of waves, but to orchestrate the quantum state transitions of individual photons in a nonreciprocal manner. This represents a paradigm shift from classical wave control to direct quantum state control via a sculpted electromagnetic dispersion landscape. The foundation of this methodology is a space-time-modulated superconducting metasurface built from Josephson Field-Effect Transistors (JoFETs). Unlike classical components, this system leverages the intrinsic, sharp nonlinearity of the Josephson effect. The key engineered parameter is the effective permeability, which is dynamically shaped by applying a unidirectional traveling wave voltage. This creates a spatiotemporally modulated potential landscape that fundamentally breaks Lorentz reciprocity, establishing a nonreciprocal foundation not merely for wave scattering, but for photon interactions themselves.

Figures~\ref{Fig:dispersion} and~\ref{Fig:sch} show the operation principle of nonreciprocal photon blockade based on nonlinear space-time Josephson metamaterial. The most profound insight lies in how the engineered band structure enforces a directional photon blockade. The system's dispersion relation, derived from a singular matrix condition, functions as a map of allowed quantum state transitions. A critical design choice synchronizes the modulation frequency with the incident photon frequency. At this point, the dispersion diagram reveals a stark nonreciprocity: all higher-order space-time harmonics for forward propagation converge at a single wavenumber, creating an "energy sink." A forward-propagating photon is strongly coupled into this dense manifold of higher-energy states, leading to its absorption and a blockade of transmission. In contrast, for backward propagation, the harmonics are spatially separated, creating no such pathway; the photon passes through unimpeded. The dispersion is thus engineered to be absorptive and "sticky" in one direction, and transparent and "slippery" in the other. This engineered dispersion directly enables the titular phenomenon. The photon blockade is not caused by a single saturated emitter, but by the collective, nonreciprocal dispersion of the entire metasurface funneling photon energy directionally. The functional payoff is a nonreciprocal quantum absorber, a device where the sculpted dispersion diagram, formalized by a directionally biased interaction Hamiltonian, acts as a programmable quantum state router. This work ultimately elevates dispersion engineering from a tool for beam steering to a fundamental principle for designing quantum matter, opening a new frontier for integrated, magnet-free nonreciprocal elements in quantum information processing.

\subsubsection{Quantum State Transfer Between Frequency-Distinct Qubits}~\label{Sec:QST}

\begin{figure}
	\begin{center}	
		\subfigure[]{\label{Fig:disp_forw}
			\includegraphics[width=0.45\columnwidth]{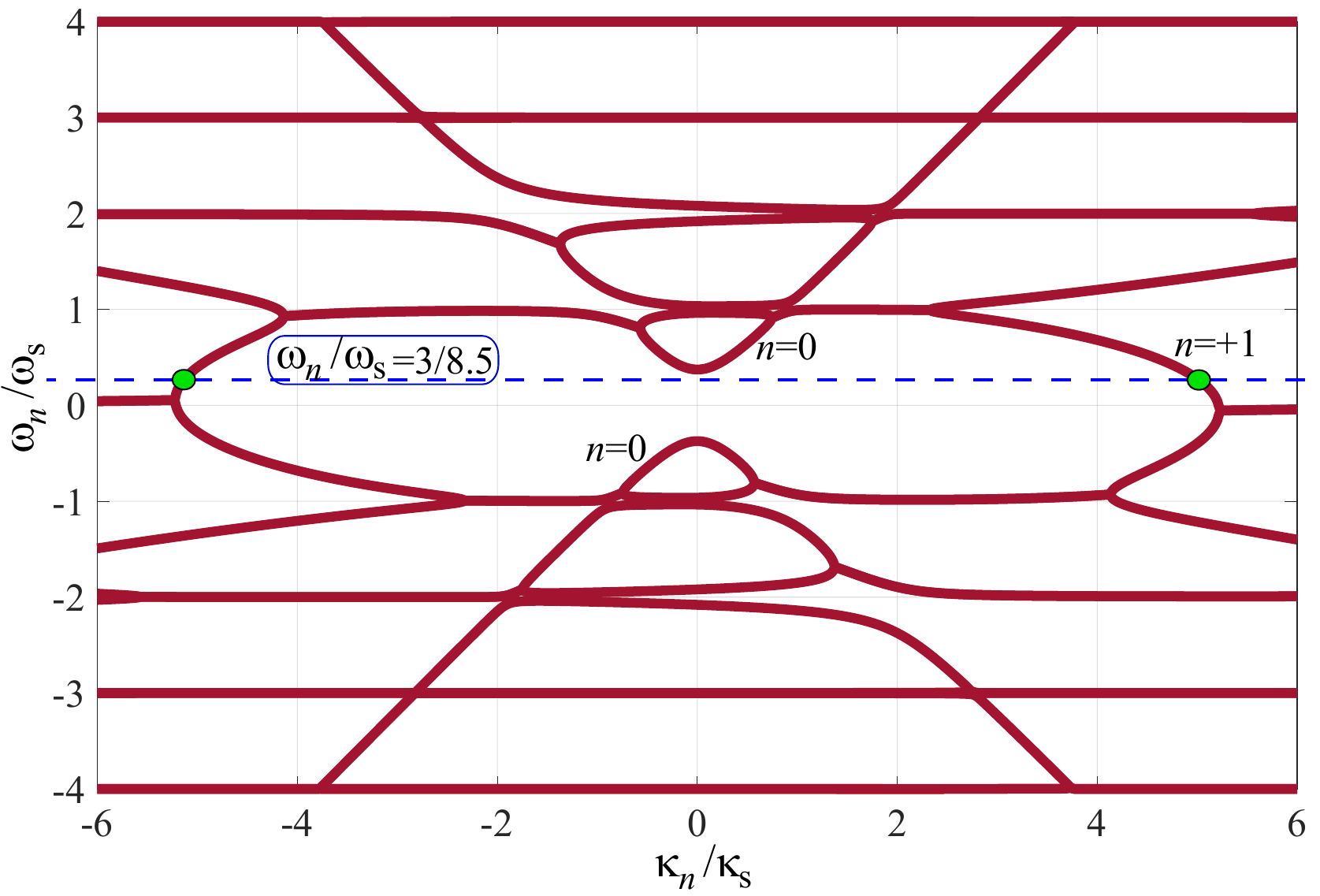}} 
		\hfil
		\subfigure[]{\label{Fig:iso_forw}
			\includegraphics[width=0.35\columnwidth]{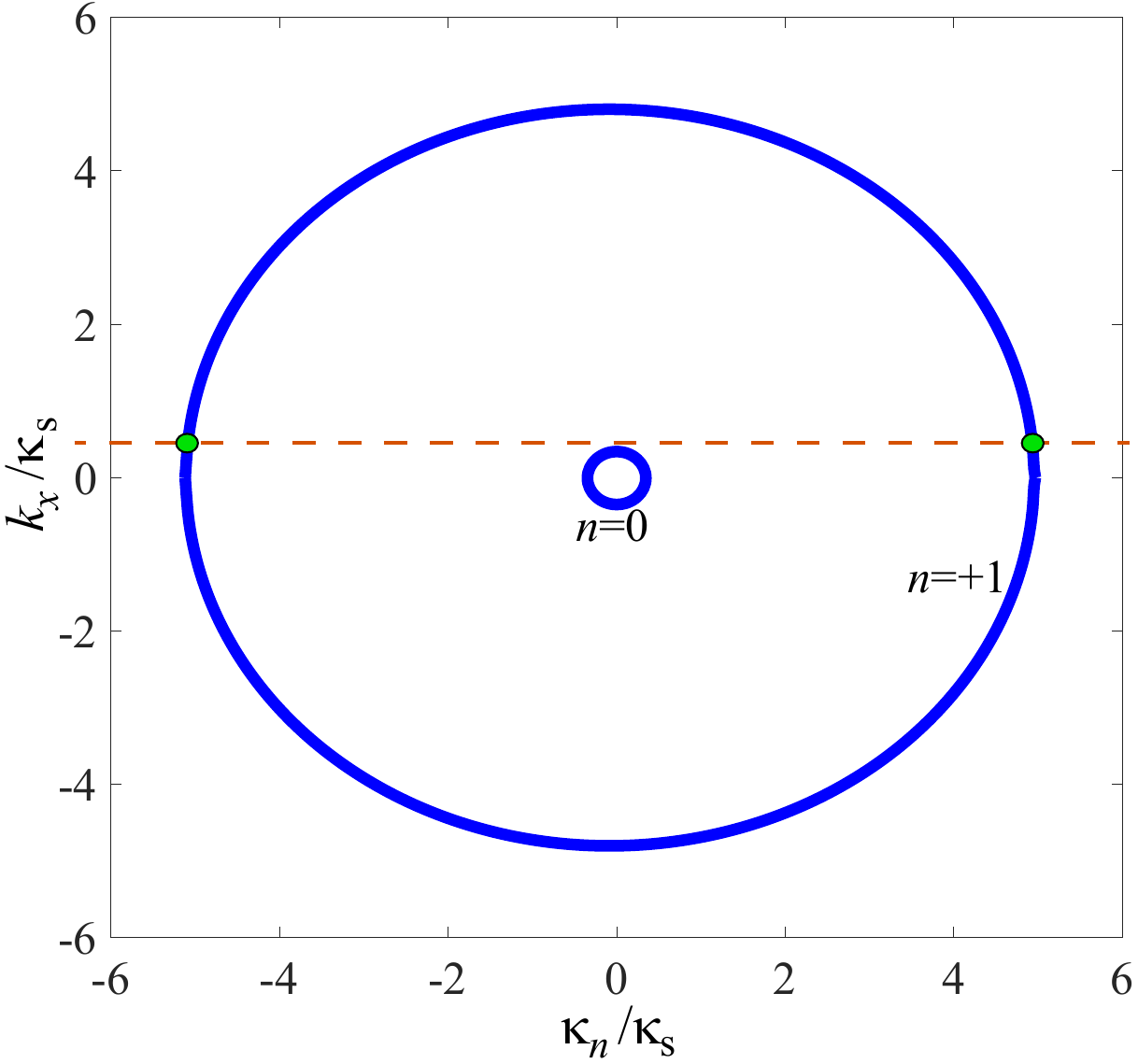}}
		\hfil		
		\subfigure[]{\label{Fig:disp_backw}
			\includegraphics[width=0.45\columnwidth]{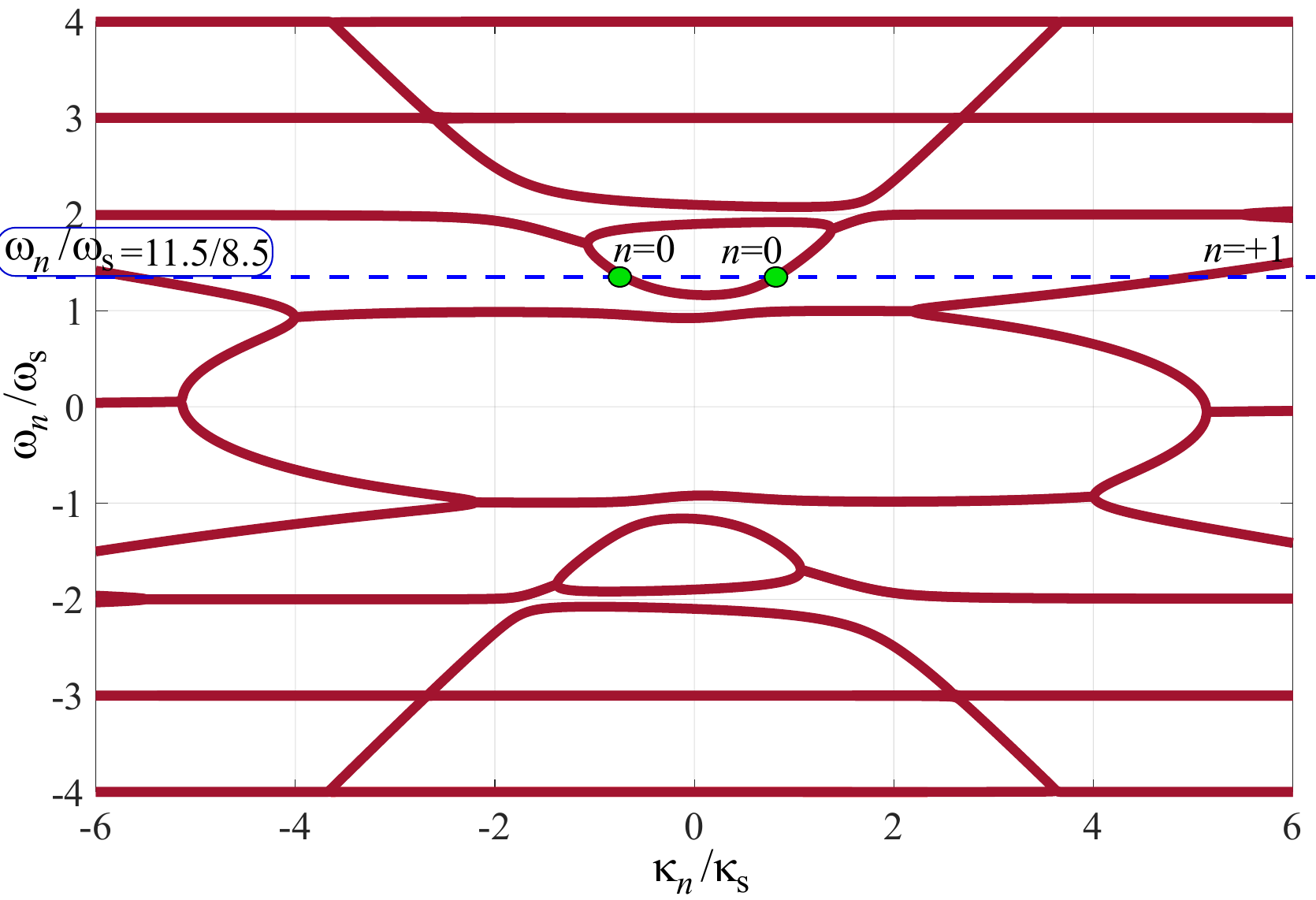}} 
		\hfil
		\subfigure[]{\label{Fig:iso_backw}
			\includegraphics[width=0.35\columnwidth]{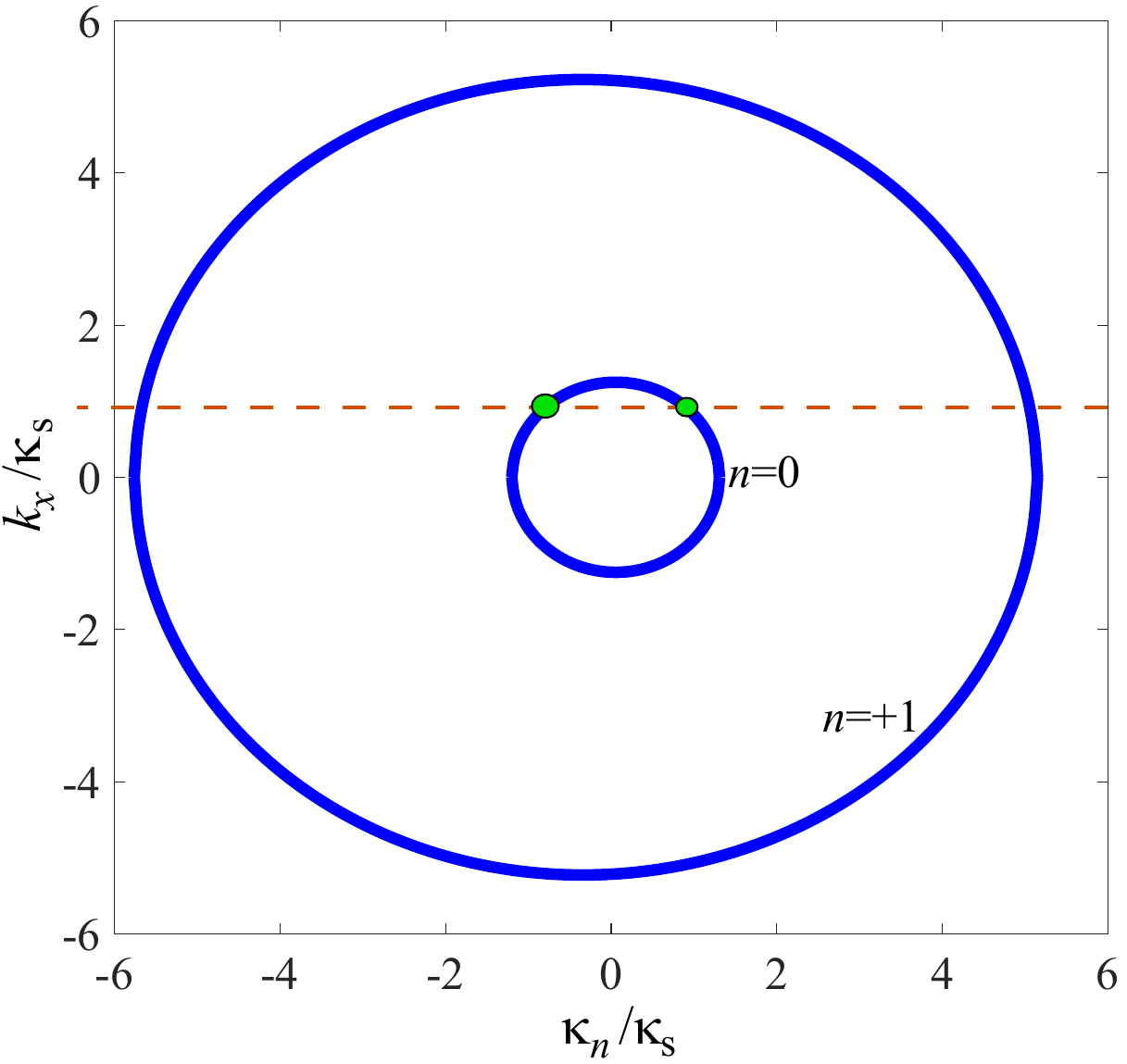}}
		\hfil
		\subfigure[]{\label{Fig:sch_ent}
			\includegraphics[width=0.95\columnwidth]{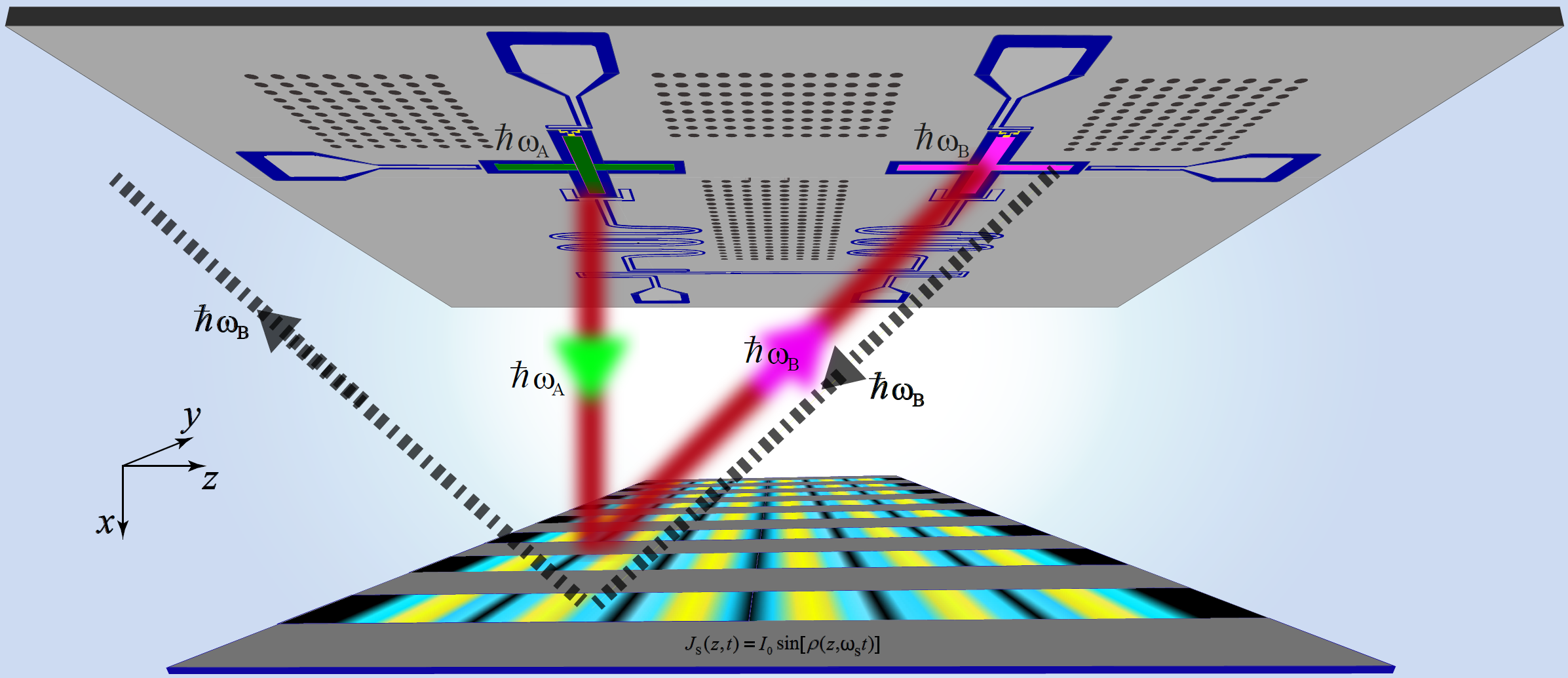}} 
		\caption{Dispersion tailoring of a spatiotemporal superconducting metasurface leading to nonreciprocal quantum state transfer between two frequency-distinct qubits~\cite{taravati2025_entangle}.
			(a) Dispersion diagram illustrating excitation of the first higher-order mode ($n=+1$) at $\omega_0/\omega_\text{s}=3/8.5$.
			(b) Corresponding isofrequency dispersion diagram.
			(c) Dispersion diagram showing excitation of the fundamental mode ($n=0$) at $\omega_0/\omega_\text{s}=11.5/8.5$.
			(d) Isofrequency dispersion diagram highlighting excitation of the fundamental mode ($n=0$).
			(e) Schematic representation of the metasurface-enabled quantum state transfer functionality~\cite{taravati2025_entangle}.} 
		\label{fig:3}
	\end{center}
\end{figure}

The core functionality of the proposed spatiotemporal metasurface in Figs.~\ref{Fig:disp_forw} to~\ref{Fig:sch_ent} hinges on sophisticated dispersion engineering~\cite{taravati2024efficient,taravati2025_entangle}, which is visualized and analyzed through band structure and isofrequency diagrams. The band structure, or dispersion diagram, is derived from the condition that the system's matrix determinant must be zero, a relationship that defines the allowed frequencies and wavevectors for wave propagation. In the forward excitation case, the input wave at a lower frequency is designed to couple exclusively to the first higher-order harmonic band (n=+1), resulting in an efficient and pure frequency up-conversion, while other harmonic paths remain inactive due to a lack of phase-matching. While the band structure shows all possible states, the isofrequency diagrams in Figs.~\ref{Fig:iso_forw} and~\ref{Fig:sch_ent} provide a complementary, operational view by slicing this structure at a constant frequency. These diagrams plot the allowed wavevectors for a single frequency, effectively showing the possible directions in which a wave at that frequency can propagate. They are instrumental in demonstrating the nonreciprocal behavior of the metasurface. For example, an isofrequency diagram for the up-converted frequency shows that a wave at this frequency, when incident from a specific angle in the backward direction, couples to a fundamentally different wavevector state (the fundamental n=0 mode) than it did during its generation in the forward direction. This asymmetry means that the path a wave takes through the system's momentum space depends entirely on its initial direction, visually illustrating the broken time-reversal symmetry that is the hallmark of nonreciprocity.

Ultimately, the synergy between these diagrams encapsulates the advanced dispersion engineering. By tailoring the spatiotemporal modulation, a synthetic material is created with a custom dispersion landscape where the pathways for energy flow are direction-dependent. This engineered control over the frequency-wavevector relationship is what enables the critical achievement: efficient, high-ratio, and spurious-free quantum state conversion in one direction only. It allows the metasurface to act not as a simple mixer, but as a directional router in the joint frequency-momentum domain, facilitating the nonreciprocal entanglement between frequency-distinct qubits that is the paper's central breakthrough.

\section{Conclusion}
The exploration of space-time metamaterials reveals that their behavior is not merely a collection of novel effects but is governed by a unified principle: the mastery of dispersion. By treating the dispersion relation as a design canvas rather than a fixed characteristic, we can sculpt wave-matter interaction with unparalleled precision. This paper has demonstrated that through deliberate spatiotemporal modulation—whether in linear systems, nonlinear Josephson arrays, or other platforms, we can open and close momentum bandgaps, steer energy into specified space-time harmonics, and break Lorentz reciprocity at a fundamental level. However, the journey from theoretical principle to robust technological application is paved with challenges. Future efforts must focus on overcoming losses, enhancing bandwidth, and developing efficient algorithms for the inverse design of ST modulation patterns that yield a desired dispersion profile. The transition to integrated systems, particularly at optical frequencies and in quantum circuits, will require novel materials and fabrication techniques. Furthermore, the exploration of non-sinusoidal and non-periodic modulation schemes promises an even richer landscape of wave control beyond the Floquet paradigm. In conclusion, dispersion engineering provides the essential language and toolkit for designing the next generation of wave-based devices. As we continue to chart the dispersion diagrams of these dynamic media, we move closer to fully unleashing their potential, paving the way for smarter radio-frequency systems, more efficient photonic processors, and fundamentally new components for quantum information science. The era of space-time metamaterials is not on the horizon, it is being built, one dispersion curve at a time.

\bibliographystyle{IEEEtran}
\bibliography{Taravati_Reference.bib}

\end{document}